\documentclass[12pt,a4paper]{article}

\usepackage{geometry}
\usepackage{amsthm}   
\usepackage{graphicx}
\usepackage{amsmath}
\usepackage{natbib}

\usepackage{amssymb}
\usepackage{psfrag,epsf}
\usepackage{enumerate}
\usepackage{hyperref}    
\usepackage{comment}    
\usepackage{bbm} 

\usepackage{booktabs}
\usepackage{multirow}
\usepackage{adjustbox}
\usepackage{makecell}
\usepackage{caption}

\usepackage[noend]{algpseudocode}
\usepackage{algorithm}
\algnewcommand{\IfThenElse}[3]{
	\State \algorithmicif\ #1\ \algorithmicthen\ #2\ \algorithmicelse\ #3}
\algnewcommand{\StateFor}[2]{
	\State  #1\ \algorithmicfor\ #2}

\usepackage{NewCommand}      

\usepackage[perpage]{footmisc} 

\title{Simultaneous false discovery proportion bounds via knockoffs and closed testing}
\author{Jinzhou Li, Marloes H. Maathuis, Jelle J. Goeman}
\date{\today}

\begin{document}

\maketitle

\begin{abstract}
	We propose new methods to obtain simultaneous false discovery proportion bounds for knockoff-based approaches.
	We first investigate an approach based on Janson and Su's $k$-familywise error rate control method and interpolation. We then generalize it by considering a collection of $k$ values, 
	and show that the bound of Katsevich and Ramdas is a special case of this method and can be uniformly improved.
	Next, we further generalize the method by using closed testing
	with a multi-weighted-sum local test statistic.
	This allows us to obtain a further uniform improvement and other generalizations over previous methods. We also develop an efficient shortcut for its implementation.
	We compare the performance of our proposed methods in simulations and apply them to a data set from the UK Biobank.
\end{abstract}

\section{Introduction}  \label{sec:intro} 

\subsection{Simultaneous inference framework}
Many modern data analysis tasks are of an exploratory nature.
In such cases, the simultaneous inference framework \citep[e.g.,][]{genovese2006exceedance, goeman2011multiple} can be more suitable than the commonly used false discovery rate (FDR) control framework \citep[e.g.,][]{benjamini1995controlling, benjamini2001control}.
Specifically, for a pre-set nominal FDR level, the output of an FDR control method is a rejection set $R$.
In order for the FDR control guarantee (see~\eqref{FDRguarantee} below) to hold,
users are not allowed to set the nominal FDR level in a post-hoc manner or to alter $R$ in any way.
For example, they cannot change to a smaller (or larger) FDR level even if the currently used one seems to return too many (or too few) rejections. Moreover, they cannot remove any rejection from $R$, not even if it violates their scientific expertise.
The simultaneous inference framework does not suffer from these issues. It allows users to freely check any set of their interest, and it returns corresponding high probability false discovery proportion (FDP) bounds (see~\eqref{simuFDPbounds} below) for the given sets.
As such, the simultaneous inference framework brings more flexibility and is very well suited for exploratory research.

Formally, let $\mN \subseteq [p] = \{1,\dots,p\}$ be the index set of true null hypotheses among $p$ null hypotheses.
The FDP of a rejection set $R\subseteq [p]$ is defined as 
\begin{align*}
	\FDP(R) = \frac{|R \cap \mN|}{ \max\{1, |R|\} },
\end{align*}
where $|\cdot|$ is the cardinality of a set. 
For a given data set and a fixed $\alpha \in (0,1)$, 
the FDR control framework aims to produce a rejection set 
$R$ such that
\begin{align} \label{FDRguarantee}
\FDR(R)  = \Exs [\FDP(R)] \leq \alpha.   
\end{align}
The simultaneous inference framework, on the other hand, aims to obtain a function 
$\barFDP: 2^{[p]} \rightarrow [0,1]$, where $2^{[p]}$ denotes the power set of $[p]$, such that
\begin{align} \label{simuFDPbounds}
\Prob( \FDP(R) \leq \barFDP(R), \forall R \subseteq [p]) \geq 1-\alpha.
\end{align}
When \eqref{simuFDPbounds} is satisfied, we say that $\barFDP(\cdot)$ is a simultaneous FDP upper bound.
Every FDP upper bound can be translated to a corresponding true discovery lower bound, and the other way around (see Appendix~\ref{supp:recapJelle2021} for more details).
In this paper we state the results in terms of FDP upper bounds. 

Existing methods for obtaining simultaneous error bounds 
are mostly based on p-values. 
For example, \cite{genovese2004stochastic}, \cite{meinshausen2006false} and \cite{hemerik2019permutation} propose 
simultaneous FDP bounds for sets of the form $\whR_t = \{i \in [p]: p_i \leq t \}$, where $p_i$ is the p-value for the $i$-th hypothesis, while
\cite{genovese2006exceedance}, \cite{goeman2011multiple}, \cite{goeman2019simultaneous}, \cite{blanchard2020post} and \cite{vesely2021permutation}
propose simultaneous FDP bounds for arbitrary sets $R \subseteq [p]$.

\subsection{Knockoff framework}

For error-controlled variable selection problems, \cite{barber2015controlling} and \cite{candes2018panning} introduced a so-called knockoff framework that does not resort to p-values.
The fundamental idea of the knockoff method is to create synthetic variables (i.e., knockoffs) that serve as negative controls.
By using the knockoffs together with the original variables as input of an algorithm (e.g., lasso, random forests, or neural networks),
one obtains knockoff statistics $W_1, \dots, W_p$, which possess a 
so-called coin-flip property (see Definition~\ref{def:coin-flip}).
Based on this property, variable selection methods were developed to control the FDR \citep{barber2015controlling}
or the $k$-familywise error rate ($k$-FWER, \cite{LJ-WS:2016}).

The knockoff method has been extended to many settings, 
including group variable selection  \citep{dai2016knockoff}, multilayer variable selection \citep{katsevich2019multilayer}, 
high-dimensional linear models \citep{barber2016knockoff},
Gaussian graphical models \citep{li2021ggm}, multi-environment settings \citep{li2021searching}, and time series settings \citep{chi2021high}.
The knockoff approach can be especially useful for high-dimensional settings, where obtaining valid p-values is very challenging.
For example, it has been successfully applied to real applications with high-dimensional data \citep[e.g.,][]{sesia2020multi, He2021.12.06.471440, Sesiae2105841118}.

Most existing knockoff-based methods aim to control the FDR.
Motivated by the advantages of simultaneous FDP bounds and the usefulness of knockoff methods, we consider the problem of obtaining such bounds for knockoff-based approaches. 
We focus on the variable selection setting, but the proposed methods can be applied to other settings as long as the coin-flip property (see Definition~\ref{def:coin-flip}) holds.
To the best of our knowledge, only \cite{katsevich2020simultaneous} investigated this problem so far.
They actually proposed a general approach to obtain simultaneous FDP bounds for a broad range of FDR control procedures. As a special case, they obtain simultaneous FDP bounds for knockoff-based methods.

\subsection{Outline of the paper and our contributions}

We start by investigating a simple approach 
(see Section~\ref{sec:simpleApproach}) 
that obtains simultaneous FDP bounds based on a $k$-FWER controlled set \citep{LJ-WS:2016} and interpolation \citep{blanchard2020post, goeman2021only}.
We found that there are two issues with this approach: (i)
it is unclear how to choose the tuning parameter $k$, which can heavily affect the results, 
and (ii) the FDP bounds for certain sets can be very conservative due to the nature of interpolation.

These issues motivated us to consider several values of $k$ at the same time.
Specifically, we first get a collection of sets $\whS_1, \dots, \whS_m$, such that the $k_i$-FWER of set $\whS_i$ is controlled jointly for all $i\in [m]$. Next, we obtain simultaneous FDP bounds by using interpolation based on these sets.
This idea is a special case of a more general procedure called joint familywise error rate control \citep{blanchard2020post}. Related ideas can be traced back to \cite{genovese2006exceedance} and \cite{van2004augmentation}.

Our proposed method (see Section~\ref{sec:JKI}) 
is uniformly better than the above simple approach using only one $k$-FWER controlled set,
and it largely alleviates the above two issues.
Issue (i) regarding the choice of tuning parameters is not fully resolved,
in the sense that one still needs to choose a sequence of $k$ values, for which the optimal choice depends on the data distribution.
We therefore suggest some reasonable choices and also propose a two-step approach to obtain the tuning parameters.
In addition, we prove that the simultaneous FDP bound from \cite{katsevich2020simultaneous} is a special case of our method, and uniformly better bounds can be obtained.

Next, in Section~\ref{sec:closedtesting}, we turn to the closed testing framework.
In particular, we propose a multi-weighted-sum local test statistic based on knockoff statistics.
We prove that all previously mentioned approaches are special cases (or shortcuts) of this method,
and we derive uniform improvements and generalizations over them. 
The closed testing procedure is computationally intractable in its standard form, so we also develop an efficient and general shortcut for its implementation.

In Section~\ref{sec:simu}, we evaluate the numerical performance of the proposed methods in simulation studies and apply them to a data set from the UK Biobank.
We close the paper with a discussion and potential future research directions in Section \ref{sec:disc}.

\section{Preliminaries} \label{sec:preliminaries}

Throughout the paper, vectors are denoted by boldface fonts, and we consider $\alpha \in (0,1)$, $m \geq 1$, $\bv=(v_1, \dots, v_m)$ with $0<v_1<\dots<v_m$, and $\bk = (k_1, \dots, k_m)$ with $0<k_1 \leq \dots \leq k_m$.


\subsection{Knockoff framework and the coin-flip property} \label{sec:reviewKnockoffStat}

We first review the knockoff framework for variable selection.
That is, we consider a response $Y$ and covariates $\bX=(X_1,\dots,X_p)$. 
We say that $X_i$ is a null variable if $Y \independent X_i | \bX_{-i}$, where $\bX_{-i} = \{X_1, \dots,X_p\}\setminus\{X_i\}$,
and the goal is to find non-null variables with error control. 

Knockoff methods yield knockoff statistics $\bW = (W_1, \dots, W_p)$ based on samples from $X$ and $Y$, where non-null variables tend to have large and positive $W_i$ values. Hence, one can select variables by choosing
$\{i\in [p]: W_i > T\}$ for some threshold $T$. In addition, knockoff methods are designed so that the following coin-flip property holds \citep{barber2015controlling, candes2018panning}.
\begin{definition}\label{def:coin-flip}
	For knockoff statistic vector $\bW$, let $D_i = \text{sign}(W_i)$.
	Let $\mN \subseteq [p]$ be the index set of null variables.
	We say that $\bW$ possesses the \textbf{coin-flip property} if conditional on $(|W_1|, |W_2|, \dots, |W_p|)$, the
	$D_i$'s are i.i.d.\ with distribution $\Prob(D_i=1) = \Prob(D_i=-1) = 1/2$ for $i \in \mN$.
\end{definition}

The coin-flip property is key to obtain valid error control in the knockoff framework. For p-value based approaches, the analogous property is that a null p-value is stochastically larger than a uniform random variable on $[0,1]$.  

In this paper, we do not specify the data generating model or the knockoff method.
Instead, we work directly on the knockoff statistic vector $\bW$, which is assumed to possess the coin-flip property.
Our methods can be applied as long as this assumption holds.
For example, the same methodology can be directly extended to the group variable selection setting, where the coin-flip property holds at the group level \citep{dai2016knockoff}.

Throughout, we assume without loss of generality that $|W_1| > |W_2| > \dots > |W_p| > 0$.
Otherwise, one can pre-process $\bW$ by relabeling the indices, discarding the zero ones, and breaking ties. 
These pre-processing steps have no influence on the remaining part of the paper because the pre-processed $\bW$ still possesses the coin-flip property.

From a more abstract point of view, the considered setting can be reformulated as a ``pre-ordered" case as discussed in \cite{katsevich2020simultaneous}.
In this scenario, we are given a sequence of ordered hypotheses $H_1, \dots, H_p$, 
each associated with a test statistic $D_i$. These test statistics possess the property that the null $D_i$'s are independent coin flips. 
The goal is to obtain simultaneous FDP bound \eqref{simuFDPbounds}.

\subsection{$k$-FWER control via knockoffs} \label{sec:reviewJS}

We now review the $k$-FWER control method using knockoffs by \cite{LJ-WS:2016}. This method is a crucial component of our starting point for the interpolation-based method proposed in Section~\ref{sec:simpleApproach}.

Specifically, for $v\geq 1$, \cite{LJ-WS:2016} defined the discovery set
\begin{align} \label{setS}
	\whS(v) = \{i \in [p]: W_i \geq \whT(v) \},
\end{align}
where the threshold
\begin{align}\label{threshold-kfwer}
\whT(v) = \max \{ |W_i|: |\{j \in [p]: |W_j|\geq |W_i| \text{ and } W_j < 0 \}| =v \}
\end{align}
with $\whT(v) = \min_{i \in [p]} \{ |W_i|\}$ if $|\{j \in [p]: W_j < 0 \}| <v$ (please note that $|\cdot|$ is used to denote absolute value for numbers and cardinality for sets).
For given $k \geq 1$ and $\alpha \in (0,1)$, they proposed 
\begin{equation}\label{v:OPT-kfwer}
\begin{aligned} 
v^{\JS}(k) 
&= \argmax \Bigg\{v \in [p]: \sum_{i=k}^{\infty} 2^{-i-v} \begin{pmatrix} i+v-1 \\ i \end{pmatrix} \leq \alpha \Bigg\},
\end{aligned}
\end{equation}
and their discovery set is then $\whS^{\JS}(k) = \whS \left( v^{\JS}(k) \right)$.

Equation~\eqref{v:OPT-kfwer} is based on the fact that $|\whS(v) \cap \mN|$ is stochastically dominated by a negative binomial random variable $N(v)$ with distribution NB$(v,1/2)$ \citep{LJ-WS:2016}. 
This implies that the $k$-FWER of $\whS^{\JS}(k)$ is controlled, since
\begin{equation} \label{KFWER-nonstrict}
\Prob \left( |\whS^{\JS}(k) \cap \mN| \leq k-1 \right) 
\geq \Prob \left( N \left(v^{\JS}(k) \right) \leq k-1 \right) \geq 1-\alpha.
\end{equation}

\section{Simultaneous FDP bounds via knockoffs and joint $k$-FWER control with interpolation} \label{sec:interpolation}

\subsection{$K$-FWER control with interpolation} \label{sec:simpleApproach}
We now combine the $k$-FWER controlled set from \cite{LJ-WS:2016} with interpolation.
Here interpolation means that, for a $k$-FWER controlled set $\whS(v)$ and arbitrary sets $R$, we can bound $|R\cap \mN|$ based on $\whS(v)$.
In particular, when $|\whS(v) \cap \mN| \leq k-1$, we have for any $R \subseteq [p]$,
\begin{equation}\label{FDbound-interpolation}
\begin{aligned}
|R \cap \mN|
&= |R \cap \whS(v) \cap \mN| + | (R \backslash \whS(v)) \cap \mN| \\
&\leq \min \{|R|, |\whS(v) \cap \mN| + |R \backslash \whS(v)| \} \\
&\leq \min \{ |R|, k-1 + |R \backslash \whS(v)| \}.
\end{aligned}
\end{equation}
Hence
\begin{align*}
 P(|R\cap \mN|\le \min\{|R|, k-1+|R \backslash \whS(v)|\}, \forall  R \subseteq [p])
 \geq P(|\whS(v) \cap \mN| \le k-1)
 \geq 1-\alpha.
\end{align*}
By dividing by $\max \{1, |R| \}$ on both sides within the above probability,
we obtain the following simultaneous FDP bound
\begin{align}\label{FDPbound-KFWER}
\barFDP_{k,v}(R) = \frac{ \min \{ |R|, k-1 + |R \backslash \whS(v)| \} }{\max \{1, |R| \}}, 
\quad R\subseteq [p].
\end{align}

Interpolation is actually a more general technique \citep{blanchard2020post, goeman2021only}, which reduces to the simplified form \eqref{FDbound-interpolation} in our special case (see also Appendix~\ref{supp:LemmaInterpolation}).

If $|\whS(v)| < k-1$, one can always obtain a larger discovery set $\whS'(v)$ with $|\whS'(v)|=k-1$ by adding the variables corresponding to the next largest positive $W_i$'s, while maintaining $k$-FWER control.
One might expect that the simultaneous FDP bound~\eqref{FDPbound-KFWER} based on $\whS'(v)$ might be better (i.e., smaller).
However, because $k-1 + |R \backslash S| \geq k-1 + |R| - |S| \geq |R|$ for any set $S$ with $|S| \leq k-1$, 
the FDP bound~\eqref{FDPbound-KFWER} based on $\whS'(v)$ and $\whS(v)$ are both equal to $1$, meaning that only trivial FDP bound can be obtained by interpolation when the base set is not large enough.
We will therefore use $\whS(v)$ for simplicity.

By using the $k$-FWER controlled set $\whS^{\JS}(k)$ from \cite{LJ-WS:2016} in \eqref{FDPbound-KFWER}, we obtain the following simultaneous FDP bound:
\begin{align}\label{FDPbound-KFWER-JS}
\barFDP^{\JS}_{k}(R) = \frac{ \min \{ |R|, k-1 + |R \backslash \whS^{\JS}(k)| \} } {\max \{1, |R| \}}, 
\quad R\subseteq [p].
\end{align}

\begin{figure}[h!]
	\centering
	\includegraphics[width=10cm, height=10cm]{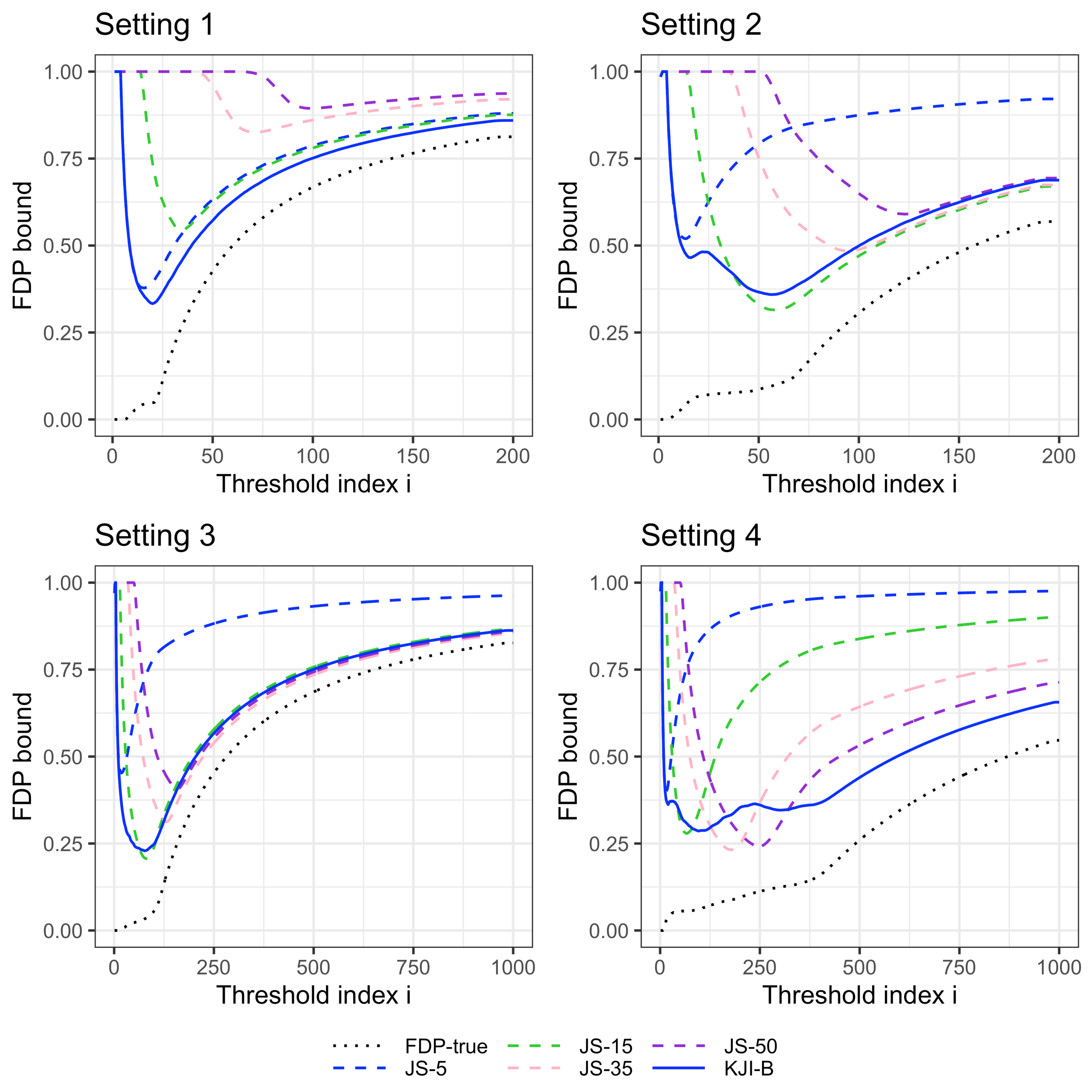}
	\caption{Simultaneous FDP bound $\barFDP^{\JS}_{k}(\cdot)$ with $k \in \{5,15,35,50\}$ (denoted by JS-5, JS-15, JS-35, and JS-50, respectively) and our later proposed Algorithm~\ref{Algo-BasedOnSimuKFWER} with the Type-B $\bv$ described in \eqref{4-types-of-v} and $\bk$ obtained by using Algorithm~\ref{Algo-getTuningParameterK} (denoted by KJI-B). The black dotted line indicates the true FDP. All results are average values over $200$ replications.}
	\label{Fig:illustration-Algo2-vary-k}
\end{figure}

As mentioned in the introduction, the simultaneous FDP bound $\barFDP^{\JS}_{k}(\cdot)$ 
has two issues:
(i) it is not clear a priori how to choose the tuning parameter $k$, and this choice can heavily affect the bound, and 
(ii) the bound can be very conservative for sets $R$ that are very different from $\whS^{\JS}(k)$.
To illustrate these two issues, 
we present Figure~\ref{Fig:illustration-Algo2-vary-k} showing the simultaneous FDP bound $\barFDP^{\JS}_{k}(\cdot)$ for sets $\whR_i = \{j \leq i: W_j > 0 \}$, $i \in [p]$, in four simulation settings (see Appendix~\ref{supp:simu-detail-Fig1} for more details). 
To see issue (ii), note that the FDP bounds strongly depend on the threshold index $i$, and their minima are reached around $i$ such that $\whR_i$ equals $\whS^{\JS}(k)$.

\subsection{Joint $k$-FWER control with interpolation}\label{sec:JKI}
The above two issues motivate the following new method. 
The idea is that since FDP bounds based on different $k$ can be better in different settings, we use many $k$-FWER controlled sets to construct them. As we mentioned in Section~\ref{sec:intro}, this idea is a special case of a more general method proposed by \cite{blanchard2020post}.

Formally, for $\whS(v_i)$ defined by formula~\eqref{setS}, 
we say that joint $\bk$-FWER control holds if
\begin{equation} \label{Joint-KFWER-guarantee}
\Prob\left( |\whS(v_i) \cap \mN| \leq k_i-1, \forall i \in [m]\right) 
\geq 1-\alpha.
\end{equation}
We then obtain a simultaneous FDP bound
\begin{align} \label{FDPbound-jointKFWER}
\barFDP^m_{(\bk,\bv)}(R) = \min_{i\in [m] } \barFDP_{k_i,v_i}(R),
\quad  R \subseteq [p],
\end{align}
where $\barFDP_{k_i,v_i}(R)$ is defined by \eqref{FDPbound-KFWER}.
$\barFDP^m_{(\bk,\bv)}(\cdot)$ is a valid simultaneous FDP bound because
\begin{align*}
\Prob\left( \FDP(R) \leq \barFDP^m_{(\bk,\bv)}(R), \forall R \subseteq [p]\right)
\geq \Prob\left( |\whS(v_i) \cap \mN| \leq k_i-1, \forall i \in [m]\right)
\geq 1-\alpha.
\end{align*}

The previous FDP bound $\barFDP^{\JS}_{k}(\cdot)$ is a special case of  $\barFDP^m_{(\bk,\bv)}(\cdot)$ with $m=1$, $k_1=k$ and $v_1=v^{\JS}(k)$. 
That is, $\barFDP^{\JS}_{k}(\cdot) = \barFDP_{k, v^{\JS}(k)}(\cdot)$.
By the definition~\eqref{FDPbound-jointKFWER}, 
we can see that $\barFDP^m_{(\bk,\bv)}(\cdot)$ with $m>1$, $k_1=k$ and $v_1=v^{\JS}(k)$ is uniformly better (i.e., smaller or equal) than $\barFDP^{\JS}_{k}(\cdot)$.


To obtain vectors $\bv$ and $\bk$ satisfying the joint $\bk$-FWER control guarantee~\eqref{Joint-KFWER-guarantee},
we generalize ideas from \cite{LJ-WS:2016}.
In particular, we use the following lemma:
\begin{lemma} \label{lemma-NB}
	Let $\mN \subseteq [p]$ be the set of null variables.
	Then, for positive integers $v_1< \dots< v_m$ and $k_1< \dots< k_m$, we have
	\begin{equation}\label{boundNB} 
		\begin{aligned}
			\Prob( |\whS(v_i) \cap \mN| \leq k_i-1, \forall i \in [m])
			\geq \Prob( N^{p}(v_i) \leq k_i-1, \forall i \in [m]),
		\end{aligned}
	\end{equation}
	where $N^{p}(v_1), \dots, N^{p}(v_m)$ are early-stopped negative binomial random variables defined on one sequence of Bernoulli trials of length $p$.
\end{lemma}
Here, an early-stopped negative binomial random variable $N^{p}(v)$ is formally defined as  $N^{p}(v) = |\{j\leq \min(U(v),p): B_j = 1 \}|$, where $U(v) = \min \{ i \geq 1: |\{j\leq i: B_j = -1 \}| = v \}$ and $B_1, B_2, \dots$ are i.i.d.\ Bernoulli random variables taking values $\{+1,-1\}$.

Based on Lemma~\ref{lemma-NB}, if vectors $\bv$ and $\bk$ satisfy
\begin{align} \label{NB-inequality}
\Prob( N^{p}(v_i) \leq k_i-1, \forall i \in [m]) \geq 1-\alpha,
\end{align}
then the joint $\bk$-FWER control guarantee~\eqref{Joint-KFWER-guarantee} holds, and 
$\barFDP^m_{(\bk,\bv)}(\cdot)$ is a simultaneous FDP bound.

In the $k$-FWER control case of \cite{LJ-WS:2016}, we have
$p=\infty$, $m=1$ and $\bk=k$ is user-chosen in inequality~\eqref{NB-inequality}. The optimal $v$ can then be naturally obtained via optimization~\eqref{v:OPT-kfwer}, because it is clear that the largest $v$ leads to the most discoveries.
In our case of getting simultaneous FDP bounds with $m>1$, 
things are more complicated. 
We will discuss the choice of $\bv$ and $\bk$ in more detail in next subsection.

For clarity, we present the simultaneous FDP bound $\barFDP^m_{(\bk,\bv)}(\cdot)$ in Algorithm~\ref{Algo-BasedOnSimuKFWER}.
As a quick illustration, we present the performance of this algorithm
in Figure~\ref{Fig:illustration-Algo2-vary-k}, where the tuning parameter $\bv$ is taken as the Type-B $\bv$ in \eqref{4-types-of-v}, and $\bk$ is obtained by our later proposed Algorithm~\ref{Algo-getTuningParameterK}.
One can see that our proposed method
generally performs well in all four settings.
\begin{algorithm}[h]
	\caption[] {\textbf{: Simultaneous FDP bounds via knockoffs and joint $k$-FWER control with interpolation} } \label{Algo-BasedOnSimuKFWER}
	\textbf{Input}:  $(R, \bW)$, where $R\subseteq [p]$ is the set of interest, $\bW \in \bbR^p$ is the vector of knockoff statistics.
	
	\textbf{Parameter}: $(\alpha, \bv, \bk)$, where $\alpha$ is the nominal level for the simultaneous FDP control,
	$\bv$ and $\bk$ are vectors satisfying inequality~\eqref{NB-inequality}.  
	
	\textbf{Output}: FDP upper bound for $R$, which holds simultaneously for all $R \subseteq [p]$.
	\begin{algorithmic}[1]
		\State For each $i \in [m]$, obtain
  \begin{align*}
\barFDP_{k_i,v_i}(R) = \frac{ \min \{ |R|, k_i-1 + |R \backslash \whS(v_i)| \} }{\max \{1, |R| \}},
\end{align*}
  where $\whS(v_i) = \{ j \in [p]: W_j \geq \whT(v_i) \}$ with
  $\whT(v_i) = \max \{ |W_j|: |\{l \in [p]: |W_l|\geq |W_j| \text{ and } W_l < 0 \}| = v_i \}$ and $\whT(v_i) = \min_{j \in [p]} \{ |W_j|\}$ if $|\{j \in [p]: W_j < 0 \}| <v_i$.
		\State Return $\min_{i\in [m] } \barFDP_{k_i,v_i}(R)$.
	\end{algorithmic}
\end{algorithm}

\subsection{Tuning parameters $\bv$ and $\bk$ and uniform improvement of \cite{katsevich2020simultaneous}} \label{sec:TuningParameters}
The vectors $\bv$ and $\bk$ satisfying inequality~\eqref{NB-inequality} are not unique, and the best $\bv$ and $\bk$ for the FDP bound $\barFDP^m_{(\bk,\bv)}(\cdot)$ depend on the unknown distribution of $\bW$.
In this paper, we first choose $\bv$, then obtain $\bk$ such that \eqref{NB-inequality} holds.
In particular, we consider the following four types of $\bv$ vectors (ordered in a way that $v_i$ grows ever faster):
\begin{equation}\label{4-types-of-v}
	\begin{aligned} 
		&\text{Type-A: } \ v_i = i,
		\qquad \qquad \qquad \qquad \qquad \qquad \ \
		\text{Type-B: }  \ v_i = \lfloor i^2/2 \rfloor, \\
		&\text{Type-C: }  \ v_1 = 1, v_2=2, v_i=v_{i-1}+v_{i-2},
		\quad \quad
		\text{Type-D: }  \ v_i = 2^{i-1}.
	\end{aligned}
\end{equation}
We always use $v_1=1$ because it gives the best FDP bounds when the obtained knockoff statistic vector is optimal, namely, when all true $W_i$'s are positive and have larger absolute values than any null $W_i$.

For a given $\bv$, one may obtain $\bk$ numerically using a greedy approach with the constraint that inequality~\eqref{NB-inequality} holds (e.g., based on Monte-Carlo simulations).
We will follow this idea and proceed by connecting to, then improve upon, the only existing simultaneous FDP bound of \cite{katsevich2020simultaneous}.

The interpolation version of the FDP bound presented in \cite{katsevich2020simultaneous} is:
\begin{align} \label{KR-FDPbound-interpolation}
\barFDP^{\KR}(R) = \frac{ \min_{i \in [p]} \{|R|, |R \backslash \whS_i | + \lfloor c(\alpha) \cdot(1+ i - |\whS_i| ) \rfloor \} } {\max\{1, |R|\}},
\end{align}
where $\whS_i = \{  j\leq i: W_j > 0 \}$ and $c(\alpha) = \frac{\log(\alpha^{-1})}{\log(2-\alpha)}$. 
\cite{katsevich2020simultaneous} mentioned that one can apply interpolation to their original bound to obtain a better bound, but they did not present the formula.
We give the explicit formula in~\eqref{KR-FDPbound-interpolation} to ensure that our new methods are compared with the best version of \cite{katsevich2020simultaneous}, i.e., the version with interpolation.
The derivation of~\eqref{KR-FDPbound-interpolation} can be found in Appendix~\ref{supp:KR-Interpolation}.


For a given $\bv$,
we first derive a valid $\bk$ based on $\barFDP^{\KR}(\cdot)$ and Theorem~\ref{Thm1:Jelle2021} (Theorem 1 in \cite{goeman2021only}, see Appendix~\ref{supp:recapJelle2021}).
Specifically, let
\begin{align} \label{cj-KR}
	k^{raw}_{v_i}= \min_{j\geq 1} \{c_j: j-c_j+1=v_i\} 
	\quad \text{with} \quad
	c_j = \left\lfloor \frac{c(\alpha) \cdot (1+ j )}{1+c(\alpha)} \right\rfloor  + 1,
\end{align}
where $\lfloor x \rfloor$ denotes the largest integer smaller than or equal to $x$ and $c(\alpha) = \frac{\log(\alpha^{-1})}{\log(2-\alpha)}$. 
Then, as the following Proposition~\ref{prop:KR-NB} shows, $\bk^{raw}=(k^{raw}_{v_1}, \dots, k^{raw}_{v_m})$ is valid.
\begin{proposition} \label{prop:KR-NB}
	Let $N^{p}(v_1), \dots, N^{p}(v_m)$ be early-stopped negative binomial random variables defined on one sequence of Bernoulli trials of length $p$. Then
		\begin{align}\label{NB-inequality-KR}
		\Prob \left(N^{p}(v_i) \leq k^{raw}_{v_i}-1, \forall i \in [m]\right) \geq 1-\alpha.
		\end{align}
\end{proposition}

By comparing formulas \eqref{KR-FDPbound-interpolation} and \eqref{FDPbound-jointKFWER}, the relationship between simultaneous FDP bounds $\barFDP^{\KR}(\cdot)$ and $\barFDP^m_{(\bk,\bv)}(\cdot)$ is not immediately clear. Perhaps surprisingly, as we show in the following Proposition~\ref{prop:KR-Algo1}, $\barFDP^{\KR}(\cdot)$ 
is in fact a special case of  $\barFDP^m_{(\bk,\bv)}(\cdot)$.
\begin{proposition} \label{prop:KR-Algo1}
	For $\bv=(1,\dots,p)$ and $\bk^{raw}=(k^{raw}_1,\dots,k^{raw}_p)$, we have
		\begin{align*}
		\barFDP^{\KR}(R) = \barFDP^p_{(\bk^{raw},\bv)}(R), \quad R\subseteq [p].
		\end{align*}
\end{proposition}

One value of the above result is that it unifies $\barFDP^{\KR}(\cdot)$, the only existing method for obtaining simultaneous FDP bounds for knockoff approaches, within our method.
  More importantly, it enables us to uniformly improve $\barFDP^{\KR}(\cdot)$. 
  Specifically, for a vector $\bv=(v_1,\dots,v_m)$, using $\bk^{raw}=(k^{raw}_{v_1}, \dots, k^{raw}_{v_m})$ to obtain $\barFDP^p_{(k^{raw},v)}(\cdot)$ is not optimal, that is, $\Prob \left(N^{p}(v_i) \leq k^{raw}_{v_i}-1, \forall i \in [m]\right)$ and $1-\alpha$ are not close,
so that the $\alpha$ level is not exhausted.


To improve this, we propose a two-step approach.
Let $j^*_i = \argmin_{j\geq 1} \{c_j: j-c_j+1=v_i\} $.
In the first step, we refine $k_{v_i}^{raw}$ by replacing $c(\alpha)$ in $c_{j^*_i}$  with the smallest constant such that inequality \eqref{NB-inequality} holds.
In most cases, however, the first step will still not exhaust the $\alpha$ level due to the discrete nature of the target probability.
Hence in the second step, we greedily update $k_i$ under the constraint~\eqref{NB-inequality}.
Here we choose to update $k_i$ in an increasing order of $i$ because the probability discretization is coarser for smaller $v_i$ and finer for larger $v_i$,
so updating in such an order is helpful to exhaust the $\alpha$ level
(see also \cite{blanchard2020post} for a closely related principle called $\lambda$-calibration).
In the practical implementation, we update $c(\alpha)$ in a discrete manner in the first step, and we use Monte-Carlo simulation to approximate the target probability $\Prob(N^{p}(v_1) \leq k_1-1, \dots, N^{p}(v_m) \leq k_m-1)$ because an exact calculation is computationally infeasible for large $m$. We summarize this two-step approach in Algorithm~\ref{Algo-getTuningParameterK}.

\begin{algorithm}[h]
	\caption[] {\textbf{: Two-step approach to obtain $k_1, \dots, k_m$} } \label{Algo-getTuningParameterK}
	\textbf{Input}:  $(v_1,\dots,v_m, \alpha, \delta, p)$, where $v_1 < \dots < v_m$, $\alpha \in (0,1)$ is the nominal level for the simultaneous FDP control, $\delta>0$ is the step size used in the first step, and $p$ is the parameter in inequality~\eqref{NB-inequality}.
	
	\textbf{Output}: $k_1, \dots, k_m$.
	\begin{algorithmic}[1]
		\State 
		For each $i \in [m]$, obtain $j^*_i = \argmin_{j\geq 1} \{c_j: j-c_j+1=v_i\} $, where
		$c_j$ is defined by \eqref{cj-KR}. 
		Refine $c(\alpha)$ by using step size $\delta$: find the largest $N \geq 0$ such that
		inequality~\eqref{NB-inequality} holds with $(v_1,\dots,v_m)$ and $(k_1^{step1}, \dots, k_m^{step1})$,
		where $k^{step1}_i = \left\lfloor \frac{c^{step1} \cdot (1+ j^*_i )}{1+c^{step1}} \right\rfloor  + 1$ and $c^{step1} = c(\alpha) - N\delta$.
		\State Update $k^{step1}_1, \dots, k^{step1}_m$ in a greedy way: find the smallest $k_1^{step2} \leq k_1^{step1}$ such that 
		inequality~\eqref{NB-inequality} holds with $(v_1,\dots,v_m)$ and $(k_1^{step2}, k_2^{step1},\dots, k_m^{step1})$. Then keep $k^{step2}_1$ and find the smallest $k^{step2}_2 \leq k_2^{step1}$ such that
		inequality~\eqref{NB-inequality} holds with $(v_1,\dots,v_m)$ and $(k_1^{step2}, k_2^{step2},\dots, k_m^{step1})$. Iterate this process until we get all $k^{step2}_1,\dots,k^{step2}_m$, and return them as output.
	\end{algorithmic}
\end{algorithm}

To illustrate the improvement of using Algorithm~\ref{Algo-getTuningParameterK} over $\bk^{raw}$, 
we present the corresponding vectors $\bk^{raw}$, $\bk^{step1}$ and $\bk^{step2}$ based on $\bv=(1,\dots,p)$ in the plot (1) of Figure~\ref{Fig:compare-4-vk} (see Appendix~\ref{supp:simu-Algo2-4-v} for the plots of other types of $\bv$).
One can see that $k_i^{step2}$ is smaller than $k_i^{step1}$ and $k_i^{raw}$ for the same $v_i$, and hence it will lead to better FDP bounds.
We also present the corresponding target probabilities for the four types of $\bv$ in Table~\ref{tab:alpha-level}. One can see that $\bk^{step2}$ exhausted the $\alpha$-level (up to numerical precision) while $\bk^{step1}$ and $\bk^{raw}$ didn't.
\begin{center}
	\begin{tabular}[t!]{l|c|c|c|c}
		\toprule
	$\bv$ & Type-A & Type-B & Type-C & Type-D \\ 
		\hline 
		$\bk^{raw}$ & 0.9632 (0.0006) & 0.9639 (0.0006) & 0.9633 (0.0005) & 0.9635 (0.0007)  \\
		\hline
		$\bk^{step1}$ & 0.9560 (0.0007) & 0.9585 (0.0006) & 0.9568 (0.0006) & 0.9574 (0.0007) \\ 
		\hline
		$\bk^{step2}$ &  0.9500 (0) & 0.9500 (0)  & 0.9500 (0)  & 0.9500 (0) \\ \bottomrule
	\end{tabular}
	\captionof{table}{The empirical target probabilities $\Prob(N^{p}(v_i) \leq k_i-1, \forall i \in [m])$ for $\bk^{raw}$, $\bk^{step1}$ and $\bk^{step2}$ based on four types of $\bv$, with standard deviations in parentheses. Here we use $p=1000$ and let $v_i < 150$. The results are based on $100$ replications.}\label{tab:alpha-level}
\end{center}

Different types of $\bv$ vectors lead to different reference sets and $\bk$ vectors, so the resulting FDP bounds differ as well.
As illustration, we present the $\bk$ vectors for the four types of $\bv$ using Algorithm~\ref{Algo-getTuningParameterK} in plots (2) and (3) of Figure~\ref{Fig:compare-4-vk}.
Their corresponding FDP bounds $\barFDP^p_{(\bk,\bv)}(\cdot)$ are examined in Section~\ref{sec:simu}. We will see that there is no uniformly best one.
\begin{figure}[h!]
	\centering
	\includegraphics[width=13.5cm, height=5cm]{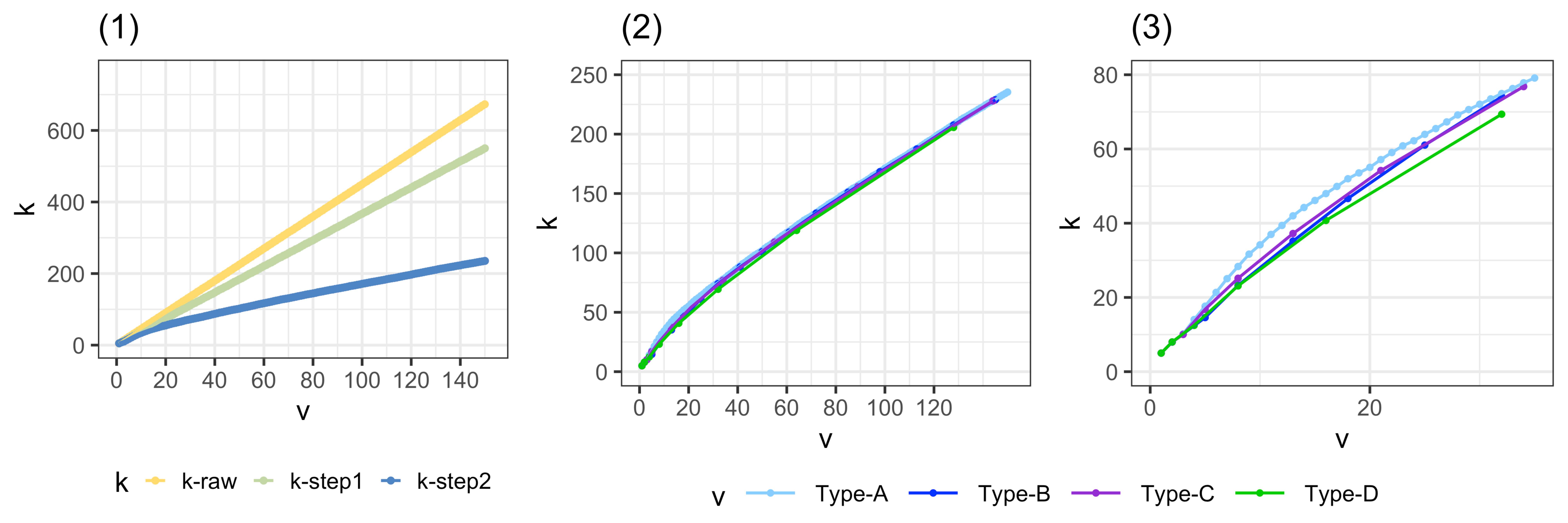}
	\caption{Plot (1): The vectors $\bk^{raw}$, $\bk^{step1}$ and $\bk^{step2}$ for a $\bv$ vector of Type-A in \eqref{4-types-of-v}.
		Plot (2): The output vector $k$ by using Algorithm~\ref{Algo-getTuningParameterK} with $\bv$ vectors of Types A-D in \eqref{4-types-of-v}. 
		Plot (3): The zoom-in version of Plot (2). 
		For all three plots, we use $p=1000$ and let $v_i < 150$. 
		All results are average values over 100 replications. 
	}
	\label{Fig:compare-4-vk}
\end{figure}


From a practical point of view, we find that the Type-B $\bv$ consistently performs well in different scenarios (see Section~\ref{sec:simu}),
so we recommend it as a default choice in practice.
The data-dependent optimal choice of $\bv$ remains an open question for future research.
For the choice of the largest value $v_m$,
based on both its role in Algorithm~\ref{Algo-BasedOnSimuKFWER} and \ref{Algo-getTuningParameterK} and our experience from simulations,
we found that this value does not have a strong influence as long as it is relatively large (e.g., $p/3$), so one may set it according to the affordable computational expense.

Lastly, by applying Algorithm~\ref{Algo-getTuningParameterK} to $\bv=(1,2,\dots,p)$, 
we can obtain an uniformly better FDP bound than $\barFDP^{\KR}(\cdot)$, meaning that it is at least as good as $\barFDP^{\KR}(\cdot)$ and can never worse in all cases. 
As we will see in Section~\ref{sec:simu}, the obtained FDP bound is actually strictly better in many simulations.
\begin{proposition} \label{prop:improveKR}
    For $\bv=(1,\dots,p)$, let $\bk$ be the output of Algorithm~\ref{Algo-getTuningParameterK}. Then,
	\begin{align*}
		\barFDP^p_{(\bk,\bv)}(R) \leq \barFDP^{\KR}(R), \quad R\subseteq [p].
	\end{align*}
\end{proposition}

The FDP bound of \cite{katsevich2020simultaneous} was obtained using a general martingale approach which can be applied to knockoff settings as special cases. Our method makes more targeted use of the Bernoulli trials, which are fundamental for the knockoff-based approach. 
In that sense, it may not come as a surprise that we are able to obtain better FDP bounds in this specific context.
A natural follow-up question is: Can we further improve $\barFDP^p_{(\bk,\bv)}(\cdot)$?
As we will show in Section~\ref{sec:closedtesting}, the answer is yes. 
The FDP bound $\barFDP^p_{(\bk,\bv)}(\cdot)$ is still not admissible \citep{goeman2021only},
and a further uniform improvement can be obtained by connecting it to the closed testing framework.

\section{Simultaneous FDP bounds via knockoffs and closed testing} \label{sec:closedtesting}
In this section, we turn to the closed testing framework and propose a method to obtain simultaneous FDP bounds by combining knockoffs and closed testing.
In particular, we propose a local test based on the knockoff statistics for closed testing.
We show that Algorithm~\ref{Algo-BasedOnSimuKFWER} is a special case (or an exact shortcut) of this closed testing based approach, and that it can be uniformly improved. 
We also present other generalizations based on closed testing and develop a shortcut for their implementations.

\subsection{Preliminaries on closed testing} \label{sec:reviewClsoedtesting}

The closed testing procedure \citep{marcus1976closed} was originally proposed for familywise error rate control.
\cite{goeman2011multiple} observed that it can also be used to obtain simultaneous FDP bounds.

Consider hypotheses $H_1, \dots, H_p$, where in the variable selection setting $H_i$ means that the $i$-th variable is null.
The closed testing procedure consists of three steps:
\begin{itemize}
	\item[1.] Test all intersection hypotheses $H_I = \cap_{i\in I} H_{i}$ for $I \subseteq [p]$ at level $\alpha$.
	Throughout the paper, we refer to such tests as local tests, in contrast to the closed tests described in step 2. 
	We use $\phi_I \in \{0,1\}$ to denote the local test result of $H_{I}$, with $1$ indicating rejection.
	For convenience, we sometimes say that a set $I$ is rejected to mean that $H_I$ is rejected. We use the convention that $\phi_{\emptyset}=0$, i.e., $H_{\emptyset}$ is always accepted.
	\item[2.] Obtain closed testing results based on the local test results: for all $I \subseteq [p]$, $H_I$ is rejected by closed testing if all supersets $J$ of $I$ are locally rejected. 
	We use $\phi^{[p]}_I \in \{0,1\}$ to denote the closed testing result of  $H_I$:
	\[
	\phi^{[p]}_I = \min \{ \phi_J: I \subseteq J \subseteq [p] \}. 
	\]
	\item[3.] For any set $R\subseteq [p]$, let
	\begin{align}\label{closedtestingFDbound}
		t^{[p]}(R) = \max_{I \in 2^R} \{|I|: \phi^{[p]}_I =0\}
	\end{align}
	be the size of the largest subset of $R$ that is not rejected by closed testing. 
	Then 
	\begin{align}\label{closedtestingFDPbound}
		\barFDP^{ct}(R) = \frac{t^{[p]}(R)} {\max\{1,|R|\} }, \quad R \subseteq [p]
	\end{align}
	is a simultaneous FDP bound. 
\end{itemize}
The first two steps guarantee that the FWER is controlled, in the sense that 
$P(\phi_I^{[p]}=0, \forall I\subseteq \mN) \ge 1-\alpha$. To see this, note that $P(\phi_{\mN}=0)\ge 1-\alpha$ for any valid local test, and by construction $\phi_{\mN}=0$ implies that $\phi_I^{[p]}=0$ for all $I\subseteq \mN$. 

Step 3 is the new step from \cite{goeman2011multiple} which gives us a simultaneous FDP bound.
To see this, note that $\phi_{\mN}=0$ implies $\phi_{R \cap \mN}^{[p]}=0$ for all $R\subseteq [p]$, and hence $|R \cap \mN | \le t^{[p]}(R)$ for all $R\subseteq [p]$. Therefore, for any valid local test, we have
$\Prob( |R \cap \mN| \leq t^{[p]}(R), \forall R\subseteq [p]) \geq
\Prob( \phi_{\mN}=0 ) \geq 1-\alpha$. This leads to a simultaneous FDP bound by dividing by $\max\{1,|R|\}$ on both sides inside the probability.

This closed testing procedure is computationally intractable in its standard form as there are $2^p$ hypotheses to test in step 1.
In some cases, however, there exist shortcuts to derive the closed testing results efficiently \citep[see, e.g.,][]{goeman2011multiple, 10.1093/biomet/asz082, goeman2021only, tian2021large, vesely2021permutation}.

\subsection{The multi-weighted-sum local test statistic and connections to other methods} \label{sec:closedtestingLocalTest}

For variable selection problems, the null hypothesis with respect to a set $I \subseteq [p]$ is
\[
H_{I}: \text{the } i \text{-th variable is null}, \ \forall i \in I.
\]
For a knockoff statistic vector $\bW$ possessing the coin-flip property, 
$D_i= \text{sign}(W_i)$'s are i.i.d.\ with distribution $\Prob(D_i=1) = \Prob(D_i=-1) = 1/2$ for $i \in I$ under the null hypothesis $H_{I}$.
Based on this, we propose the following multi-weighted-sum test statistic to locally test $H_{I}$ in the closed testing framework:
\begin{align} \label{formula:weighted-sum-test}
\whL^{I}_i = \sum_{j \in I} w^I_{i,j} \Ind_{D_j=1}, \quad i \in [m],
\end{align}
where $w^I_{i,j}$ are tuning parameters. 
Let $z^I_1, \dots, z^I_m$ be the corresponding critical values such that under $H_I$, 
\begin{align} \label{formula:criticalvalueLocaltest}
\Prob \left(\whL^{I}_i \leq z^I_i-1, \forall i \in [m] \right) \geq 1 - \alpha.
\end{align}
We locally reject $H_I$ if there exists some $i\in[m]$ such that $\whL^{I}_i \geq z^I_i$.
Note that because the distribution of $D_i$'s under $H_{I}$ is known, these critical values can be numerically approximated by first simulating i.i.d.\ $D_i$ with $\Pr(D_i=1) = \Pr(D_i=-1) = 1/2$,
and then approximating the critical values in a greedy manner. For a concrete implementation example, please refer to \url{https://github.com/Jinzhou-Li/KnockoffSimulFDP}.
We would like to mention that the test statistic \eqref{formula:weighted-sum-test} would inherently lack power at any reasonable significance level if the cardinality of $I$ is small. For example, in the case of $|I|=1$, one can only accept $H_I$ for all significance levels $\alpha \in (0,0.5)$. This is due to the discrete nature of the test statistic, which requires accumulating enough evidence (thus necessitating a large $|I|$) in order to reject the null hypothesis. This limitation is unavoidable but not particularly detrimental to our problem and goals.

Given a concrete local test, the remaining key issue for obtaining FDP bounds using closed testing is the computation.
This is because the standard form require conducting $2^p-1$ tests in the first step, which grows exponentially with respect to $p$.
As a result, an efficient implementation method, referred to as a ``shortcut", is required.
A shortcut yields the same results as the standard closed testing procedure but without the need to test all $2^p-1$ intersection hypotheses.
For example, in terms of familywise
error rate control, the Holm method is the shortcut of a closed testing procedure that uses the Bonferroni correction to test intersection hypotheses in the first step.
In the following, we show that Algorithm~\ref{Algo-BasedOnSimuKFWER} is actually an exact shortcut for closed testing using local test statistic~\eqref{formula:weighted-sum-test} with $w^I_{i,j}=\Ind_{r^I_j > |I|-b^I_i}$, that is,
\begin{align} \label{formula:weighted-sum-test-forEquivalencePart}
\whL_i^{I}(b^I_i) = \sum_{j \in I} \Ind_{r^I_j > |I|-b^I_i} \Ind_{D_j=1}, \quad i=1,\dots, m,
\end{align}
where $b^I_i$'s are tuning parameters and $r^I_j$ is the local rank of $|W_j|$ in $\{|W_j|, j\in I \}$. 
For example, $r^I_1=3, r^I_3=2, r^I_4=1$ for $I=\{1,3,4\}$.
Note that the test statistic $\whL_i^{I}(b^I_i)$ counts the number of positive $W_j$ among the first $b^I_i$ based on the local rank for $j \in I$.

Specifically, the following Theorem~\ref{thm:Algo-BasedOnJointKFWER-as-CL} shows that the output of Algorithm~\ref{Algo-BasedOnSimuKFWER}
(i.e., our previously proposed FDP bound $\barFDP^m_{(\bk,\bv)}(\cdot)$)
coincides with the FDP bound based on closed testing, provided the latter is used with local test statistic~\eqref{formula:weighted-sum-test-forEquivalencePart} with $b^I_i=k_i-v_i+1$ and critical values $z^I_i =k_i$.


\begin{theorem} \label{thm:Algo-BasedOnJointKFWER-as-CL}
	For $\bv=(v_1, \dots, v_m) $ and $\bk=(k_1, \dots, k_m)$,
	let $\barFDP^{ct}(R)$ (see \eqref{closedtestingFDPbound}) be the FDP bound based on closed testing using test statistic~\eqref{formula:weighted-sum-test-forEquivalencePart} with $b^I_i=k_i+v_i-1$ and critical values $z^I_i =k_i$ to locally test $H_I$.
	Then 
	\[
	\barFDP^{ct}(R) = \barFDP^m_{(\bk,\bv)}(R), \quad R \subseteq [p].
	\]
\end{theorem}

Proposition~\ref{prop:KR-Algo1} and Theorem~\ref{thm:Algo-BasedOnJointKFWER-as-CL}
directly imply that the interpolation version of the FDP bound from \cite{katsevich2020simultaneous} 
is also a special case of closed testing:
\begin{corollary} \label{coro:KR-connect-CL}
	Let $\barFDP^{ct}(R)$ (see \eqref{closedtestingFDPbound}) be the FDP bound based on closed testing using test statistic~\eqref{formula:weighted-sum-test-forEquivalencePart} with
	$m=|I|$, $b^I_i=k^{raw}_{i}+i-1$ and critical values $z^I_i = k^{raw}_{i}$ to locally test $H_I$,
	then
	\[
	\barFDP^{ct}(R) = \barFDP^{\KR}(R), \quad R \subseteq [p].
	\]
\end{corollary}

We point out that instead of using Theorem~\ref{thm:Algo-BasedOnJointKFWER-as-CL}, one can also apply a general method from \cite{goeman2021only} 
to connect the simultaneous FDP bounds $\barFDP^{\KR}(\cdot)$ (and $\barFDP^m_{(\bk,\bv)}(\cdot)$) to closed testing (see Appendix~\ref{supp:applyJelle2021ToKR}).
This approach, however, only results in $\barFDP^{ct}(R) \leq \barFDP^{\KR}(R)$ 
(see Proposition~\ref{prop:KR-CL-JellePaper} in Appendix~\ref{supp:KR-JellePaper}). 
Our proof of Theorem~\ref{thm:Algo-BasedOnJointKFWER-as-CL} is more targeted to the knockoff setting we consider here and does not rely on the method from \cite{goeman2021only}.
As a result, it leads to the more precise characterization $\barFDP^{ct}(R) = \barFDP^{\KR}(R)$ and unifies all simultaneous FDP bounds into the closed testing framework.

Finally, as a side result, we show a connection between the $k$-FWER control method of \cite{LJ-WS:2016} and closed testing.
Specifically, for a given $k$, the method of \cite{LJ-WS:2016} outputs a $k$-FWER controlled set $\whS^{\JS}(k)$ (see Section~\ref{sec:reviewJS}).
The following Proposition~\ref{prop:Algo-JS-as-CL} shows that in the case of $|\whS^{\JS}(k)| \geq k-1$, the same $k$-FWER guarantee for $\whS^{\JS}(k)$ is obtained by closed testing using local test statistic~\eqref{formula:weighted-sum-test-forEquivalencePart} with $m=1$, $b^I_1=k-v+1$ and $z^I_1=k$.
\begin{proposition}\label{prop:Algo-JS-as-CL}
	For the closed testing procedure using test statistic~\eqref{formula:weighted-sum-test-forEquivalencePart} with $m=1$, $b^I_1=k+v-1$ and critical value $z^I_1=k$ to locally test $H_I$, let $t_{\alpha}( \whS^{\JS}(k) )$ be the corresponding false discovery upper bound defined by \eqref{closedtestingFDbound}.
	Then $t_{\alpha}( \whS^{\JS}(k) ) = |\whS^{\JS}(k)|$ when $|\whS^{\JS}(k)| < k-1$ and $t_{\alpha}( \whS^{\JS}(k) ) = k-1$ when $|\whS^{\JS}(k)| \geq k-1$.
\end{proposition}

\subsection{Uniform improvement, generalization and shortcut} \label{sec:UniformlyImprovementClosedTesting}

Based on Theorem~\ref{thm:Algo-BasedOnJointKFWER-as-CL},
a uniform improvement of $\barFDP^m_{(\bk,\bv)}(\cdot)$ can be achieved.
In particular, when locally testing the null hypothesis $H_{I}$ using test statistic~\eqref{formula:weighted-sum-test-forEquivalencePart},
instead of using the same tuning parameter $b^I_i=k_i+v_i-1$ and critical value $z^I_i =k_i$ for all $I \subseteq [p]$,
we can use the adapted $b^I_i=k^{|I|}_i+v_i-1$ and $z^I_i =k^{|I|}_i$,
where $k^{|I|}_i$ is obtained by using Algorithm~\ref{Algo-getTuningParameterK} with $|I|$, instead of $p$, as the last input value.
Note that the resulting local test is valid because of \eqref{fact1} in Appendix~\ref{supp:Afact}.
As $k^{|I|}_i \leq k_i$, the new local test is uniformly better than the previous one,
in the sense that it must locally reject $H_{I}$ if the previous local test rejects.
This justifies the uniform improvement of closed testing over the previous FDP bound $\barFDP^m_{(\bk,\bv)}(\cdot)$.
As we show in Section~\ref{sec:simu:Algo1AndCT}, such improvement is strict in some settings.

Aside from uniform improvement, 
the closed testing framework also enables generalizations of $\barFDP^m_{(\bk,\bv)}(\cdot)$
by using other local tests.
As a first attempt, we incorporate the local rank $r^I_j$ in the weights of test statistic \eqref{formula:weighted-sum-test}. In particular, we use $w^I_{i,j}= r^I_j \Ind_{r^I_j > |I|-b^I_i}$ instead of $w^I_{i,j}= \Ind_{r^I_j > |I|-b^I_i}$.
In simulations, we find that this local test generally performs worse than the method without including local rank, but there also exist settings where it performs better, see Appendix~\ref{sec:simu:CTandRankCT} for details.

In addition, the closed testing framework provides a direct way to incorporate background knowledge. 
For instance, one may use a large weight $w^I_{i,j}$ in test statistic \eqref{formula:weighted-sum-test} if one expects that the $i$-th variable is non-null. 
Moreover, it is important to remark that since all proposed procedures are valid conditional on the absolute statistics $|W_1|, \ldots, |W_p|$, the user may observe these before selecting the final procedure to use.

For the above uniformly improved 
(i.e., using local test statistic~\eqref{formula:weighted-sum-test-forEquivalencePart} with $b^I_i=k^{|I|}_i+v_i-1$ and critical value $z^I_i =k^{|I|}_i$) 
and rank-generalization versions (i.e., using local test statistic~\eqref{formula:weighted-sum-test} with $w^I_{i,j}= r^I_j \Ind_{r^I_j > |I|-b^I_i}$, $b^I_i=k^{|I|}_i+v_i-1$ and critical values calculated based on inequality~\eqref{formula:criticalvalueLocaltest}),
the analytical shortcut $\barFDP^m_{(\bk,\bv)}(\cdot)$ (see Theorem~\ref{thm:Algo-BasedOnJointKFWER-as-CL}) for obtaining FDP bound based on closed testing is not valid anymore.
Hence, a new shortcut is needed for their practical implementations.
To this end, we propose a general shortcut which is valid when the following two conditions hold:
\begin{itemize}
	\item[(C1)] The critical values $z^I_1, \dots, z^I_m$ depend on $I$ only through $|I|$.
	\item[(C2)] 
	For any two sets $I_1 = \{u_1,\dots, u_s\}$ and $I_2 = \{v_1,\dots, v_s\}$ of the same size,
	there exists a permutation $\pi: [p] \rightarrow [p]$ 
	such that if $\pi^{I_1}_{(j)} \leq \pi^{I_2}_{(j)}$ for all $j=1, \dots, s$, 
	then $\whL^{I_1}_i \leq \whL^{I_2}_i$ for all $i \in [m]$,
	where $\pi^{I}_{(1)} < \dots < \pi^{I}_{(s)}$ are ordered values of $\{\pi(i), i\in I \}$.
\end{itemize}

Condition (C2) pertains to a form of monotonicity of the test statistic, which, in turn, leads to computational reductions.
These two conditions ensure the following for a null hypothesis $H_I$ with $|I|=s$:
instead of checking whether all subsets of $I$ of size $t \in \{ s, \dots, 1 \}$ are rejected by closed testing, 
it is sufficient to check only one set $B$ of each size $t$
(see Algorithm~\ref{Algo-ct-shortcut} for the definition of set $B$). 
Moreover, instead of checking whether all supersets of $B$ of size $r \in \{ 0, \dots, p-t \}$ are locally rejected, we again only need to check one set of each size. 
This brings us the necessary computational reduction in the implementation of closed testing.

Note that these two conditions are on the local test method itself, rather than the data, so one can check whether their methods satisfy these conditions.
In particular, two examples where conditions (C1) and (C2) hold are the above mentioned uniformly improved and rank-generalization versions.
Specifically, one such permutation that satisfying condition (C2) can be obtained by letting $\pi(i)=l$ such that $W_i=W_{(l)}$ in the sorted $W_{(1)} < \dots < W_{(p)}$.

We summarize the shortcut of getting FDP bounds using closed testing in Algorithm~\ref{Algo-ct-shortcut}, and show its validity in Appendix~\ref{supp:justificationShortcut}.
The computational complexity of Algorithm~\ref{Algo-ct-shortcut} is $O(p^2)$ in the worst case, which is larger than the previous analytical shortcut 
Algorithm~\ref{Algo-BasedOnSimuKFWER}.
We think that this is a fair price to pay for a more general shortcut, in the sense that it can be applied to more types of local tests.
In practice, one may first use Algorithm~\ref{Algo-BasedOnSimuKFWER}, and only use the uniformly improved version based on closed testing (with the shortcut provided in Algorithm~\ref{Algo-ct-shortcut}) if the result is not already satisfactory.
\begin{algorithm}[h!]
	\caption[] {\textbf{: A shortcut for closed testing} } \label{Algo-ct-shortcut}
	\textbf{Input}:  $(R, \pi)$, where $R \subseteq [p]$ is any non-empty set of size $s$, 
	and $\pi$ is a permutation satisfying condition (C2).
	
	\textbf{Output}: FDP upper bound for $R$, which holds simultaneously for all $R \subseteq [p]$.
	\begin{algorithmic}[1]
		\For {$t=s,\dots,1$}
		\newline 
		\-\hspace{0.5cm} Let $B = \left\{ \pi^{-1} \left( \pi^{R}_{(1)} \right), \dots, \pi^{-1} \left( \pi^{R}_{(t)} \right) \right\} \subseteq R$
		(i.e., the index set of the $t$ smallest 
		\newline 
		\-\hspace{0.5cm} 
		$\pi(i)$ with $i \in R$).
		\For {$r=0,\dots,p-t$}
		\newline 
		\-\hspace{1cm} 
		\textbf{if} $r=0$, \textbf{then} let $U_{t+r} =B$,
		\textbf{else}
		Let $ U_{t+r} = B \cup B^c_r$, where
		\newline 
		\-\hspace{1cm} 
		$B^c_r = \left\{ \pi^{-1} \left( \pi^{B^c}_{(1)} \right), \dots, \pi^{-1}\left( \pi^{B^c}_{(r)} \right) \right\}$
		(i.e., the index set of the $r$ smallest $\pi(i)$ 
		\newline 
		\-\hspace{1cm} 
		with $i \in B^c$).
		\newline
		\-\hspace{1cm} 
		For $j\in [m]$, let $\whL^{U_{t+r}}_j$ be the test statistic related to $H_{U_{t+r}}$ and $z_{j}^{U_{t+r}}$ be the 
		\newline 
		\-\hspace{1cm} 
		corresponding critical value.
		\newline
		\-\hspace{1cm} \textbf{if} $\whL^{U_{t+r}}_j < z_{j}^{U_{t+r}}$ for all $j\in [m]$, \textbf{then return }$ t/s$.
		\EndFor
		\EndFor
		\State \textbf{Return} $0$.
	\end{algorithmic}
\end{algorithm}

\section{Simulations and a real data application} \label{sec:simu}
We now examine the finite sample performance of our proposed methods in a range of settings and compare them to the method of \cite{katsevich2020simultaneous}.
Since comparing FDP bounds for all sets $R \subseteq [p]$ is computationally infeasible, we follow \cite{katsevich2020simultaneous} and focus on the nested sets $\whR_i=\{ j \leq i: W_j > 0 \}$, $i \in [p]$.
All simulations were carried out in R, and the code is available at \url{https://github.com/Jinzhou-Li/KnockoffSimulFDP}.

\subsection{Comparison of Algorithm~\ref{Algo-BasedOnSimuKFWER} and the method of \cite{katsevich2020simultaneous}} \label{sec:simu:Algo1AndKR}

We consider the following methods:
\begin{itemize}
	\item KR: The simultaneous FDP bound from \cite{katsevich2020simultaneous} (see \eqref{KR-FDPbound-interpolation}). 
	We use the R code from \url{https://github.com/ekatsevi/simultaneous-fdp} for its implementation.
	\item KJI:
 Our proposed simultaneous FDP bounds based on Knockoffs and Joint $k$-familywise error rate control with Interpolation (see Algorithm~\ref{Algo-BasedOnSimuKFWER}). 
	We implement four types of tuning parameters $(\bv,\bk)$, where $\bv$ is as described in \eqref{4-types-of-v} and $\bk$ is obtained using Algorithm~\ref{Algo-getTuningParameterK}, and we denote them as KJI-A, KJI-B, KJI-C, and KJI-D, respectively. 
\end{itemize}
All methods use the same knockoff statistic vectors $\bW$.
We consider the following two settings:
\begin{itemize}
	\item [(a)] 
	Linear regression setting: We first generate $X \in \bbR^{n \times p}$ by drawing 
	$n$ i.i.d.\ samples from the multivariate Gaussian distribution $N_p(0,\Sigma)$ with $(\Sigma)_{i,j}=0.6^{|i-j|}$.
	Then we generate $\beta \in \bbR^{p \times 1}$ by randomly sampling
	$s \times p$ non-zero entries from $\{1, \dots, p\}$ with sparsity level $s \in \{0.1, 0.2, 0.3\}$, and we set the signal amplitude of the non-zero $\beta_j$'s to be $a / \sqrt{n}$, where $n$ is the sample size and the amplitude $a \in \{6,8,10\}$.
	Finally, we sample $Y \sim N_n(X\beta,I)$. 
	The design matrix $X$ is fixed over different replications in the simulations.
	Based on the generated data $(X, Y)$, we obtain $\bW$ by using the fixed-X ``sdp" knockoffs and the signed maximum lambda knockoff statistic.
	Similar settings and the same knockoff statistics are considered in \cite{barber2015controlling} and \cite{LJ-WS:2016}.
	\item [(b)] Logistic regression setting:
	For $i=1,\dots,n$, we first sample $x_i \sim N_p(0,\Sigma)$ with $(\Sigma)_{i,j}=0.6^{|i-j|}$. Then we sample $y_i \sim  \text{Bernoulli} \left(\frac{1}{1+ e^{-x_i^T \beta}} \right)$, where $\beta$ is generated as in (a), except now we consider amplitude $a \in \{8,10,12 \}$.
	Based on the generated data $(x_i,y_i)$, $i=1,\dots,n$, we obtain $W$ by using the ``asdp" model-X knockoffs and the Lasso coefficient-difference statistic with 10-fold cross-validation.
	Similar settings and the same knockoff statistics are considered in \cite{candes2018panning}.
\end{itemize}
We take $p=200$ and $n=500$ and use the four types of $v$ described in \eqref{4-types-of-v} with $v_m < 65$. The reported FDP bounds are average values over $200$ replications. 
We also tried the case of $p=1000$ and $n=2500$, and a high-dimensional linear case with $p=800$ and $n=400$.
The simulation results convey similar information, see Appendix~\ref{supp:simu-KR and KJI}.

\begin{figure}
	\centering
	\includegraphics[width=8.5cm, height=8.5cm]{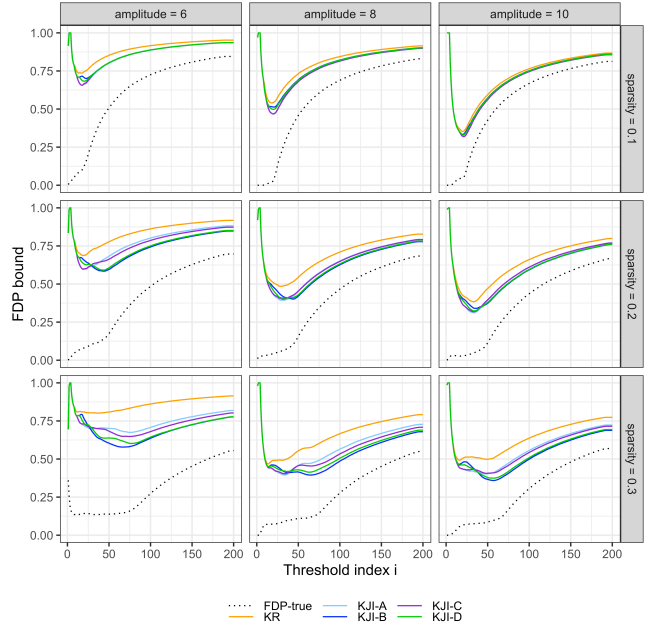}
	\caption{Simultaneous FDP bounds of KR and KJI 
	in the linear regression setting with $p=200$ and $n=500$. The dashed black line indicates the true FDP. All FDP bounds are average values over $200$ replications.}
	\label{Fig:linear-p200}
\end{figure}
\begin{figure}
	\centering
	\includegraphics[width=8.5cm, height=8.5cm]{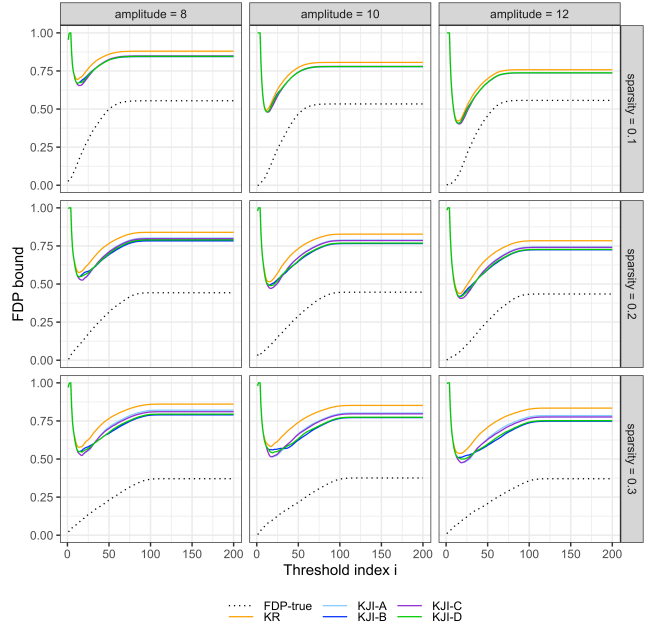}
	\caption{Simultaneous FDP bounds of KR and KJI 
	in the logistic regression setting with $p=200$ and $n=500$. The dashed black line indicates the true FDP. All FDP bounds are average values over $200$ replications.}
	\label{Fig:logistic-p200}
\end{figure}
As expected, the simultaneous FDP guarantee~\eqref{simuFDPbounds} holds for all methods,
see Figure~\ref{Fig:error1} in Appendix~\ref{supp:simu-KR and KJI}.
Figures~\ref{Fig:linear-p200} and \ref{Fig:logistic-p200} show the FDP bounds of KR and KJI with the four types of $(\bv,\bk)$ in the linear and logistic regression settings, respectively.
One can see that KJI-A gives better FDP bounds for all $\whR_i$ than KR over all settings, which verifies Proposition~\ref{prop:improveKR}.
The improvement is larger in denser settings with a smaller signal.
All methods give similar FDP bounds in settings with sparsity level $0.1$ and large amplitude.
The reason is that in such settings, almost all true $W_i$'s are positive and have larger absolute values than any null $W_i$,
so the FDP bounds based on the reference set $\whS(v_1=1)$ and interpolation are the best possible one can obtain.
We also observe that KJI with different types of $(\bv,\bk)$ can give better FDP bounds for different $\whR_i$ under different settings, and there is no uniformly best one.
Finally, we would like to mention that the tightness of FDP bounds depends on the distribution of knockoff vector $W$, which in turn, is influenced by the data distribution, the methodology used to construct knockoffs, and the choice of feature statistics.

\subsection{Comparison of Algorithm~\ref{Algo-BasedOnSimuKFWER} and the closed testing based approach} \label{sec:simu:Algo1AndCT}

We now compare KJI described in Section~\ref{sec:simu:Algo1AndKR} to the following method:
\begin{itemize}
	\item KCT: Our proposed simultaneous FDP bounds based on Knockoffs and Closed Testing using local test statistic \eqref{formula:weighted-sum-test-forEquivalencePart} with $b^I_i=k^{|I|}_i+v_i-1$ and the critical value $z^I_i =k^{|I|}_i$.
	We use the same four types of $v$ as for KJI and obtain $k^{|I|}$ by Algorithm~\ref{Algo-getTuningParameterK} with input $(v_1,\dots,v_m, \alpha, \delta=0.01, |I|)$,
	and we denote them as KCT-A, KCT-B, KCT-C and KCT-D.
	We use Algorithm~\ref{Algo-ct-shortcut} as shortcut for the implementation of closed testing.
\end{itemize}

We applied KJI and KCT in the same linear and logistic regression settings as in Section~\ref{sec:simu:Algo1AndKR} with $p=200$ and $n=500$.
As expected, the FDP bounds of KCT are indeed smaller than or equal to that of KJI for the same tuning parameter vector $v$.
In most cases, however, the corresponding FDP bounds are identical, 
and the improvement is typically very tiny in the cases where they are not identical (see Appendix~\ref{supp:simu-KJI and KCT}).

To understand this, it is helpful to think of the closed testing equivalent form of KJI (see Theorem~\ref{thm:Algo-BasedOnJointKFWER-as-CL}).
Even though the local tests of KCT are uniformly more powerful than the local tests of KJI, and hence can make more local rejections,
these local rejections do not necessarily lead to more rejections by closed testing,
especially in the sparse settings we consider here.
Therefore the final FDP bounds can be the same.

These simulation results also carry a valuable implication.
Specifically,
given that \cite{goeman2021only} proved that only closed testing methods are admissible and KJI is a special case of closed
testing (see Theorem~\ref{thm:Algo-BasedOnJointKFWER-as-CL}),
the observation that the uniform improvement is very small implies that KJI is nearing optimal, and that further substantial uniform improvements are unlikely.



\begin{figure}[h!]
	\centering
	\includegraphics[width=8.5cm, height=8.5cm]{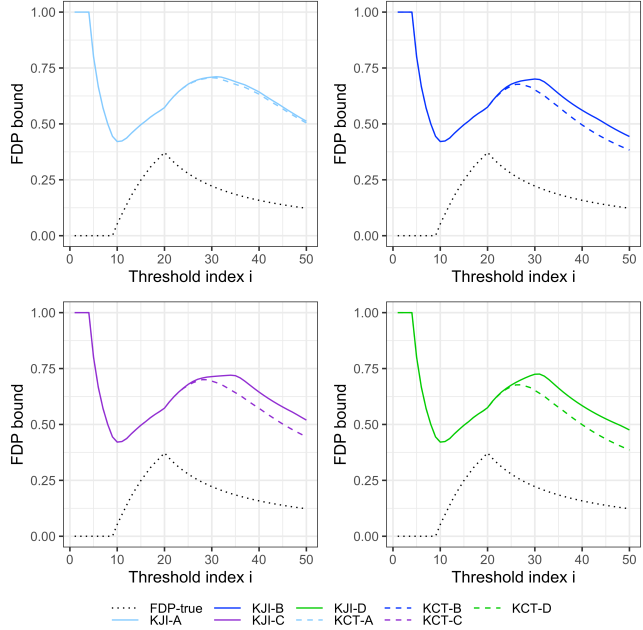}
	\caption{Simultaneous FDP bounds of KJI and KCT 
	in the simulation setting where the knockoff statistic vector $W \in \bbR^{50}$ is directly generated.
	The dashed black line indicates the true FDP. 
	All FDP bounds are average values over $200$ replications.}
	\label{Fig:ct}
\end{figure}

At last, to verify our claim in Section~\ref{sec:UniformlyImprovementClosedTesting} that the uniform improvement by KCT over KJI can be non-trivial in certain cases, 
we consider a simulation setting that directly generates the knockoff statistic vector $W$.
Specifically, we consider a dense setting with $p=50$ and null variable set $\mN = \{10,\dots,20 \}$.
For the knockoff statistics, we set $|W_i|=50-i+1$, 
$\text{sign}(W_i)=1$ for $i \not\in\mN$
and $\text{sign}(W_i) \stackrel{i.i.d.\ }{\sim} \{-1,1\}$ with the same probability $1/2$ for $i\in\mN$.
It is clear that the generated $W$ satisfies the coin-flip property,
so it is a valid knockoff statistic vector.
Figure~\ref{Fig:ct} shows the simulation results in this setting.
We can see that the improvements of KCT over KJI can be large for some $\whR_i$. 

\subsection{Real data application} \label{sec:realdata}
We now apply our methods to genome-wide association study (GWAS) datasets from the UK Biobank \citep{bycroft2018uk}.
A GWAS data set typically consists of genotypes and traits of different individuals, and the goal is to identify genotypes that are conditionally dependent on a trait given the other genotypes (i.e., variable selection).
The traits considered in our analysis are height, body mass index, platelet count, systolic blood pressure, cardiovascular disease, hypothyroidism, respiratory disease and diabetes.

We follow the analysis of \cite{katsevich2020simultaneous} (see \url{https://github.com/ekatsevi/simultaneous-fdp})
and use the knockoff statistics $\bW=(W_1,\dots,W_p)$ constructed by \cite{sesia2020multi} (see \url{https://msesia.github.io/knockoffzoom/ukbiobank.html}).
We assume that the model-X assumption holds, as do previous analyses in the literature.
For each set $\whR_i = \{ j \leq i: W_j > 0 \}$, we compute simultaneous FDP bounds using KR and KJI with the four types of tuning parameter vectors $\bv$ (see \eqref{4-types-of-v}) with $v_m < 1200$ (see also Section~\ref{sec:simu:Algo1AndKR}). Since we do not have the true FDP, we estimate it by $\widehat{\FDP}(\whR_i) = \frac{1 + \{ j \leq i, W_j < 0 \} }{\max\{|\whR_i|, 1\} }$.
The latter also leads to a corresponding FDR control method: taking the largest set $\whR_i$ such that $\widehat{\FDP}(\whR_i) \leq q$ provides FDR control at level $q\in(0,1)$ \citep{barber2015controlling}.

\begin{figure}
	\centering
	\includegraphics[width=12cm, height=6.5cm]{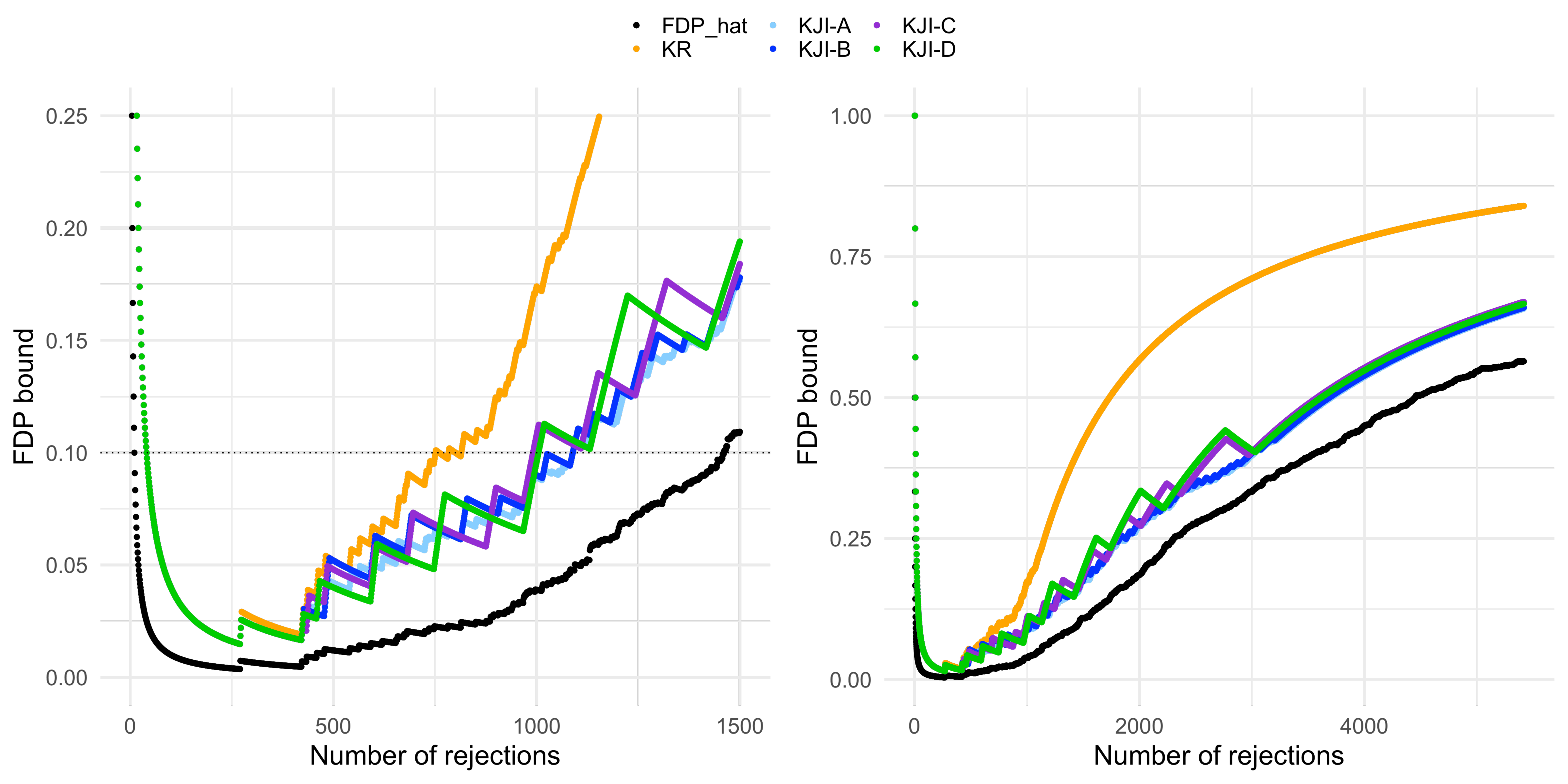}
	\caption{Simultaneous FDP bounds computed by KR and KJI versus the number of rejections for each considered set $\whR_i$, for the trait platelet count of the UK Biobank data. The black dotted line is the estimated FDP.
		The left plot is the zoom-in version of the right plot. 
		}
	\label{Fig:RealData}
\end{figure}

Figure~\ref{Fig:RealData} shows the results for the trait platelet count, showing the simultaneous FDP bound versus the number of rejections of each considered set.
We see that KJI with all four types of $\bv$ gives smaller bounds than KR.
One may also notice that the simultaneous FDP bounds based on the different $\bv$ vectors are quite similar, especially for sets with more than about $3000$ rejections.
The latter can be explained by the fact that the largest reference sets for the four types of $\bv$ vectors are similar, and hence the simultaneous FDP bounds based on interpolation beyond these sets are also similar.

\begin{figure}
	\centering
	\includegraphics[width=8.5cm, height=8.5cm]{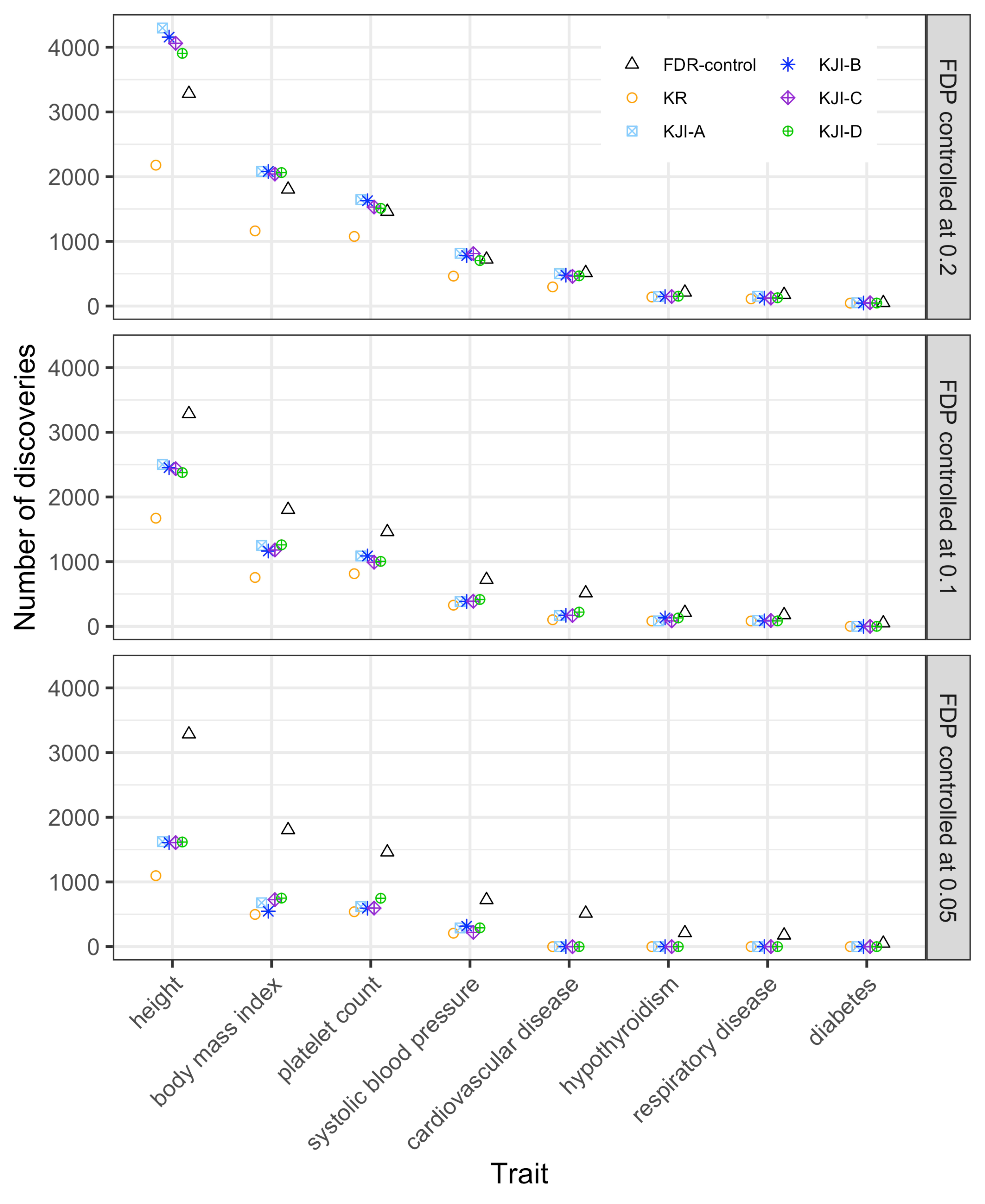}
	\caption{The number of discoveries obtained for all traits of the UK Biobank data, when controlling the FDR at level $0.1$, or the simultaneous FDP at levels 0.05, 0.1 and 0.2 by using KR and KJI. 
	}
	\label{Fig:RealDataPlot2}
\end{figure}

Figure~\ref{Fig:RealDataPlot2} shows the number of discoveries for all traits, when controlling the FDR at level $0.1$ or the simultaneous FDP at levels $0.05$, $0.1$, and $0.2$.
We see that the four versions of KJI yield similar numbers of discoveries in most cases, and that these numbers are generally larger than the number of discoveries of KR.

To illustrate the flexibility of using simultaneous FDP bounds, we take a closer look at the results for the trait height in Figure~\ref{Fig:RealDataPlot2}. 
We see that KJI-B returns $4157$ discoveries when controlling the simultaneous FDP at level $0.2$. Thus, with probability larger than 0.95, KJI-B guarantees more than $0.8 \cdot 4157 = 3325$ true discoveries among those $4157$ discoveries. If this set is somehow too large for practical purposes, one may switch to lower levels like $0.1$ or $0.05$, without losing control. For example, at level $0.05$, we obtain a set of $1607$ discoveries, meaning that, with probability larger than $0.95$, we are guaranteed more than $0.95 \cdot 1607  = 1526$ true discoveries among these.
One can also look at other levels or sets, and the simultaneous FDP bounds remain valid. This flexibility is extremely useful in exploratory data analysis. 

FDR control methods do not possess this kind of flexibility. For example, considering different levels will break the FDR control. Moreover, FDR control is only in expectation. In the data example, we get $3284$ discoveries when controlling the FDR at level $0.1$, meaning that we are guaranteed more than $0.9 \cdot 3248 = 2955$ true discoveries among these, in the expectation sense.

\section{Discussion} \label{sec:disc}

We studied the problem of obtaining simultaneous FDP bounds for knockoff-based approaches. 
We first proposed a method based on joint $k$-FWER control and interpolation (Algorithm~\ref{Algo-BasedOnSimuKFWER}),
and we showed that it is a generalization of the only existing approach for this problem by \cite{katsevich2020simultaneous}, in the sense that the latter is a special case of Algorithm~\ref{Algo-BasedOnSimuKFWER} with a specific choice of tuning parameters.
We also suggested other choices for the tuning parameters, and showed that one of them was uniformly better.
Next, we proposed a method based on closed testing
and showed that Algorithm~\ref{Algo-BasedOnSimuKFWER} is a special case (or exact shortcut) of this method.
Using closed testing, we showed that Algorithm~\ref{Algo-BasedOnSimuKFWER} can be further uniformly improved, and that other generalizations can be derived.
We also developed a new shortcut for the implementation of this closed testing based method.

Closed testing is a very general and powerful methodology in which one has the flexibility to choose different local tests. 
Different local tests can be better in different settings.
One natural problem is to obtain optimal data-dependent local tests, while maintaining valid simultaneous FDP bounds.
Sample-splitting is a possible solution, but it can be sub-optimal, especially in cases where the sample size is not large. This is an interesting problem for future research. 

The key assumption underlying our methods is that the knockoff statistics satisfy the coin-flip property. In some cases, however, this assumption does not hold.
Examples are Gaussian graphical models \citep{li2021ggm} where the knockoff statistic matrix for all edges does not possess this property, 
or the model-X knockoff setting when the model-X assumption does not hold, but an estimated distribution of the covariates is used \citep{barber2020robust}.
Developing simultaneous FDP bounds for such settings is another interesting direction for future work.

\section*{Acknowledgments}
We are grateful to the anonymous reviewers for many helpful comments and suggestions, which greatly helped to improve the manuscript.
J.L. gratefully acknowledges the support by the SNSF Grant P500PT-210978.

\bibliographystyle{apalike}
\bibliography{Reference}

\addtocontents{toc}{\vspace{.5\baselineskip}}
\addtocontents{toc}{\protect\setcounter{tocdepth}{1}}
\appendix
\newpage
\setcounter{page}{1}

\begin{center}
{\large\bf SUPPLEMENTARY MATERIAL}
\end{center}

The supplementary material consists of the following five sections.
	\begin{description}
		\item[A] Validity of Algorithm~\ref{Algo-ct-shortcut}
		\item[B] Two facts about early-stopped negative binomial and truncated binomial statistics
		\item[C] Connecting the FDP bounds from \cite{katsevich2020simultaneous} and closed testing by using the method of \cite{goeman2021only}
		\item[D] Main proofs
		\item[E] Supplementary materials for simulations
	\end{description}

\section{Validity of Algorithm~\ref{Algo-ct-shortcut}} \label{supp:justificationShortcut}


Let $R \subseteq [p]$ be of size $s_R$ and let $I \subseteq R$ be of size $s_I$.
To obtain the false discovery bound of $R$ using closed testing,
we need to find its largest subset whose related null hypothesis is not rejected by closed testing.
To check whether a null hypothesis $H_I$ is rejected by closed testing, we need to check whether all superset of $I$ are locally rejected.
In the following, we show that if conditions (C1) and (C2) in Section~\ref{sec:UniformlyImprovementClosedTesting} hold, then
the following two claims (a) and (b) are true. 
As a result, we only need to check one set for each size when: (i) searching for the largest subset whose related null hypothesis is not rejected by closed testing; (ii) checking whether the null hypothesis $H_I$ is rejected by closed testing.
These then guarantee that Algorithm~\ref{Algo-ct-shortcut} is valid.

\begin{itemize}

\item[(a)] 
	Let $\pi^{I^c}_{(1)} < \dots < \pi^{I^c}_{(p-s_I)}$ be the ordered values of $\{\pi(i), i\in I^c \}$, where $I^c =  [p]\backslash I$.
	For any $t \in \{ 0, \dots, p-s_I \} $, let $U_{s+t} = I \cup I^c_t$ of size $s_I+t$, where $I^c_t = \{ \pi^{-1}( \pi^{I^c}_{(1)} ), \dots, \pi^{-1}( \pi^{I^c}_{(t)} ) \}$ (i.e., the index set of the $t$ smallest $\pi(i)$ with $i \in I^c$) for $t > 0$ and $I^c_t =\emptyset$ for $t=0$.
 Then, there exists a superset of $I$ with size $s_I+t$ that is not locally rejected if and only if $U_{s+t}$ is not locally rejected. 
	\item[(b)] 
	Let $\pi^{R}_{(1)} < \dots < \pi^{R}_{(s_R)}$ be the ordered values of $\{\pi(i), i\in R\}$.
	For any $t \in \{ 1,\dots,s_R \} $, let $B =\{ \pi^{-1}( \pi^{R}_{(1)} ), \dots, \pi^{-1}( \pi^{R}_{(t)} ) \} \subseteq R$ of size $t$ (i.e., the index set of the $t$ smallest $\pi(i)$ with $i \in R$). 
 Then, there exists a subset of $R$ with size $t$ that is not rejected by closed testing if and only if $B$ is not locally rejected
\end{itemize}

For (a):
the claim holds when $t=0$, now assume $t \in \{ 1, \dots, p-s_I \}$.
\begin{itemize}
	\item[] $\Longleftarrow$: This direction clearly holds.
\item[] $\Longrightarrow$: It is sufficient to show that if $U_{s+t}$ is locally rejected, then every supersets of $I$ with size $s_I+t$ is locally rejected.
If $U_{s+t}$ is locally rejected, then
$\whL^{U_{s_I+t}}_{i^*} \geq z^{s_I+t}_{i^*}$ for some ${i^*} \in[m]$.
Let $V_{s_I+t}  = I \cup J$ be any superset of $I$ with size $s_I+t$, where $J \subseteq I^c$ is of size $t$.
	Denote $U_{s_I+t} = \{u_1, \dots, u_{s_I+t} \}$ and $V_{s_I+t} = \{v_1, \dots, v_{s_I+t} \}$.
	By definition of $U_{s_I+t}$ and $V_{s_I+t} $, it is clear that $\pi(u_i) \leq \pi(v_i)$ for all $i=1\dots, s_I+t$. 
	Let $\whL^{U_{s_I+t}}_i$ and $\whL^{V_{s_I+t}}_i$ be the test statistics of $H_{U_{s_I+t}}$ and $H_{V_{s_I+t}}$, respectively.
	Then by (C2), we have $ \whL^{V_{s_I+t}}_{i^*} \geq \whL^{U_{s_I+t}}_{i^*} \geq z_{i^*}^{s_I+t}$, which implies that $H_{V_{s_I+t}}$ is locally rejected by (C1).
\end{itemize}

For (b): the claim holds when $t=s_R$, now assume $t \in \{ 1,\dots,s_R-1 \}$.
\begin{itemize}
	\item[] $\Longleftarrow$: This direction clearly holds.
	\item[] $\Longrightarrow$: It is sufficient to show that if $B$ is rejected by closed testing, then all subsets of $R$ of size $t$ are rejected by closed testing.
	Let $J \subseteq R$ be any subset of $R$ with size $t$.
	By claim (a), it is sufficient to show that for any $r \in \{0, \dots, p-t \} $,
	$ V_{t+r} = J \cup J^c_r$ is locally rejected, where $J^c = [p]\backslash J$, $J^c_r = \{ \pi^{-1}( \pi^{J^c}_{(1)} ), \dots, \pi^{-1}( \pi^{J^c}_{(r)} ) \}$ for $r>0$, and $J^c_r = \emptyset$ for $r=0$. 
	
	$B$ is rejected by closed testing means that all supersets of $B$ are locally rejected. 
	So for any $r \in \{ 0, \dots, p-t \}$ and $ U_{t+r} = B \cup B^c_r$, 
	there exists some $j_r \in [m]$ such that $\whL^{U_{t+r}}_{j_r}  \geq z_{j_r}^{t+r}$.
	Here $B^c = [p]\backslash B$, $B^c_r = \{ \pi^{-1}( \pi^{B^c}_{(1)} ), \dots, \pi^{-1}( \pi^{B^c}_{(r)} ) \}$ for $r>0$, and $B^c_r =\emptyset$ for $r=0$.
	
	Denote $U_{t+r} = \{u_1, \dots, u_{t+r} \}$ and $V_{t+r} = \{v_1, \dots, v_{t+r} \}$ such that 
	$\pi(u_1) < \dots < \pi(u_{t+r})$ and $\pi(v_1) < \dots < \pi(v_{t+r})$.
	By the definition of $U_{t+r}$ and $V_{t+r} $, it is clear that $\pi(u_i) \leq \pi(v_i)$ for all $i \in [t+r]$. 
	Hence by (C2), we have $\whL_{j_r}^{V_{t+r}} \geq \whL_{j_r}^{U_{t+r}} \geq z_{j_r}^{t+r}$, which then implies that $V_{t+r}$ is locally rejected by (C1).
\end{itemize}

\section{Two facts about early-stopped negative binomial and truncated binomial statistics} \label{supp:Afact}

We present the following two facts which will be used in many proofs to connect 
early-stopped negative binomial and truncated binomial test statistics.
Specifically, for any positive integers $v, k$ and a sequence $X_1, \dots, X_p$, where $X_i \in \{-1,+1\}$, let 
\begin{align*}
t_v = \min \{ i \in [p]: |\{j\leq i: X_j = -1 \}| = v \},
\end{align*}
and $t_v = p$ if $|\{i \leq [p]: X_i = -1 \}| < v$. Let
\begin{align*} 
N^{p}(v) &= |\{i \leq t_{v}: X_i = 1 \}|
\end{align*}
be the early-stopped negative binomial variable, and let
\begin{align*}
B^{p}(k+v-1)= |\{i \leq \min(p, k+v-1): X_i = 1 \}|
\end{align*}
be the truncated binomial variable.
Then the following events are identical:
\begin{align} \label{fact1}
N^{p}(v) \geq k \Longleftrightarrow B^{p}(k+v-1) \geq k
\end{align}
and
\begin{align} \label{fact2}
N^{p}(v) \leq k-1 \Longleftrightarrow B^{p}(k+v-1) \leq k-1.
\end{align}

\section{Connecting the FDP bounds from \cite{katsevich2020simultaneous} and closed testing by using the method of \cite{goeman2021only}} \label{supp:applyJelle2021ToKR}

The purpose of this appendix is to: (i) provide the explicit formula for the interpolation version of the FDP bound from \cite{katsevich2020simultaneous}, which we utilized in Section~\ref{sec:interpolation}; (ii) demonstrate the application of the method from \cite{goeman2021only} to $\barFDP^{\KR}(\cdot)$, as discussed in the paragraph under Corollary~\ref{coro:KR-connect-CL} in Section~\ref{sec:closedtestingLocalTest}.
Before presenting these results, we provide a brief review of some results from \cite{goeman2021only} in Appendix~\ref{supp:recapJelle2021} and introduce two new lemmas in Appendix~\ref{supp:LemmaInterpolation}.


\subsection{A brief recap of \cite{goeman2021only}} \label{supp:recapJelle2021}

Following \cite{goeman2021only}, we present the results in terms of the true discovery guarantee defined by \eqref{TDguarantee}. All results can be easily translated to equivalent statements in terms of a simultaneous FDP bound, as we will show.

Assume that we have data $X$ from some distribution $P_{\theta}$ with $\theta \in\Theta$,
and we formulate hypotheses for $\theta$ in the form $\theta \in H \subseteq \Theta$.
We are interested in testing hypotheses $(H_i)_{i\in I}$, where $I \subseteq C \subseteq \bbN$ is finite.
Here $C$ plays the role of maximal set and will only be important later.
For any rejection set $R \subseteq I$, let $R_1 \subseteq R$ be the index set of false hypotheses (i.e., true discoveries).
We say that a random function (depending on data $X$) $\bd^I: 2^I \rightarrow \bbR$ has $(1-\alpha)$-\textit{true discovery guarantee} on $I$ if for all $\theta \in \Theta$, we have 
\begin{equation} \label{TDguarantee}
\Prob_{\theta} ( \bd^I(R) \leq |R_1| \text{ for all } R \subseteq I ) \geq 1-\alpha.
\end{equation}
We will usually suppress the dependence on $\alpha$ when discussing true discovery guarantees.

A true discovery guarantee and a simultaneous FDP bound are equivalent, 
in the sense that if we have a simultaneous FDP bound $\barFDP(R)$ satisfying~\eqref{simuFDPbounds}, 
then 
\begin{align}\label{TrueDiscovery-baesd-on-FDP}
\bd^I(R) = (1 - \barFDP(R) ) |R| 
\end{align}
satisfies the true discovery guarantee~\eqref{TDguarantee}.
Conversely, if $\bd^I(R)$ satisfies the true discovery guarantee~\eqref{TDguarantee},
then
\[
\barFDP(R) = \frac{|R| - \bd^I(R) }{\max \{1,|R| \} }
\]
satisfies~\eqref{simuFDPbounds}. 
Hence we can derive results in terms of $\bd^I(R)$, and then translate these to equivalent statements in terms of $\barFDP(R)$.

The \textit{interpolation} $\barbd^I$ of $\bd^I$ is defined as 
\begin{equation} \label{Interpolation}
\barbd^I(R) = \max_{U \in 2^I} \left\{ \bd^I(U) - |U \backslash R| + \bd^I(R \backslash U) \right\}.
\end{equation}
Lemma 2 in \cite{goeman2021only} shows that if $\bd^I$ has the true discovery guarantee~\eqref{TDguarantee}, then also $\barbd^I$.
It is clear that $\barbd^I \geq \bd^I$ by taking $U=R$ in \eqref{Interpolation}, which implies that interpolation will never result in worse bounds.

Interpolation may not be a one-off process.
We call $\bd^I$ \textit{coherent} if $\bd^I = \barbd^I$,
that is, the interpolation of $\bd^I$ does not bring further improvement.
The following lemma provides a convenient way to check for coherence.
\begin{lemma}[Lemma 3 in \cite{goeman2021only}] \label{Lemma3-Jelle2021}
	$\barbd^I$ is coherent if and only if for every disjoint $U,V \subseteq I$, we have 
	\[
	\bd^I(U) + \bd^I(V) \leq \bd^I(U \cup V) \leq 	\bd^I(U) + |V|.
	\]
\end{lemma}

We can embed procedure $\bd^I$ into a stack of procedures $\bd = (\bd^I)_{I \subseteq C, |I| \leq \infty}$, where we may have some maximal family $C \subseteq \bbN$. 
For example, when $\bd^I$ is based on equation~\eqref{TrueDiscovery-baesd-on-FDP} with the FDP bound obtained by our Algorithm~\ref{Algo-BasedOnSimuKFWER} applying to $(W_i)_{i\in I}$, then we can get a stack of procedures $( \bd^{I} )_{ I \subseteq [p]}$ (in this case $C = [p]$).

We call a stack of procedures $\bd$ a \textit{monotone procedure} if the following three criteria are fulfilled.
\begin{itemize}
	\item[1.] \textit{true discovery guarantee}: $\bd^I$ has true discovery guarantee for all finite $I \subseteq C$.
	\item[2.] \textit{coherent}: $\bd^I$ is coherent for all finite $I \subseteq C$.
	\item[3.] \textit{monotonicity}: $\bd^I(R) \geq \bd^J(R)$ for every finite $R \subseteq I \subseteq J \subseteq C $
\end{itemize}

Now we define a closed testing procedure. Let $\phi_R \in \{0,1\}$ be the local test for testing $H_R$, with $\phi_R=1$ indicating rejection. 
Choosing a local test for every finite $R \subseteq C$ will yield a suite of local tests $\phi = (\phi_R)_{R\subseteq C, |R|<\infty}$.
The true discovery guarantee procedure based on $\phi$ using closed testing (cf. Section~\ref{sec:reviewClsoedtesting}) is 
\[
\bd^I_{\phi}(R) = \min_{U \in 2^R} \{ |R \backslash U|: \phi^{I}_{U}=0 \},
\]
where 
\[
\phi^{I}_{R} = \min \{ \phi_{U}: R \subseteq U \subseteq I \}
\]
indicates whether $H_R$ is rejected by closed testing.

The main result of \cite{goeman2021only}
shows that any monotone procedure $\bd$ is either equivalent to a closed testing procedure with the explicit local test defined in \eqref{Thm1JelleLocaltest}, or it can be uniformly improved by that procedure.
\begin{theorem}[Theorem 1 in \cite{goeman2021only}]\label{Thm1:Jelle2021}
If $\bd^I$ has $(1-\alpha)$-true discovery guarantee, then for every finite $R \subseteq C$,
	\begin{align}\label{Thm1JelleLocaltest}
		\phi_{R} = \Ind{ \{ \bd^{R}(R) > 0 \} }
	\end{align}
	is a valid local test of $H_R$. That is,
	\[
	\sup_{\theta \in H_R} \Prob_{\theta} (\phi_R=1)\leq \alpha.
	\]
 
In addition, if $\bd = (\bd^I)_{I \subseteq C, |I| \leq \infty}$ is a monotone procedure, then for the suite $\phi = (\phi_R)_{R\subseteq C, |R|<\infty}$ and all $R \subseteq I \subseteq C$ with $|I| < \infty$, we have
	\[
	\bd_{\phi}^I(R) \geq \bd^I(R).
	\]
\end{theorem}

\subsection{Two lemmas about interpolation and coherence} \label{supp:LemmaInterpolation}

We refer to interpolation at several places in the paper. 
This technique has been used in \cite{goeman2021only} and \cite{blanchard2020post}. We now start from the interpolation formula~\eqref{Interpolation} of \cite{goeman2021only}. 
The following lemma shows that, for certain true discovery guarantee procedures given by \eqref{Joint-KFWER-TDguarantee}, their interpolation can be
simplified to \eqref{interpolation-specific}. 
This result is equivalent to the Proposition 2.5 of \citep{blanchard2020post}.
For $m=1$, $K_1=\whS(v)$ and $k_1=k-1$, \eqref{interpolation-specific} is equivalent to the last equation of \eqref{FDbound-interpolation}, even though the latter could be derived directly as we show in \eqref{FDbound-interpolation}.
In Appendix~\ref{supp:KR-Interpolation}, we apply the following lemma to obtain the interpolation version of the FDP bound in \cite{katsevich2020simultaneous}.

\begin{lemma}\label{Lemma:simplifiedInterpolation}
	For $m \geq 1$ and a set $I$, let $R\subseteq I$, $K_1 \subseteq K_2 \subseteq \dots \subseteq K_m \subseteq I$
	be nested sets, $k_i \in \bbZ_{>0}$ and $|K_i| \geq k_i$, $i \in [m]$.
	For a true discovery guarantee procedure of the form
	\begin{align} \label{Joint-KFWER-TDguarantee}
	\bd^{I}(R) = \begin{cases}
	|K_i| - k_i, & \text{if $R=K_i$ for some $1\leq i \leq m$},\\
	0, & \text{otherwise},
	\end{cases}
	\end{align}
	its interpolation is
	\begin{align}\label{interpolation-specific}
		\barbd^I(R) 
		= \max_{i\in[m]} \{|R \cap K_i| - k_i  \}  \vee 0.
	\end{align}
\end{lemma}

\begin{proof}
	By definition of $\barbd^I(R) $ (see \eqref{Interpolation}), we need to show that 
	\[
	\max_{U \in 2^I} \left\{ \bd^I(U) - |U \backslash R| + \bd^I(R \backslash U) \right\}
	= \max_{i\in[m]} \{|R \cap K_i| - k_i  \}  \vee 0.
	\]
	
	For any $R \subseteq I$.
	We consider three cases of $U \in 2^I$.
	
	(1) $U \in 2^I$ such that for all $i\in[m]$, $U \neq K_i$ and $R \backslash U \neq K_i$, 
	then we have
	\[
	\bd^I(U) - |U \backslash R| + \bd^I(R \backslash U) = - |U \backslash R|.
	\]
	
	(2) $U \in 2^I$ such that for some $i\in[m]$, $R \backslash U = K_i$. 
	Then $U \neq K_j$ for any $j\in[m]$ because $K_1 \subseteq \dots \subseteq K_{m}$. And we have that $K_i = R \cap K_i$ as $K_i \subseteq R$. Hence 
	\begin{align*}
	\bd^I(U) - |U \backslash R| + \bd^I(R \backslash U)
	&= - |U \backslash R| + \bd^I(K_i) \\
	&= - |U \backslash R| + |K_i| - k_i \\
	&= - |U \backslash R| + |R \cap K_i| - k_i 
	\end{align*}
	
	(3) $U \in 2^I$ such that for some $i\in[m]$, $U = K_i$. 
	Then $R \backslash U \neq K_j$ for any $j\in[m]$ because $K_1 \subseteq \dots \subseteq K_{m}$. Hence 
	\begin{align*}
	\bd^I(U) - |U \backslash R| + \bd^I(R \backslash U)
	&= \bd^I(K_i) - |K_i \backslash R| \\
	&=  |K_i| - k_i - |K_i \backslash R| \\
	&= |R \cap K_i| - k_i
	\end{align*}
	
	In addition, since $\barbd^I(R) \geq \bd^I(R) \geq 0$, we have
	\[
	\barbd^I(R) = \max_{U \in 2^I} \left\{ \bd^I(U) - |U \backslash R| + \bd^I(R \backslash U) \right\}
	= \max_{i\in[m]} \{ |R \cap K_i| - k_i\}  \vee 0,
	\]
	which proves the desired result.
\end{proof}


Next lemma shows that $\barbd^I$ defined in  \eqref{interpolation-specific} is coherent.
This lemma will be used in the proof of Proposition~\ref{prop:KR-CL-JellePaper}

\begin{lemma} \label{lemma:coherent-aspecialcase}
	For $m \geq 1$ and a set $I$, let $R\subseteq I$, $K_1 \subseteq K_2 \subseteq \dots \subseteq K_m \subseteq I$
	be nested sets, and $k_i \in \bbZ_{>0}$, $i \in [m]$.
	Let
	\[
	\barbd^I(R) = \max_{i \in [m]} \{|R \cap K_i| - k_i\} \vee 0.
	\]
	Then $\barbd^I(R)$ is coherent. 
\end{lemma}

\begin{proof}

 We show that $\barbd^{I}$ is coherent by applying Lemma~\ref{Lemma3-Jelle2021}. We first show that 
	\[
	\bd^I(U) + \bd^I(V) \leq \bd^I(U \cup V).
	\]

We either have (i) $\barbd^I(U)=0$ or/and $\barbd^I(V)=0$ or (ii) $\barbd^I(U)> 0$ and $\barbd^I(V)> 0$.


In case (i), it is clear that $\bd^I(U) + \bd^I(V) \leq \bd^I(U \cup V)$ by definition.

In case (ii), there exist $i_1$ and $i_2$ such that
\begin{align*}
    \barbd^I(U)=|U \cap K_{i_1}| - k_{i_1} >0
    \quad \text{and} \quad
    \barbd^I(V)=|V \cap K_{i_2}| - k_{i_2} >0,
\end{align*}
and we have 
\begin{align*}
    \barbd^I(U \cup V) = \max_{i\in [m]} \{ |U \cap K_i| + |V \cap K_i| - k_i  \} > 0.
\end{align*}
Under case (ii),
we either have $i_1=i_2$ or $i_1 \neq i_2$.
If $i_1=i_2$, $\bd^I(U) + \bd^I(V) \leq \bd^I(U \cup V)$ clearly holds.
If $i_1 \neq i_2$, we assume without loss of generality that $i_1<i_2$, 
	then we have
	\begin{align*}
		 |U \cap K_{i_2}| \geq |U \cap K_{i_1}| \geq |U \cap K_{i_1}| - k_{i_1}.
	\end{align*}
	since the $K_i$'s are nested. Therefore, 
	\begin{align*}
	\bd^I(U \cup V) \geq |U \cap K_{i_2}| + |V \cap K_{i_2}| - k_{i_2} 
	\geq |U \cap K_{i_1}| - k_{i_1} + |V \cap K_{i_2}| - k_{i_2} = \barbd^I(U) + \barbd^I(V)
	\end{align*}
	
	Next, we show that 
	\[
	\bd^I(U \cup V) \leq \bd^I(U) + |V|.
	\]
	Since
	\[
	|U \cap K_i| + |V \cap K_i| - k_i
	\leq 
	|U \cap K_i| - k_i + |V|,
	\]
	we have 
	\[
	( |U \cap K_i| + |V \cap K_i| - k_i ) \vee 0
	\leq 
	\left( ( |U \cap K_i| - k_i ) \vee 0 \right) + |V|,
	\]
	which implies the desired result.
	So $\barbd^{I}$ is coherent by Lemma~\ref{Lemma3-Jelle2021}. 
\end{proof}

\subsection{Interpolation version of the FDP bound in \cite{katsevich2020simultaneous}} \label{supp:KR-Interpolation}

Let $\bW=(W_1, \dots, W_p)$, with $|W_1| > |W_2| > \dots > |W_p| > 0$, be a knockoff statistic vector satisfying the coin-flip property, and 
let $\whS_i=\{ j\leq i: W_j > 0 \}$.
For nested sets $\whS_i$, $i \in [p]$,
\cite{katsevich2020simultaneous} proposed a simultaneous FDP bound (cf. $\widebar{V}_{\text{knockoff}}(\mathcal{R}_k)$ on the page $6$ of \cite{katsevich2020simultaneous})
$$\barFDP^{\KR_\text{ori}}(R) = \frac{\widebar{V}(\whS_i)}{\max \{1, |R| \}},$$ 
where
\begin{align} \label{KR-FDbound}
\widebar{V}(\whS_i) = \lfloor c(\alpha) \cdot(1+i-|\whS_i|) \rfloor
\quad
\text{with}
\quad 
c(\alpha) = \frac{\log(\alpha^{-1}) }{\log(2-\alpha) }.
\end{align}

The above bound can be uniformly improved and extended to all sets using interpolation \citep{goeman2021only,blanchard2020post}.
\cite{katsevich2020simultaneous} mentioned this, but did not present the formula for the interpolation version of the bound.
Here we give the explicit formula.

Specifically, 
we first transform the false discovery bound~\eqref{KR-FDbound} to its equivalent true discovery guarantee procedure following the equation (4) of \cite{goeman2021only}:
\begin{align} \label{KR-originalTDprocedure}
\bd^{[p]}_{\KR}(R) = \begin{cases}
|\whS_i| - \widebar{V}(\whS_i), & \text{if $R=\whS_i$ for some $1\leq i \leq p$},\\
0, & \text{otherwise}.
\end{cases}
\end{align}
Then, by Lemma~\ref{Lemma:simplifiedInterpolation}, the interpolated version of $\bd^{[p]}_{\KR}$ is
\begin{align} \label{KR-interpolation}
\barbd^{[p]}_{\KR}(R) = \max_{i\in[p]} \{|R \cap \whS_i| - \widebar{V}(\whS_i)  \}  \vee 0,
\end{align}
which is equivalent to the following simultaneous FDP bound (cf. formula~\eqref{KR-FDPbound-interpolation})
\[
\barFDP^{\KR}(R) = \frac{\min_{i\in[p]} \{ |R|, |R \backslash \whS_i | + \widebar{V}(\whS_i) \} } {\max \{1, |R| \}}, \quad \forall R \subseteq [p].
\]
Note that $\barFDP^{\KR}(\cdot)$ is better than the original FDP bound $\barFDP^{\KR_\text{ori}}(\cdot)$, in the sense that
\[
\barFDP^{\KR}(R) \leq \barFDP^{\KR_\text{ori}}(R), \quad \forall R \subseteq[p],
\]
because the interpolation $\barbd^{[p]}_{\KR}(R) \geq \bd^{[p]}_{\KR}(R)$.

\subsection{Applying the method of \cite{goeman2021only} to 
$\barFDP^{\KR}(\cdot)$} \label{supp:KR-JellePaper}

Finally, we apply Theorem~\ref{Thm1:Jelle2021} (Theorem 1 in \cite{goeman2021only}) to connect the simultaneous FDP bounds $\barFDP^{\KR}(\cdot)$ to closed testing. Compared to Corollary~\ref{coro:KR-connect-CL} in the main paper, here we can only obtain $\leq$ relation.




\begin{proposition} \label{prop:KR-CL-JellePaper}
	Let $\barFDP^{ct}(R)$ (see \eqref{closedtestingFDPbound}) be the FDP bound based on closed testing using test statistic~\eqref{formula:weighted-sum-test-forEquivalencePart} with
	$m=|I|$, $b^I_i=k^{raw}_{i}+i-1$ and critical values $z^I_i = k^{raw}_{i}$ to locally test $H_I$,
	then
	\[
	\barFDP^{ct}(R) \leq \barFDP^{\KR}(R), \quad R \subseteq [p].
	\]
\end{proposition}

\begin{proof}
	
	We will apply Theorem~\ref{Thm1:Jelle2021} for the proof. We proceed by the following four main steps.
	
	\textbf{Step 1: Define the stack of procedures $\barbd_{\KR} = (\barbd^{I}_{\KR})_{I \subseteq [p]}$.}

	For any $I \subseteq [p]$ and $i \leq |I|$, 
	let $\whS^I_i=\{ j \in I: W_j = W^I_l \text{ such that } l\leq i \text{ and } W^I_l > 0 \}$,
	where $W^I_l$ denotes the $W_j$ whose absolute value $|W_j|$ is the $l$-th largest in $ \{|W_j|: j \in I \}$.
	
	Based on $\{W_i: i \in I \}$, the corresponding true discovery control procedure (cf.\ \eqref{KR-interpolation} in Appendix~\ref{supp:KR-Interpolation}) of \cite{katsevich2020simultaneous} is
	\begin{align*} 
	\barbd^{I}_{\KR}(R) = \max_{i=1,\dots,|I|} \{|R \cap \whS^I_i| - \widebarV(\whS^I_i) \}  \vee 0, \quad \forall R \subseteq I,
	\end{align*}
	where $\widebar{V}(\whS^I_i) = \lfloor c(\alpha) \cdot(1+i-| \whS^{I}_i |) \rfloor$ and 
	$c(\alpha) = \frac{\log(\alpha^{-1})}{\log(2-\alpha)}$.

\textbf{Step 2: Show that $\barbd_{\KR} = (\barbd^{I}_{\KR})_{I \subseteq [p]}$ is a monotone procedure.}

For every $I \subseteq [p]$, it is clear that $\barbd^{I}_{\KR}$ has true discovery guarantee as $\{W_i: i \in I \}$ still possesses the coin-flip property. By Lemma~\ref{lemma:coherent-aspecialcase}, $\barbd^{I}_{\KR}$ is coherent, so it is enough to show the monotonicity. 
	
	For any $R \subseteq I \subseteq J \subseteq [p]$
	and any $j_1 \in [|J|]$.
	If $\whS^{J}_{j_1} \cap I = \emptyset$, we have $|R \cap \whS^{J}_{j_1}| - \widebarV(\whS^{J}_{j_1}) = - \widebarV(\whS^{J}_{j_1})\leq 0$. 
	Otherwise, if $\whS^{J}_{j_1} \cap I \neq \emptyset$, 
 let $i^* = \max \{\whS^{J}_{j_1} \cap I \}$ and let $j_2$ be the index such that $W^I_{j_2}=W_{i^*}$.
        By definition, we have
	$\whS^{J}_{j_1} \cap I = \whS^{I}_{j_2} = \whS^{I}_{j_2} \cap I$
	and $j_2-|\whS^{I}_{j_2}| \leq j_1-|\whS^{J}_{j_1}|$.
	Hence, 
	$|R \cap \whS^{I}_{j_2}| = |R \cap \whS^{I}_{j_2} \cap I| = |R \cap \whS^{J}_{j_1} \cap I| = |R \cap \whS^{J}_{j_1}|$ 
	and $\widebarV(\whS^{I}_{j_2}) \leq \widebarV(\whS^{J}_{j_1})$, 
	which implies that 
 $|R \cap \whS^{J}_{j_1}| - \widebarV(\whS^{J}_{j_1}) \leq |R \cap \whS^{I}_{j_2}| - \widebarV(\whS^{I}_{j_2}) $.
	Therefore,
	\[
	\barbd^{J}_{\KR}(R) =
	\max_{j_1=1,\dots,|J|} \{|R \cap \whS^{J}_{j_1}| - \widebarV(\whS^{J}_{j_1})  \}  \vee 0
	\leq 
	\max_{j_2=1,\dots,|I|} \{|R \cap \whS^{I}_{j_2}| - \widebarV(\whS^{I}_{j_2}) \} \vee 0 
	=\barbd^{I}_{\KR}(R),
	\]
	which shows that the monotonicity holds, so $\barbd_{\KR}$ is a monotone procedure.
	
	\textbf{Step 3: Apply Theorem~\ref{Thm1:Jelle2021}.}
	
	By Theorem~\ref{Thm1:Jelle2021}, for suite $\phi = (\phi_U)_{U\subseteq [p]}$ with local test
	\[
	\phi_U = \Ind{ \{ \barbd^U_{\KR}(U) > 0 \} },
	\]
	we have
	\[
	\bd_{\phi}^I(R) \geq \barbd^I_{\KR}(R), \quad \forall R \subseteq I \subseteq [p],
	\]
	where $\bd_{\phi}^I(R)$ is the true discovery guarantee procedure based on closed testing with suite $\phi$.
	Taking $I=[p]$, the above is equivalent to
	\[
	\barFDP^{ct}(R) \leq \barFDP^{\KR}(R),
	\]
	where $\barFDP^{ct}(R)$ is the simultaneous FDP bound of the closed testing using local test $\phi_U$
	for testing local hypothesis $H_U$.
	
	\textbf{Step 4: Show that the local test $\phi_U$ is equivalent to the one described in the proposition.}
	
	The local test
	$\phi_U = \Ind{ \{ \barbd^U_{\KR}(U) > 0 \} }$ means that we reject $H_U$ if and only if 
	\begin{align*}
	\barbd^U_{\KR}(U) > 0 
	&\Longleftrightarrow \max_{i=1,\dots,|U|} \{|U \cap \whS_i^U| - \widebarV(\whS_i^U) \}  > 0 \\
	&\Longleftrightarrow \text{there exists some $i=1,\dots,|U|$ such that } |\whS_i^U| > \widebarV(\whS_i^U) \text{ (since $\whS_i^U \subseteq U$)}.
	\end{align*}
	Let $\whL^{U}(i)$ be defined as in \eqref{formula:weighted-sum-test-forEquivalencePart}, so $|\whS_i^U| = \whL^{U}(i) $. Hence, we have 
	\begin{align*}
	|\whS_i^U| > \widebarV(\whS_i^U) 
	&\Longleftrightarrow |\whS_i^U| > \lfloor c(\alpha)\cdot(1+i-|\whS^{U}_i|) \rfloor 
	\Longleftrightarrow |\whS_i^U| > c(\alpha)\cdot(1+i-|\whS^{U}_i|) \\
	&\Longleftrightarrow |\whS_i^U| > \frac{c(\alpha) \cdot (1+i)}{1+c(\alpha)}
	\Longleftrightarrow |\whS_i^U| \geq \left\lfloor \frac{c(\alpha) \cdot (1+i)}{1+c(\alpha)} \right\rfloor + 1 \\
	&\Longleftrightarrow \whL^{U}(i) \geq \left\lfloor \frac{c(\alpha)\cdot (1+i)}{1+c(\alpha)} \right\rfloor + 1 = c_{i}.
	\end{align*}

For any positive integer $l$, there must exist some $i$ such that $i-c_i+1 = l$ because $i-c_i+1 = \left\lceil \frac{1}{1+c(\alpha)} (1+i) \right\rceil - 1$ and $\frac{1}{1+c(\alpha)} \in \left(0,\frac{1}{2} \right)$ for any $\alpha \in (0,1)$.
So we can define $k^{raw}_{i}= \min_{j\geq 1} \{c_j: j-c_j+1=i\} $ and let $b^I_i=k^{raw}_{i}+i-1$. Then,
	\begin{align*}
	&\text{ there exists some $i=1,\dots,|U|$ such that } \whL^{U}(i) \geq c_{i} \\
	\Longleftrightarrow
	&\text{ there exists some $i=1,\dots,|U|$ such that } \whL^{U}(b^I_i) \geq k^{raw}_{i}.
	\end{align*}
	Therefore, the local test $\phi_U$ is equivalent to a test based on the test statistic~\eqref{formula:weighted-sum-test-forEquivalencePart} with $m=|U|$, $b^U_i=k^{raw}_{i}+i-1$ and critical values $z^U_i=k^{raw}_{i}$.
\end{proof}

\section{Main proofs} \label{supp:OtherProofs}

\subsection{Proof of Lemma~\ref{lemma-NB}} \label{supp:prooflemma-NB}
\begin{proof}
	This proof follows the proof idea of Lemma 2 in \cite{LJ-WS:2016}.
	
	Let $m_0=|\mN|$ be the number of null variables. 
	Denote the random variables obtained by taking the signs of $W_i$, $i \in \mN$, by
	$B_1, \dots, B_{m_0}$.
	By the coin-slip property,
	$B_1, \dots, B_{m_0}$ are i.i.d.\ random variables with distribution $\Prob(B_i=+1)=\Prob(B_i=- 1)=1/2$.
	In the same probability space,
	we can generate i.i.d.\ $B_{m_0+1}, \dots$ from the same distribution to form an i.i.d.\ Bernoulli sequence $B_1, \dots, B_{m_0}, B_{m_0+1}, \dots$ taking values $\{+1,-1\}$.
	
	Let $U(v) = \min \{ i \geq 1: |\{j\leq i: B_j = -1 \}| = v \}$.
	Define $N^{m_0}(v)= |\{j\leq \min(U(v),m_0): B_j = 1 \}|$
	and $N^{p}(v) = |\{j\leq \min(U(v),p): B_j = 1 \}|$, which follow the early-stopped negative binomial distribution defined on one sequence of Bernoulli trials of length $m_0$ and $p$, respectively.
	Then, by definition,
	for any $v \geq 1$, we have
	\[
	|\whS(v) \cap \mN| \leq N^{m_0}(v) \leq N^{p}(v), 
	\]
	where the first inequality is due to the fact that 
	some $W_i$, $i \in [p] \setminus \mN$, might be negative, so the corresponding threshold
	$\whT(v)$ 
	will lead to an $\whS(v)$ 
	such that $|\whS(v) \cap \mN| \leq N^{m_0}(v)$.
	Hence the event $\{ N^{p}(v_1) \leq k_1-1, \dots, N^{p}(v_m) \leq k_m-1 \}$ implies the event $\{ |\whS(v_1) \cap \mN| \leq k_1-1, \dots, |\whS(v_m) \cap \mN| \leq k_m-1 \}$, which leads to the desired inequality.
	
\end{proof}

\subsection{Proof of Proposition~\ref{prop:KR-NB}} \label{supp:proofprop-KR-NB}

The following proof is essentially based on the first claim of Theorem~\ref{Thm1:Jelle2021}.
Here, for simplicity, we proceed directly based on the simultaneous FDP guarantee, rather than converting it to its equivalent true discovery guarantee and then applying Theorem~\ref{Thm1:Jelle2021}.


\begin{proof}
	We first define two integers $i_1 = \min \{i \geq 1: i - c_i +1 =1 \}$ 
	and $i_m = \min \{i \geq 1: i - c_i +1 =v_m \}$. 
	Note that they are well-defined because 
	\begin{align*} 
		i-c_i+1 
		= i - \left\lfloor \frac{c(\alpha) \cdot (1+i)}{1+c(\alpha) } \right\rfloor
		= \left\lceil i - \frac{c(\alpha) }{1+c(\alpha)} (1+i) \right\rceil 
		= \left\lceil \frac{1}{1+c(\alpha)} (1+i) \right\rceil - 1,
	\end{align*}
    and $\frac{1}{1+c(\alpha)} \in \left(0,\frac{1}{2} \right)$ for any $\alpha \in (0,1)$.
    So for any positive integer $l$, there must exist some $i$ such that $i-c_i+1 = l$.
	
	Let $q = \max \{i_m, p \}$ 
	and consider a knockoff statistic vector $\bW=(W_1,W_2, \dots, W_{q})$ satisfying the coin-flip property and $|W_1| > |W_2| > \dots > |W_q| > 0$.
	Since $\barFDP^{\KR}(\cdot)$ (see \eqref{KR-FDPbound-interpolation}) is a simultaneous FDP bound, we have 
	\[
	\Prob\left( \FDP(R) \leq \barFDP^{\KR}(R), \forall R \subseteq [q]\right)
	\geq 1-\alpha.
	\]
	Under the global null case ($\FDP(R)=1$ for $R \neq \emptyset$) and taking $R=[q]$, we have
	\begin{align*}
		 \Prob\left( |\whS_i | \leq  \lfloor c(\alpha) \cdot(1+ i - |\whS_i| ) \rfloor, \forall i \in [q] \right)
		= &  \Prob\left( q \leq \min_{i \in [q]} \{ |[q] \backslash \whS_i | + \lfloor c(\alpha) \cdot(1+ i - |\whS_i| ) \rfloor , q \} \right) \\
		= & \Prob\left( \FDP([q]) \leq \barFDP^{\KR}([q]) \right) \\
		\geq & 1-\alpha.
	\end{align*}
	
	Because
	\begin{align*}
	|\whS_i| \leq \lfloor c(\alpha)\cdot(1+i-|\whS_i|) \rfloor 
	&\Longleftrightarrow |\whS_i| \leq c(\alpha)\cdot(1+i-|\whS_i|) \\
	&\Longleftrightarrow |\whS_i| \leq \frac{c(\alpha) \cdot (1+i)}{1+c(\alpha) } \\
	&\Longleftrightarrow |\whS_i| \leq \left\lfloor \frac{c(\alpha) \cdot (1+i)}{1+c(\alpha) } \right\rfloor
	= c_i-1,
	\end{align*}
	we have
	\begin{equation} \label{proofstep:Sinequa}
		\begin{aligned}
			& \Prob(|\whS_{i_1}| \leq c_{i_1}-1, |\whS_{i_1+1}| \leq c_{i_1+1}-1, \dots, |\whS_{q}| \leq c_{q}-1 ) \\
			& \quad \geq  \Prob(|\whS_i| \leq c_i-1, \forall i \in [q] ) \\
			& \quad =  \Prob\left( |\whS_i | \leq  \lfloor c(\alpha) \cdot(1+ i - |\whS_i| ) \rfloor, \forall i \in [q] \right) \\
			& \quad \geq  1-\alpha.
		\end{aligned}
	\end{equation}
	
	Let
	\[
	N^{q}(v) = |\{i \in [q]: W_i \geq \whT^{q}(v) \}|,
	\]
	where $ \whT^{q}(v) = \max \{ |W_i|: |\{j \in [q]: |W_j|\geq |W_i| \text{ and } W_j < 0 \}| =v \}$	
	with $\whT^{q}(v) = |W_q|$ if $|\{i \in [q]: W_i < 0 \}| <v$. 
	Note that $N^{q}(v)$ is an early-stopped negative binomial random variable since we are under the global null case.
	For $i \geq i_1$, we have $i-c_i \geq 0$ because the sequence $i-c_i$ is monotonically increasing and $i_1-c_{i_1}=0$. 
	By the fact \eqref{fact2} in Appendix~\ref{supp:Afact},
	for $i \geq i_1$, we have
	\[
	|\whS_i| \leq c_i-1 	\Longleftrightarrow N^{q}(i-c_i+1) \leq c_i-1.
	\]
	
	Therefore, by inequality~\eqref{proofstep:Sinequa}, we have
	\begin{align*}
	&\Prob(N^{q}(i_1-c_{i_1}+1) \leq c_{i_1}-1, 
 N^{q}(i_1+1-c_{i_1+1}+1) \leq c_{i_1+1}-1,
 \dots, N^{q}(q-c_{q}+1) \leq c_{q}-1) \\
	& \quad =  \Prob(|\whS_{i_1}| \leq c_{i_1}-1, 
 |\whS_{i_1+1}| \leq c_{i_1+1}-1, \dots, |\whS_{q}| \leq c_{q}-1 ) \\
& \quad \geq 1-\alpha.
	\end{align*}

    Define $k^{raw}_i= \min_{j\geq 1}\{c_j: j-c_j+1=i\} $ for $i \in [v_m]$. 
    By the definition of $q$, 
    for $1\leq v_1 < \dots < v_m$,
    we have
	\begin{align*}
	& \Prob(N^{p}(v_1) \leq k^{raw}_{v_1}-1, N^{p}(v_2) \leq k^{raw}_{v_2}-1, \dots, N^{p}(v_m) \leq k^{raw}_{v_m}-1) \\
	& \quad \geq  \Prob(N^{p}(1) \leq k^{raw}_1-1, N^{p}(2) \leq k^{raw}_2-1, \dots, N^{p}(v_m) \leq k^{raw}_{v_m}-1) \\
	& \quad \geq  \Prob(N^{q}(1) \leq k^{raw}_1-1, N^{q}(2) \leq k^{raw}_2-1, \dots, N^{q}(v_m) \leq k^{raw}_{v_m}-1) \\
	& \quad \geq \Prob(N^{q}(i_1-c_{i_1}+1) \leq c_{i_1}-1, 
 N^{q}(i_1+1-c_{i_1+1}+1) \leq c_{i_1+1}-1,
 \dots, N^{q}(q-c_{q}+1) \leq c_{q}-1) \\
	& \quad \geq  1-\alpha.
	\end{align*}
\end{proof}

\subsection{Proof of Proposition~\ref{prop:KR-Algo1}} \label{supp:proofprop-KR-Algo1}

To prove Proposition~\ref{prop:KR-Algo1}, we use the following Lemma~\ref{lemma:equivalenceOfTwoNumbers}.
\begin{lemma} \label{lemma:equivalenceOfTwoNumbers}
	For any $\alpha \in (0,1)$ and $i \geq 1$,
	\[
	\min_{j\geq 1}\{c_j: j-c_j+1=i\} = \lfloor c(\alpha) i \rfloor + 1,
	\]
	where
	$c_j = \left\lfloor \frac{c(\alpha) \cdot (1+ j )}{1+c(\alpha)} \right\rfloor  + 1$
	and $c(\alpha) = \frac{\log(\alpha^{-1})}{\log(2-\alpha)}$. 
\end{lemma}


We give the proof of Lemma~\ref{lemma:equivalenceOfTwoNumbers}.
\begin{proof}
	For convenience of notation, let 
	\[
	\beta  = \frac{c(\alpha)}{1+c(\alpha)},
	\]
	so
	\begin{align*}
	j - c_j + 1 &= j - \left\lfloor \frac{c(\alpha) \cdot (1+ j )}{1+c(\alpha)} \right\rfloor
	= j - \left\lfloor \beta (1+ j ) \right\rfloor
	= \left\lceil j - \beta (1+ j ) \right\rceil
	= \left\lceil j (1- \beta) - \beta \right\rceil,
	\end{align*}
	and
	\[
    \min_{j \geq 1}\{c_j: j-c_j+1=i\} = \min_{j \geq 1} \{ c_j: \left\lceil j (1- \beta) - \beta \right\rceil = i \}.
	\]
	
	Note that
	\begin{align*}
	\left\lceil j (1- \beta) - \beta \right\rceil = i
	\Longleftrightarrow
	i-1 <  j (1- \beta) - \beta \leq i
	\Longleftrightarrow
	\frac{i + \beta - 1}{1- \beta} < j \leq \frac{i + \beta }{1- \beta},
	\end{align*}
	and $\frac{i + \beta - 1}{1- \beta}$ must not be an integer, 
	we have that
	\[
	\argmin \{ j \geq 1: \left\lceil j (1- \beta) - \beta \right\rceil =i \}
	= \left\lceil \frac{i + \beta - 1}{1- \beta} \right\rceil =: j^*,
	\]
	so
	\begin{align*}
	& \min_{j \geq 1} \{ c_j: \left\lceil j (1- \beta) - \beta \right\rceil = i \}
	= c_{j^*}
	= \left\lfloor \beta \left( 1+j^* \right) \right\rfloor + 1 \\
	=& \left\lfloor \beta \left(1+ \left\lceil \frac{i + \beta - 1}{1- \beta} \right\rceil \right) \right\rfloor + 1 
	= \left\lfloor \beta \left\lceil \frac{i}{1- \beta} \right\rceil \right\rfloor + 1 \\
	=& \left\lfloor \frac{c(\alpha)}{1+c(\alpha)} \left\lceil (1+c(\alpha)) i \right\rceil \right\rfloor + 1 
	\end{align*}
	
	Therefore, to show that 
	\[
	\min_{j\geq 1}\{c_j: j-c_j+1=i\} = \lfloor c(\alpha) i \rfloor + 1,
	\]
	it is sufficient to show that 
	\begin{align} \label{proof:equi-floor}
	\left\lfloor \frac{c(\alpha)}{1+c(\alpha)} \left\lceil (1+c(\alpha)) i \right\rceil \right\rfloor
	= \lfloor c(\alpha) i \rfloor.
	\end{align}
	When $c(\alpha) i$ is an integer, \eqref{proof:equi-floor} clearly holds.
	
	Now assume that $c(\alpha) i$ is not an integer.
	For convenience of notation, let
	\[
	\gamma = \frac{c(\alpha)}{1+c(\alpha)} \left\lceil (1+c(\alpha)) i \right\rceil,
	\]
	note that we have $c(\alpha) i \leq \gamma$.
	To show \eqref{proof:equi-floor}, it is equivalent to show that
	$\gamma$ and $c(\alpha) i$ have the same integer part,
	which is equivalent to
	\begin{equation*}
	\begin{aligned}
	\gamma < \left\lceil c(\alpha) i \right\rceil
	&\Longleftrightarrow
	\frac{c(\alpha)}{1+c(\alpha)} \left\lceil (1+c(\alpha)) i \right\rceil < \left\lceil c(\alpha) i \right\rceil \\
	&\Longleftrightarrow
	\frac{c(\alpha)}{1+c(\alpha)} i + \frac{c(\alpha)}{1+c(\alpha)} \left\lceil c(\alpha) i \right\rceil < \left\lceil c(\alpha) i \right\rceil \\
	&\Longleftrightarrow
	\frac{c(\alpha)}{1+c(\alpha)} i < \left\lceil c(\alpha) i \right\rceil \left( 1 -  \frac{c(\alpha)}{1+c(\alpha)} \right) \\
	&\Longleftrightarrow
	\frac{c(\alpha)}{1+c(\alpha)} i < \left\lceil c(\alpha) i \right\rceil \frac{1}{1+c(\alpha)}  \\
	&\Longleftrightarrow
	c(\alpha) i < \left\lceil c(\alpha) i \right\rceil,
	\end{aligned}
	\end{equation*}
	which clearly holds.
\end{proof}

Now we prove Proposition~\ref{prop:KR-Algo1}.
\begin{proof}[Proposition~\ref{prop:KR-Algo1}]

Let $\whS^{-}_i = \{  j\leq i: W_j < 0 \}$ and $\widebar{V}(|\whS^{-}_i|) = \lfloor c(\alpha) \cdot(1+|\whS^{-}_i|) \rfloor$. To prove $\barFDP^p_{(\bk^{raw},\bv)}(R) = \barFDP^{\KR}(R) $,
it is sufficient to show that
	\begin{align} \label{proof-3.1: target}
	\min_{i=1,\dots,p} \{ k^{raw}_i-1 + |R \backslash \whS(i)|, |R|\}
	= \min_{j=1,\dots,p} \{ \widebar{V}(|\whS^{-}_j|) + |R \backslash \whS_j |  , |R| \},
	\end{align}
 where $\whS_i = \{  j\leq i: W_j > 0 \}$ and $\whS(i)$ is defined in \eqref{setS}.
	
	Note that $\widebar{V}(i) \geq 1$ is an increasing sequence,
	and $k^{raw}_i \geq 1$ is also increasing
	as $i-c_i$ and $c_i$ are both increasing.
	Let $n^{-} $ be the total number of negatives in $(W_1, W_2, \dots, W_p)$.
	
	\textbf{In the case of $n^{-}=p$,} by definition, we have
	\[
	\min_{i=1,\dots,p} \{ k^{raw}_i-1 + |R \backslash \whS(i)|, |R|\} = \min \{ k^{raw}_1-1 + |R|, |R|\} = |R|
	\]
	and
	\[
	\min_{j=1,\dots,p} \{ \widebar{V}(|\whS^{-}_j|) + |R \backslash \whS_j |  , |R| \} = \min_{j=1,\dots,p} \{ \widebar{V}(j) + |R|  , |R| \} = |R|,
	\]
	so \eqref{proof-3.1: target} holds.
	
	\textbf{In the case of $n^{-}<p$,} to show \eqref{proof-3.1: target}, we first rewrite its both sides.
	
	For the left hand side of \eqref{proof-3.1: target}, 
	by definition and the fact that $k^{raw}_i$ is increasing, we have
	\begin{align} \label{proofstepPiece1}
		\min_{i=1,\dots,p} \{ k^{raw}_i-1 + |R \backslash \whS(i)|, |R|\} 
		=
		\min_{i=1,\dots,n^{-}+1} \{ k^{raw}_i-1 + |R \backslash \whS(i)|, |R|\}.
	\end{align}

	For the right hand side of \eqref{proof-3.1: target}, 
	let $\whS_0=\whS_0^{-}=\emptyset$
	and $a_l = \max\{j \geq 0: |\whS^{-}_j|=l\}$ 
	for $l=0,\dots,n^{-}$. Then, by definition,
	\begin{align} \label{proofstepPiece2}
	\min_{j=1,\dots,p} \{ \widebar{V}(|\whS^{-}_j|) + |R \backslash \whS_j |, |R| \}
	=
	\min_{l=0,\dots,n^{-}} \{  \widebar{V}(l) + |R \backslash \whS_{a_l} |, |R| \}
 =
	\min_{l=1,\dots,n^{-}+1} \{  \widebar{V}(l-1) + |R \backslash \whS_{a_{l-1}} |, |R| \}.
    \end{align}
	
	At last, for $u=1,\dots,n^{-}+1$, we have
	\begin{align} \label{proofstepPiece3}
		\whS(u) = \whS_{a_{u-1}} 
	\end{align}
	by definition. 
	
	Therefore, by combining \eqref{proofstepPiece1}, \eqref{proofstepPiece2} and \eqref{proofstepPiece3}, to prove \eqref{proof-3.1: target},
	it is sufficient to show that for any $u=1,\dots,n^{-}+1,$
	\[
	k^{raw}_{u}-1 = \widebar{V}(u-1),
	\]
	that is,
	\[
	\min_{v\geq 1}\{c_v: v-c_v+1=u\} - 1=  \lfloor c(\alpha) u \rfloor,
	\]
	which holds by Lemma~\ref{lemma:equivalenceOfTwoNumbers}.
\end{proof}

\subsection{Proof of Proposition~\ref{prop:improveKR}} 

\begin{proof}
	Since  $\bk$ is the output of Algorithm~\ref{Algo-getTuningParameterK},
    it is clear that $k_i \leq k_i^{raw}$ for all $i \in [p]$.
    Therefore, by Proposition~\ref{prop:KR-Algo1}, we have
    $$\barFDP^p_{(\bk,\bv)}(R) \leq \barFDP^p_{(\bk_{raw},\bv)}(R) = \barFDP^{\KR}(R), \quad \forall R\subseteq [p].$$
\end{proof}

\subsection{Proof of Theorem~\ref{thm:Algo-BasedOnJointKFWER-as-CL}}

The proof of Theorem~\ref{thm:Algo-BasedOnJointKFWER-as-CL} relies on the following result, 
which gives the explicit formula (or shortcut) for the false discovery bound of the closed testing procedure with certain local test.


\begin{theorem} \label{thm:equivalence-closedtesting}
	For $m, p \geq 1$ and any $I \subseteq [p] $,
	let $S^I_1 \subseteq \dots \subseteq S^I_{m} \subseteq I$ be a sequence of nested sets defined based on $I$.
	Denote $N^I = I \backslash S^{I}_{m}$.
	For $1 \leq k_1 \leq \dots \leq k_{m}$ and $R \subseteq I$, let 
	\begin{align*}
		t^I(R) = \min_{i \in [m] } \{|R \backslash S^I_{i}| + k_i - 1, |R| \}
	\end{align*}
	and $t^I_{ct}(R)$ be the false discovery upper bound by closed testing (see \eqref{closedtestingFDbound}) with local test 
	\begin{align*}
		\phi_R = \max_{i \in [m]} \Ind{ \{ |S^R_{i}| \geq k_i \} }.
	\end{align*}
	For any $R \subseteq I$, if
	\begin{align} \label{consition-on-set}
	R \cap S^I_i = S_i^{R \cup N^I} \subseteq S_i^{R},
	\end{align}
	then
	\begin{align*}
	t^I_{ct}(R) = t^I(R).
	\end{align*}
\end{theorem}

We will first provide the proof of Theorem~\ref{thm:Algo-BasedOnJointKFWER-as-CL}, followed by the proof of Theorem~\ref{thm:equivalence-closedtesting}.
\begin{proof}
	For any $I \subseteq [p]$,
	let $\whS^I(v_i) = \{j \in I: W_j \geq \whT^I(v_i) \}$ and
	$\whT^I(v_i) = \max \{ |W_j|: j\in I, |\{l \in I: |W_l|\geq |W_j| \text{ and } W_l < 0 \}| =v_i \}$
	with $\whT^I(v_i) = \min_{j \in I } \{ |W_j| \}$ if $ |\{j \in I: W_j < 0 \}| < v_i $.
	
	Let $S^I_i = \whS^I(v_i)$ and $N^I = I \backslash \whS^{I}_{m}$.
	By the definition of $S^I_i$, we have
	$S^I_1 \subseteq \dots \subseteq S^I_{m} \subseteq I$
	and 
	\begin{align*}
	R \cap S^I_i = S_i^{R \cup N^I} \subseteq S_i^{R}, \quad \forall R \subseteq I.
	\end{align*}
    Hence, for $1 \leq k_1 \leq \dots \leq k_{m}$,  by Theorem~\ref{thm:equivalence-closedtesting} and taking $I=[p]$,we have 
    \begin{align} \label{equi-FDbound}
    	t^{[p]}_{ct}(R) = t^{[p]}(R), \quad R \subseteq [p],
    \end{align}
    where
    \begin{align*}
    	t^{[p]}(R) = \min_{i \in [m] } \{|R \backslash S^{[p]}_{i}| + k_i - 1, |R| \}
    \end{align*}
     and $t^{[p]}_{ct}(R)$ is the false discovery upper bound by closed testing (see \eqref{closedtestingFDbound}) with local test 
	\begin{align*}
	\phi_R = \max_{i \in [m]} \Ind{ \{ |S^R_{i}| \geq k_i \} }.
	\end{align*}
	
	At last, because of the fact \eqref{fact1} in Appendix~\ref{supp:Afact},
	the above local test $\phi_R$ is equivalent to the test statistic~\eqref{formula:weighted-sum-test-forEquivalencePart} with $b^R_j=k_j+v_j-1$ and critical values $z^R_j =k_j$ for locally testing $H_R$.
	Hence, $\barFDP^{ct}(R)$ is equal to the left hand side of equality~\eqref{equi-FDbound} divided by $\max(1, |R|)$.
	In addition, note that $\barFDP^m_{(\bk, \bv)}(R)$ is equal to the right hand side of equality~\eqref{equi-FDbound} divided by $\max(1, |R|)$,
	so we have 
	\[
	\barFDP^{ct}(R) = \barFDP^m_{(\bk,\bv)}(R), \quad R \subseteq [p].
	\]
\end{proof}

Now we give the proof of Theorem~\ref{thm:equivalence-closedtesting}.
\begin{proof}

	For any $R \subseteq I$, we prove $t^I_{ct}(R) = t^I(R)$ by showing that the equality holds in both cases where  $H_{R}$ is not rejected and rejected by closed testing. Our proof is based on how closed testing works (see Section~\ref{sec:reviewClsoedtesting}), and we will use the fact that $t^I_{ct}(R)$ is the size of the largest subset of $R$ that is not rejected by closed testing.
	
	The following result, which is implied by condition~\eqref{consition-on-set}, will be used throughout the proof:
	\begin{align} \label{LocalTestInequality}
		|R \cap S^I_i| =|S_i^{R \cup N^I}| \leq |S^R_{i}|, \quad \forall R \subseteq I.
	\end{align}
	
	
	Now we start our proof.
	
	\textbf{Case 1: $H_{R}$ is not rejected by closed testing.}
	
	In this case, the largest subset of $R$ that is not rejected by closed testing is $R$ itself, so we have $t^I_{ct}(R) = |R|$.
	In addition, there must exist some $R_{\text{sup}} \supseteq R$ such that $H_{R_{\text{sup}}}$ is not rejected by the local test.
	That is, for all $i \in [m]$, we have $|S^{R_{\text{sup}}}_{i}| \leq k_i-1$.
 	Hence, for $a = \argmin_{i \in [m]} \{ |R \backslash S^I_{i}| + k_i - 1\} $,
 	we have $ |R \cap S^I_{a}| \leq  |R_{\text{sup}}  \cap S^I_{a}| \leq |S^{R_{\text{sup}}}_{a}| \leq k_{a}-1$, where the middle inequality holds due to \eqref{LocalTestInequality}.
	As a result, $t^I_{ct}(R) = |R| = |R \cap S^I_{a}| + |R \backslash S^I_{a}|=  \min \{|R \cap S^I_{a}|, k_{a}-1 \} + |R \backslash S^I_{a}| = \min \{|R|, |R \backslash S^I_{a}| +k_{a}-1 \}  = t^I(R).$
	
	\textbf{Case 2: $H_{R}$ is rejected by closed testing.}
	
	In this case, all supersets of $R$ are locally rejected.
	Hence, there must exist some $b \in [m]$ such that $|R \cap S^I_b| \geq k_b$. 
	Otherwise, the set $R \cup N^I$, which is a superset of $R$, is not rejected by the local test
	because $|S^{R \cup N^I}_{i}| = |R \cap S^I_i| \leq k_i-1$ for all $i\in[m]$, leading to a contradiction.
	
	As a result, the set 
	\[
	B=\{ i\in[m]: |R \cap S^I_i| \geq k_i \}
	\]
	is non-empty.
	We denote 
	\[
	B = \{ b_1, \dots, b_{|B|} \}
	\]
	such that $S^I_{b_1} \subseteq \dots \subseteq S^I_{b_{|B|}}$.
	Hence,
	\begin{align}\label{proof:Sb|B|}
		S^I_{b_{|B|}} = \cup_{i \in B} S^I_i
	\end{align}
    and
	\begin{align}\label{proof:increasingRcapS}
		R\cap S^I_{b_1} \subseteq \dots \subseteq R\cap S^I_{b_{|B|}}.
	\end{align}
	Let
	\begin{align}\label{proof:defBc}
		B^c=\{i\in[m]: |R \cap S^I_i| \leq k_i-1\}.
	\end{align}
	
	In the following proof, we first construct $R_{sub}$, the largest subset of $R$ that is not rejected by closed testing,
	based on which we can obtain the explicit formula of $t^I_{ct}(R)$.
	Then we prove $t^I_{ct}(R) = t^I(R)$ based on this explicit formula.
	We proceed in three main steps.
	
	\textbf{Step 1: Construct $R_{sub}$.}
	
	
	For $i=1,\dots, |B|$,
	we first construct $E_{i}$ as follows:
	Let $E_{1} = R \cap S^I_{b_{1}}$.
	In the case that $|B| \geq 2$, also let 
	$E_{i} = (R\cap S^I_{b_{i}}) \backslash (R\cap S^I_{b_{i-1}})$ for $i=2, \dots, |B|$.
	
	Then, we construct $F_i$ as follows:
	Take $F_1$ as a subset of $E_1$ such that $|F_1|=k_{b_1}-1$. 
	Note that such an $F_1$ must exist because $|E_1| = |R \cap S^I_{b_{1}}| \geq k_{b_1}$.
	In the case that $|B| \geq 2$, we construct $F_i$, $i=2, \dots, |B|$, in a recursively manner:
	Take $F_i$ as a subset of $E_i$ with size $ \min \{k_{b_i} -|\cup_{j\leq i-1} F_{j} | - 1, |E_{i}|\}$. 
	
	By construction, we have the following properties of $E_i$ and $F_i$ which we will use in the remaining proofs:
	\begin{itemize}
		\item[-] $F_i \subseteq E_i \subseteq S^I_{b_i} \subseteq S^I_{b_{|B|}}$.
		\item[-] $F_i \subseteq E_i \subseteq R$.
		\item[-] $E_{i}$'s are disjoint. 
		\item[-] For $i=1, \dots, |B|$, $\cup_{j\leq i} E_{j} = R \cap S^I_{b_i}$ (due to \eqref{proof:increasingRcapS}).
		\item[-] $F_{i}$'s are disjoint. 
		\item[-] For $i=1, \dots, |B|$, $|\cup_{j\leq i} F_{j} | =  |\cup_{j\leq i-1} F_{j} | + |F_i| \leq k_{b_i}-1$.
		\item[-] For $i=2, \dots, |B|$, $k_{b_i} -|\cup_{j\leq i-1} F_{j} | - 1 \geq k_{b_i} -k_{b_{i-1}} \geq 0$.
	\end{itemize}

	
	Finally, let 
	\begin{align} \label{proofDefRsub}
	R_{sub} = \cup_{i\leq |B|} F_i \cup (R \backslash S^I_{b_{|B|}}).
    \end{align}
	Note that 
	\begin{align} \label{proofDef|Rsub|}
	|R_{sub}| = \sum_{i=1}^{|B|} |F_i| + |R \backslash S^I_{b_{|B|}}|
	\end{align}
     because $F_i$'s and $R \backslash S^I_{b_{|B|}}$ are disjoint.
    
    \textbf{Step 2: Prove that  $R_{sub}$ is the largest subset of $R$ that is not rejected by closed testing.}

	To prove this, we first show that $H_{R_{sub}}$ is not rejected by closed testing,
	then we show that any other subset of $R$ which is strictly larger than $H_{R_{sub}}$ must be rejected by closed testing.
	
	\textbf{Step 2-1: Prove that $H_{R_{sub}}$ is not rejected by closed testing.}
	
	It is sufficient to show that the superset $R_{sub} \cup N^I$ of $R_{sub}$ is not locally rejected,
	that is, $|S^{R_{sub} \cup N^I}_{i}| \leq k_i-1$ for all $i \in [m]$.
	In the following, we prove that this inequality holds for both $i \in B^c$ and $i \in B$ cases, thus it holds for all $i \in [m]$.
	
	\textbf{In the case of $i \in B^c$}, we have $|S^{R_{sub} \cup N^I}_{i}|  = |R_{sub} \cap S^I_i| \leq |R \cap S^I_i| \leq k_i-1$,
	where the first equality is due to \eqref{LocalTestInequality}, the first inequality is due to the fact that $R_{sub} \subseteq R$,
	and the last inequality holds by the definition of $B^c$ (see~\eqref{proof:defBc}).
	
	\textbf{In the case of $i \in B$}, denote $i=b_l$.
	In the following, we consider two sub-cases that $l=|B|$ and $1 \leq l<|B|$. Note that the former case covers the scenario that $|B|=1$.
	
	\textbf{In the sub-case of $l=|B|$}, that is, $i=b_{|B|}$, we have
	\begin{align*}
	|S^{R_{sub} \cup N^I}_{i}|
	&= |R_{sub} \cap S^I_i|
	= |R_{sub} \cap S^I_{b_{|B|}}|
	= |\cup_{j\leq |B|} F_j|
	\leq k_{b_{|B|}}-1 = k_{i}-1,
	\end{align*}
     where the third equality holds by the definition of $R_{sub}$ (see \eqref{proofDefRsub}).

    \textbf{In the sub-case of $1 \leq l<|B|$}, for $l< j \leq |B|$, we have
	\begin{align} \label{proofEmptySet}
	F_{j} \cap S^I_{b_l} 
	\subseteq E_{j} \cap S^I_{b_l} 
	= E_{j} \cap S^I_{b_l} \cap R 
	= (R\cap S^I_{b_{j}}) \backslash (R\cap S^I_{b_{j-1}}) \cap (R \cap S^I_{b_l}) = \emptyset,
	\end{align}
	where the first equality holds because $E_{j}$ is a subset of $R$, the second equality holds by the definition of $ E_{j}$,
	and the last equality holds because $R \cap S^I_{b_{l}} \subseteq R\cap S^I_{b_{j-1}}$.
	Hence,
	\begin{align*}
	|S^{R_{sub} \cup N^I}_{i}|
	&= |R_{sub} \cap S^I_i|
	= |R_{sub} \cap S^I_{b_l}| 
	= |(\cup_{j\leq |B|} F_j) \cap S^I_{b_l}|
	\stackrel{\eqref{proofEmptySet}}{=} |\cup_{j \leq l} (F_j \cap S^I_{b_l})|  \\
	&\leq |\cup_{j \leq l} F_j |  \leq  k_{b_l}-1 = k_i-1.
	\end{align*}
	
	Combining all above cases, we have shown that $H_{R_{sub}}$ is not rejected by closed testing.

	\textbf{Step 2-2: Prove that any other subset of $R$ which is strictly larger than $H_{R_{sub}}$ must be rejected by closed testing.}
	
	Formally, we show that for any $R' \subseteq R$ with $|R'| \geq |R_{sub}| + 1$,
	$H_{R'}$ must be rejected by closed testing.
	That is, all its supersets must be locally rejected.
	
	First, because $\cup_{i\leq |B|} E_{i} = R \cap S^I_{b_{|B|}}$,
	such $R'$ can always be written as $R' = \cup_{i\leq |B|} R'_{i} \cup R'_{c} \subseteq R$, 
	where $R'_{i} \subseteq E_i$ and $R'_{c} \subseteq R\backslash S^I_{b_{|B|}}$.
	Note that $R'_{i}$'s are disjoint because $E_{i}$'s are disjoint.
	By this formulation of $R'$, we have $|R'| = \sum_{i=1}^{|B|} |R'_i| + |R'_{c}|$.
	
	Then, 
	we must have
	\begin{align} \label{proof:substep1}
		\sum_{i=1}^{|B|} |R'_i| \geq \sum_{i=1}^{|B|} |F_i| + 1,
	\end{align}
	otherwise if $\sum_{i=1}^{|B|} |R'_i| \leq \sum_{i=1}^{|B|} |F_i|$, 
	we have $|R'| = \sum_{i=1}^{|B|} |R'_i| + |R'_{c}| \leq \sum_{i=1}^{|B|} |F_i| + |R \backslash S^I_{b_{|B|}}|  \stackrel{\eqref{proofDef|Rsub|}}{=} |R_{sub}|$, which contradicts to $|R'| \geq |R_{sub}| + 1$.
	
	Therefore, the set 
	$G = \{j \leq |B|: \sum_{i=1}^{j} |R'_i| \geq \sum_{i=1}^{j} |F_i| + 1 \}$ must be non-empty as $|B|$ must be in it.
	As a result, we can define
	\begin{align} \label{proofDefu}
		u = \argmin G.
	\end{align}
	We will show that  $|\cup_{i\leq u} R'_{i}| \geq k_{b_u}$. 
	
	We prove two cases where $u=1$ and $u>1$.
	
	\textbf{In the case of $u=1$}, we have $| R'_{1}| \stackrel{\eqref{proof:substep1}}{\geq} |F_1| + 1 = k_{b_1}$,
	where the last equality is by the construction of $F_1$.
	
	\textbf{In the case of $u>1$}, we have 
	$|\cup_{i\leq u-1} R'_{i}| \leq |\cup_{i\leq u-1} F_{i}|$ by the definition of $u$. 
	Then, we must have $|F_u| = k_{b_u} -|\cup_{i\leq u-1} F_{i} | - 1$.
	Otherwise, by the construction of $F_u$, we have $|F_u| = |E_u|$, so $R'_u \subseteq E_u = F_u$ and $|R'_u| \leq |F_u|$.
	As a result, 
	$|\cup_{i\leq u} R'_{i}| = |\cup_{i\leq u-1} R'_{i}| + |R'_u| \leq |\cup_{i\leq u-1} F_{i}|+|F_u| = |\cup_{i\leq u} F_{i}|$,
	which contradicts to
	$|\cup_{i\leq u} R'_{i}| \geq |\cup_{i\leq u} F_{i}|+1$.
	Based on $|F_u| = k_{b_u} -|\cup_{i\leq u-1} F_{i} | - 1$,
	we then have $|\cup_{i\leq u} R'_{i}| \geq |\cup_{i\leq u} F_{i}|+1 = k_{b_u}$.
	
	Combining the above two cases, we have proved that
	\begin{align} \label{proof:substep2}
		|\cup_{i\leq u} R'_{i}| \geq k_{b_u}.
	\end{align}

	Finally, for any $R'_{sup} \supseteq R'$ and for the $u$ defined by~\eqref{proofDefu},
	we have 
	\begin{align*}
		S_{b_u}^{R'_{sup}} 
		\stackrel{\eqref{LocalTestInequality}}{\geq}
		&|R'_{sup} \cap S^I_{b_u}| \\
		\geq &|R' \cap S^I_{b_u}| \\
		= & |\cup_{i\leq |B|} R'_{i} \cap S^I_{b_u}| \\
		\geq &|\cup_{i\leq u} R'_{i} \cap S^I_{b_u}| \\
		= &|\cup_{i\leq u} R'_{i} \cap R \cap S^I_{b_u}| \\
		 = &|\cup_{i\leq u} R'_{i}| \\
		 \stackrel{\eqref{proof:substep2}}{\geq} &k_{b_u},
	\end{align*}
	where first equality is by the definition of $R'$
	and the second last equality is due to
	$ \cup_{i\leq u} R'_{i} \subseteq \cup_{i \leq u} E_{i} = R \cap S^I_{b_u}$.
	This means that all supersets of $R'$ must be locally rejected, which implies that $H_{R'}$ is rejected by closed testing. 
	Hence, we have proved that any other subset of $R$ which is strictly larger than $H_{R_{sub}}$ must be rejected by closed testing.
	
	Combining Step 2-1 and Step 2-2, we have shown that $R_{sub}$ defined by \eqref{proofDefRsub} is the largest subset of $R$ that is not rejected by closed testing.
	As a result, we have
	\begin{align} \label{proofDef|tct|}
	t^I_{ct}(R) = |R_{sub}| = \sum_{i=1}^{|B|} |F_i| + |R \backslash S^I_{b_{|B|}}|.
	\end{align}

	\textbf{Step 3: Prove $t^I_{ct}(R) = t^I(R)$ based on the explicit formula of $t^I_{ct}(R)$.}
	
	We first present the following result that will be referenced in subsequent proofs.
	For $ 0 < l \leq |B|-1$, we have 
	\begin{align*}
	(R \backslash S^I_{b_l}) \backslash (R \backslash S^I_{b_{|B|}}) 
	&= (R \cap (S^I_{b_l})^c) \cap (R \cap (S^I_{b_{|B|}})^c)^c 
	= R \cap (S^I_{b_l})^c \cap S^I_{b_{|B|}} \\
	&= (R \cap S^I_{b_{|B|}} ) \cap (R^c \cup (S^I_{b_l})^c) 
	= (R \cap S^I_{b_{|B|}} ) \backslash (R \cap S^I_{b_l})
	=\cup_{l < i \leq |B|} E_{i},
	\end{align*}
	so
	\begin{align} \label{sumOfEinequality}
	|R \backslash S^I_{b_l}| - |R \backslash S^I_{b_{|B|}}| 
	= |(R \backslash S^I_{b_l}) \backslash (R \backslash S^I_{b_{|B|}})| 
	= |\cup_{l < i \leq |B|} E_{i}|
	= \sum_{l < i \leq |B|} |E_{i}|.
	\end{align}
	
	For the convenience of notation, let 
	\[
	t^I_i(R) = \min \{ |R \backslash S^I_{i}| + k_i - 1, |R| \},
	\]
	so $t^I(R) = min_{i \in [m]} t^I_i(R)$.
	
	Below are two useful results that will be used in the remaining proofs:
	\begin{itemize}
		\item[(i)] For $i \in B^c$, we have $|R| = |R \backslash S^I_{i}| + |R \cap S^I_{i}| \leq |R \backslash S^I_{i}| + k_i -1$,
		so $t^I_i(R) = |R|$.
		\item[(ii)] For $i \in B$, we have $|R| = |R \backslash S^I_{i}| + |R \cap S^I_{i}| \geq |R \backslash S^I_{i}| + k_i$,
		so $t^I_i(R) = |R \backslash S^I_{i}| + k_i - 1$.
	\end{itemize}

	To show $t^I_{ct}(R) = t^I(R)$, it is sufficient to show that: (i) $t^I_{ct}(R) \leq t^I_{i}(R)$ for all $i \in [m]$, which implies that $t^I_{ct}(R) \leq min_{i \in [m]} t^I_i(R) = t^I(R)$,
	and (ii) $t^I_{ct}(R) = t^I_l(R)$ for some $l \in [m]$, which implies that $t^I_{ct}(R) \geq min_{i \in [m]} t^I_i(R) = t^I(R)$.
	We prove these two claims in the following.
	
	\textbf{Step 3-1: Prove  $t^I_{ct}(R) \leq t^I_{i}(R)$ for all $i \in [m]$.}
	
	\textbf{In the case of $i \in B^c$}, we have $ t^I_i(R)= |R| \geq t^I_{ct}(R)$.
	
	\textbf{In the case of $i \in B$}, we have $t^I_i(R) = |R \backslash S^I_i| + k_i-1$.
	Denote $i = b_l$. 
	
	\textbf{In the sub-case of $l=|B|$ (note that this case covers the scenario that $|B|=1$)}, we have $i = b_{|B|}$, so
	\begin{align*}
	t^I_{ct}(R) - t^I_i(R) 
	\stackrel{\eqref{proofDef|tct|}}{=} 
	\sum_{j=1}^{|B|} |F_j| + |R \backslash S^I_{b_{|B|}}| - |R \backslash S^I_{b_{|B|}}| - k_{b_{|B|}} + 1 
	= \sum_{j=1}^{|B|} |F_j| - k_{b_{|B|}} + 1
	\leq 0,
	\end{align*}
	where the last inequality holds because $\sum_{j=1}^{|B|} |F_j| = |\cup_{j\leq |B|} F_j| \leq k_{b_{|B|}} - 1$.
	
	\textbf{In the sub-case of  $0 < l \leq |B|-1$},
	we have 
	\begin{align*}
	t^I_{ct}(R) - t^I_i(R) 
	&\stackrel{\eqref{proofDef|tct|}}{=}  
	\sum_{j=1}^{|B|} |F_j| + |R \backslash S^I_{b_{|B|}}| - |R \backslash S^I_{b_l}| - k_{b_l} + 1  \\
	&\stackrel{\eqref{sumOfEinequality}}{=}
	 \sum_{j=1}^{|B|} |F_j| - k_{b_l} + 1 - \sum_{l < j \leq |B|} |E_{j}| \\
	&\leq \sum_{j\leq l} |F_j| + \sum_{l < j \leq |B|} |E_j| - k_{b_l} + 1 - \sum_{l < j \leq |B|} |E_{j}| \\
	&\leq \sum_{j\leq l} |F_j| - k_{b_l} + 1
	\leq 0.
	\end{align*}
	Combining the above cases, we have proved that $t^I_{ct}(R) \leq t^I_{i}(R)$ for all $i \in [m]$.
	
	
	\textbf{Step 3-2: Prove that there exists some $l \in [m]$ such that $t^I_{ct}(R) = t^I_l(R)$.}
	
	\textbf{In the case of $|B|=1$},
	let $l = b_1 \in B$, then
	\begin{align*}
		t^I_{ct}(R) - t^I_{l}(R) 
		&\stackrel{\eqref{proofDef|tct|}}{=}  
		 |F_1| + |R \backslash S^I_{b_{1}}|- |R \backslash S^I_{b_{1}}| - k_{b_{1}} + 1  \\
		&= |F_1| - k_{b_{1}} + 1
		=0,
	\end{align*}
	where the last equality holds because $|F_1| = k_{b_{1}} + 1$ by the construction of $F_1$.
	
	\textbf{In the case that $|B|\geq2$ and $|F_i| = |E_i|$ for all $1< i \leq |B|$},
	let $l = b_1 \in B$, 
	so $t^I_l(R) = k_{b_1} - 1 + |R \backslash S^I_{b_1}|$.
	Then,
	\begin{align*}
	t^I_{ct}(R) - t^I_l(R) 
	\stackrel{\eqref{proofDef|tct|}}{=}  & \sum_{i=1}^{|B|} |F_i| + |R \backslash S^I_{b_{|B|}}| - k_{b_1} + 1 - |R \backslash S^I_{b_1}| \\
	\stackrel{\eqref{sumOfEinequality}}{=}  & |F_1| + \sum_{1< i \leq |B|} |F_i| - k_{b_1} + 1 - \sum_{1< i \leq |B|} |E_{i}| = 0.
	\end{align*}

    \textbf{In the case that $|B|\geq2$ and $|F_j| = k_{b_j} -|\cup_{i\leq j-1} F_{i} | - 1$ for some $1< j \leq |B|$},
    let $j'$ be the largest among such $j$ and let $l = b_{j'} \in B$, so $t^I_l(R) = k_{b_{j'}} - 1 + |R \backslash S^I_{b_{j'}}|$
	and $|\cup_{i\leq {j'}} F_{i} | = \sum_{i=1}^{j'} |F_i| = k_{b_{j'} } - 1$.
	
	\textbf{In the sub-case that $j' = |B|$},
	we have
	\begin{align*}
	t^I_{ct}(R) - t^I_{l}(R) 
	&\stackrel{\eqref{proofDef|tct|}}{=}  \sum_{i=1}^{|B|} |F_i| + |R \backslash S^I_{b_{|B|}}| - k_{b_{|B|}} + 1 - |R \backslash S^I_{b_{|B|}}| \\
	&= \sum_{i=1}^{|B|} |F_i| - k_{b_{|B|}} + 1
	=0.
	\end{align*}

	\textbf{In the sub-case that $j' < |B|$},
	we have $|F_i| = |E_i|$ for all $j' < i \leq |B|$. Thus
	\begin{align*}
	t^I_{ct}(R) - t^I_{l}(R) 
	& \stackrel{\eqref{proofDef|tct|}}{=}  \sum_{i=1}^{|B|} |F_i| + |R \backslash S^I_{b_{|B|}}| - k_{b_{j'}} + 1 - |R \backslash S^I_{b_{j'}}| \\
	&= \sum_{i\leq {j'}} |F_i| + \sum_{  j'< i \leq |B| } |F_i| - k_{b_{j'}} + 1 - \sum_{j'< i \leq |B|} |E_{i}|
	=0.
	\end{align*}

    Combining all above cases, we have proved that there exists some $l \in [m]$ such that $t^I_{ct}(R) = t^I_l(R)$.

    Finally, by Step 3-1 and Step 3-2, we have
	\begin{align*}
	t^I_{ct}(R) = t^I(R).
	\end{align*}
	
\end{proof}

\subsection{Proof of Corollary~\ref{coro:KR-connect-CL}} \label{supp:proofcoro:KR-as-CL}

\begin{proof}
    By applying Proposition~\ref{prop:KR-Algo1} and Theorem~\ref{thm:Algo-BasedOnJointKFWER-as-CL},
	$\barFDP^{\KR}(\cdot)$ is equal to the FDP bound based on closed testing using local test statistic~\eqref{formula:weighted-sum-test-forEquivalencePart} with 
	$m=p$, $b^I_i=k^{raw}_{i}+i-1$ and critical values $z^I_i = k^{raw}_{i}$ (for testing null hypothesis $H_I$).
	Because $k^{raw}_{i} \geq 1$, we have $b^I_i \geq i$, 
	so using $m=|I|$ and $m=p$ give the same test result.
\end{proof}

\subsection{Proof of Proposition~\ref{prop:Algo-JS-as-CL}} \label{supp:proofprop:Algo-JS-as-CL}
\begin{proof}
	For $m=1$ and $b^I_1=k+v-1$, the local test statistic is
	\[
	\whL^{I}(k+v-1) = \sum_{i \in I} \Ind_{r^{I}_i > |I|- (k+v-1)} \Ind_{D_i=1}.
	\]
	
	In the case of $|\whS^{\JS}(k)| \leq k-1$, 
	we have $\whL^{\whS^{\JS}(k)}(k+v-1) = |\whS^{\JS}(k)| \leq k-1 < z^I_1$, so $\whS^{\JS}(k)$ is not locally rejected,
	which implies that $t_{\alpha}( \whS^{\JS}(k) ) = |\whS^{\JS}(k)|$. 
	In particular, $t_{\alpha}( \whS^{\JS}(k) ) = k-1$ when $|\whS^{\JS}(k)| = k-1$.
	
	In the case of $|\whS^{\JS}(k)| \geq k$, for any superset $I$ of $\whS^{\JS}(k)$, 
	we have that the number of positive $W_i$ before $v$ negative $W_i$ for $i \in I$ is larger than or equal to
	$k$, which is equivalent to $\whL^{I}(k+v-1) \geq k = z^I_1$ (see \eqref{fact1} in Appendix~\ref{supp:Afact}). So $\whS^{\JS}(k)$ is rejected by closed testing.
	Let $\whS' \subseteq \whS^{\JS}(k)$ with $|\whS'|=k-1$, then we have $\whL^{\whS'}(k+v-1) = |\whS'| = k-1 < z^I_1$,
	so $\whS'$ is not locally rejected, which implies that $t_{\alpha}( \whS^{\JS}(k) ) = k-1$. 
\end{proof}
\section{Supplementary materials for simulations} \label{supp:simulations}

\subsection{Simulation details of Figure~\ref{Fig:illustration-Algo2-vary-k}}\label{supp:simu-detail-Fig1}

For all settings, the knockoff statistic vector $\bW$ is generated based on the linear regression model as described in Section~\ref{sec:simu:Algo1AndKR}. The considered settings are as follows:
\begin{itemize}
	\item Setting 1: $p=200$, $n=500$, $s=0.1$, $a=10$.
	\item Setting 2: $p=200$, $n=500$, $s=0.3$, $a=10$.
	\item Setting 3: $p=1000$, $n=2500$, $s=0.1$, $a=8$.
	\item Setting 4: $p=1000$, $n=2500$, $s=0.3$, $a=8$.
\end{itemize}
For the simultaneous FDP bounds~\eqref{FDPbound-KFWER}, we use $\whS^{\JS}(k)$ obtained by a randomization scheme. This scheme is more powerful than the non-randomized version by dealing with the 
issue that the $\alpha$-level in \eqref{v:OPT-kfwer} may not be exhausted. See \cite{LJ-WS:2016} for details.

\subsection{More simulation results for $\bk^{raw}$, $\bk^{step1}$ and $\bk^{step2}$ in Algorithm~\ref{Algo-getTuningParameterK} using the four types of $\bv$}\label{supp:simu-Algo2-4-v}

\begin{figure}[h!]
	\centering
	\includegraphics[width=10cm, height=10cm]{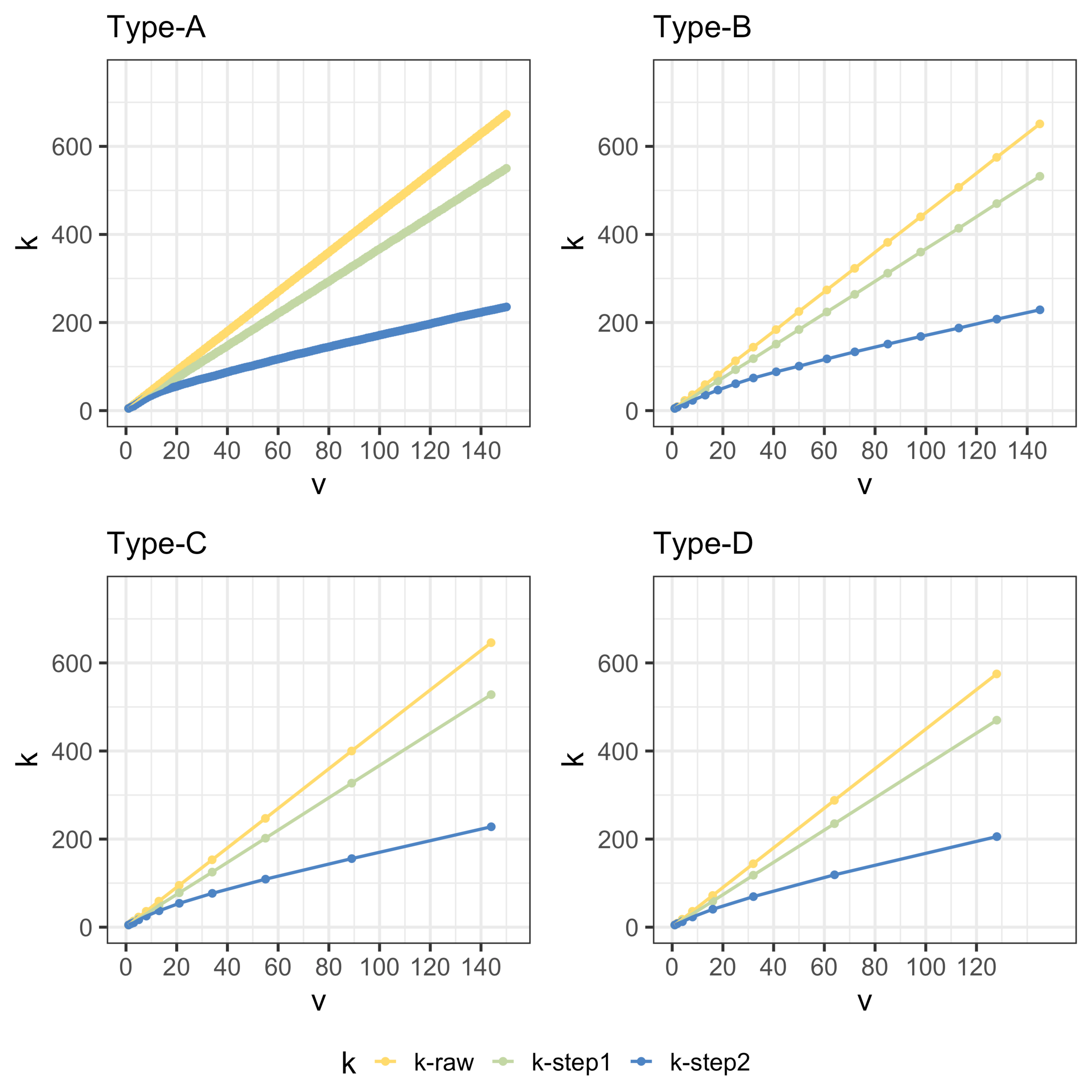}
	\caption{The $\bk^{raw}$, $\bk^{step1}$ and $\bk^{step2}$ in Algorithm~\ref{Algo-getTuningParameterK} for the four types of $\bv$ in \eqref{4-types-of-v}.}
	\label{Fig:4-types-of-v-k}
\end{figure}

In Figure~\ref{Fig:4-types-of-v-k}, we present the values of $\bk^{raw}$, $\bk^{step1}$ and $\bk^{step2}$ by Algorithm~\ref{Algo-getTuningParameterK}
for the four types of $\bv$ described in \eqref{4-types-of-v}.
One can see that for all four types, we have $\bk^{step2} \leq \bk^{step1} \leq \bk^{raw}$, which implies that using $\bk^{step2} $ will lead to better (i.e., smaller) FDP bounds.

\subsection{More simulation results for KR and KJI in Section~\ref{sec:simu:Algo1AndKR} }\label{supp:simu-KR and KJI}

\subsubsection{Empirical Type-I error}
\begin{figure}[h!]
	\centering
	\includegraphics[width=10cm, height=10cm]{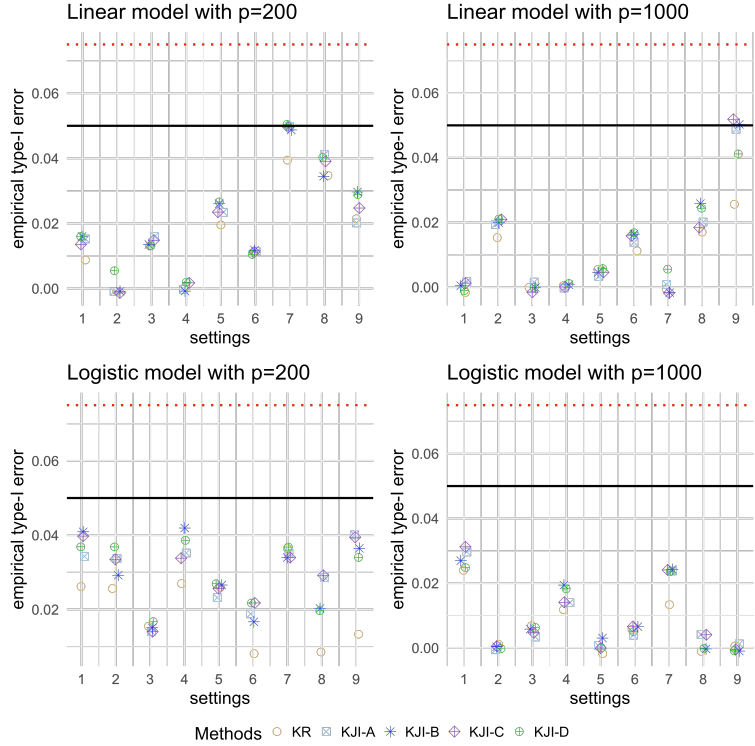}
	\caption{Empirical Type-I errors of KR and KJI in different settings. The solid black line indicates the nominal level $0.05$.
		The dotted red line indicates the $95\%$ confidence upper bound of the true Type-I error over $200$ replications.}
	\label{Fig:error1}
\end{figure}

To check the simultaneous FDP guarantee~\eqref{simuFDPbounds},
it is equivalent to check the Type-I error guarantee
\[
\Prob( \FDP(R) >\barFDP(R), \exists R \subseteq [p]) \leq \alpha,
\]
where $\alpha=0.05$ in our case.
The type-I error plots of KR and KJI are presented in Figure~\ref{Fig:error1}.
Here different settings in the x-axis correspond to different combinations of the setting parameters $a$ and $s$.
As expected, all methods control the Type-I error.

\subsubsection{Simulations with $p=1000$ and  $n=2500$}
Figures~\ref{Fig:linear-p1000} and \ref{Fig:logistic-p1000} show the FDP bounds of KR and KJI in the linear and logistic regression settings, respectively. Here $p=1000$ and  $n=2500$, and we use the tuning parameter vector $\bv$ with $v_m < 300$.

\begin{figure}
	\centering
	\includegraphics[width=8.5cm, height=8.5cm]{Plots/Supp/Linear-p1000-main.pdf}
	\caption{Simultaneous FDP bounds of KR and KJI with four types of $(\bv, \bk)$ in the linear regression setting with $p=1000$ and $n=2500$. The dashed black line indicates the true FDP. All FDP bounds are the average values over $200$ replications.}
	\label{Fig:linear-p1000}
\end{figure}

\begin{figure}
	\centering
	\includegraphics[width=8.5cm, height=8.5cm]{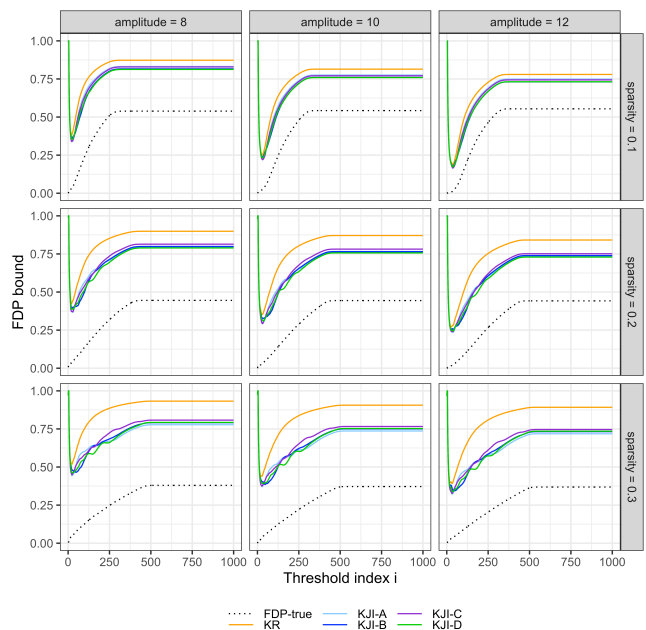}
	\caption{Simultaneous FDP bounds of KR and KJI with four types of $(\bv, \bk)$ in the logistic regression model with 
		$p=1000$ and $n=2500$. The dashed black line indicates the true FDP. All FDP bounds are the average values over $200$ replications.}
	\label{Fig:logistic-p1000}
\end{figure}

\subsubsection{High-dimensional case}
\begin{figure}
	\centering
	\includegraphics[width=8.5cm, height=8.5cm]{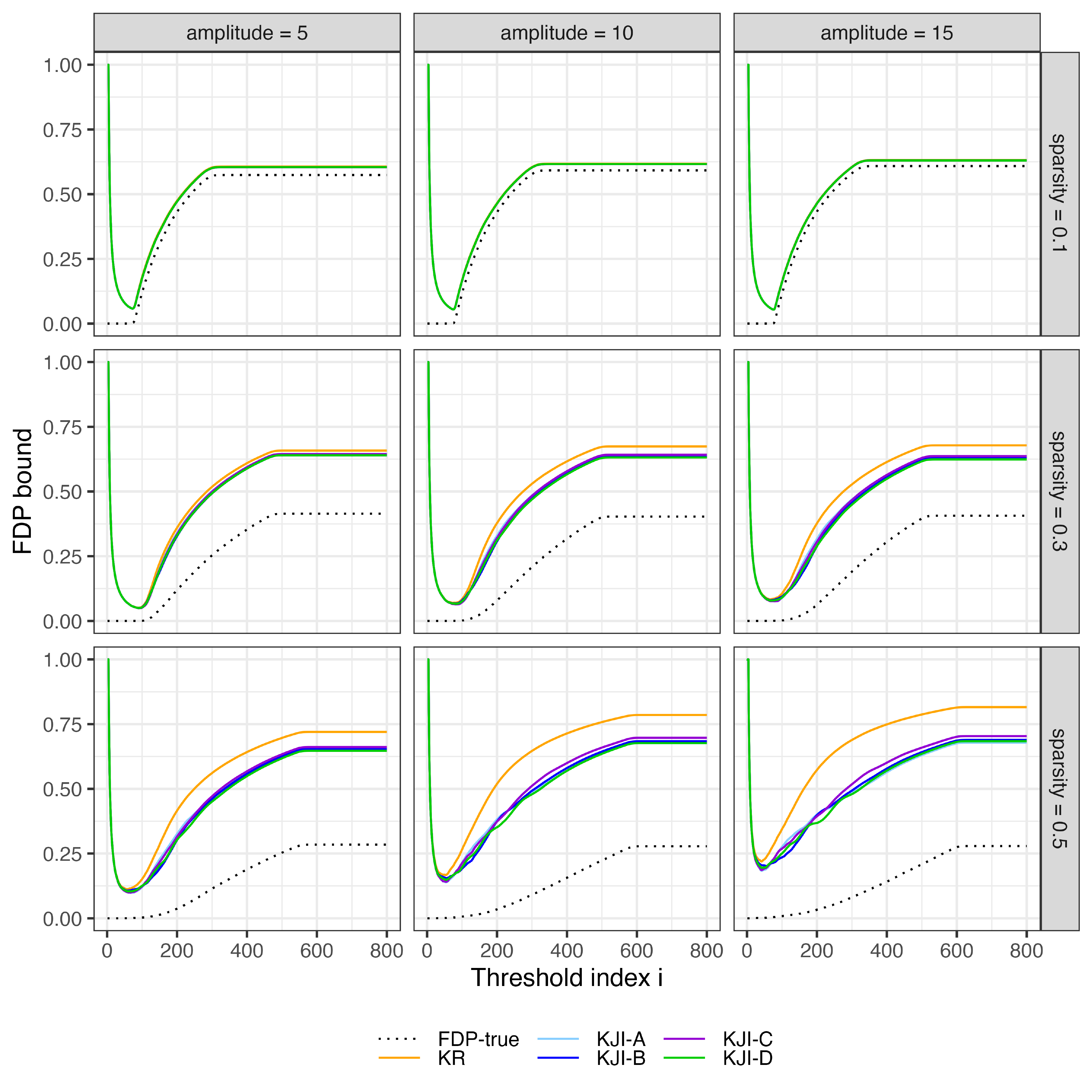}
	\caption{Simultaneous FDP bounds of KR and KJI with four types of $(\bv, \bk)$ in the high dimensional linear regression model with 
		$p=800$ and $n=400$. The dashed black line indicates the true FDP. All FDP bounds are the average values over $200$ replications.}
	\label{Fig:highDim}
\end{figure}

In the high-dimensional linear regression scenario with $p=800$ and  $n=400$,
we consider a regression coefficient vector $\beta \in \bbR^{p \times 1}$ with consecutive non-zero entries.
Specifically, we first randomly sample a starting index $u$ from $\{p(1-s) \dots, p\}$, where $s \in \{0.1, 0.3, 0.5\}$ is the sparsity level.
Then, starting from index $u$, we set the next $ps$ entries to be non-zero, with both weak and strong signals. 
In particular, the values of the first half of these non-zero entries are randomly sampled from $\text{Unif}(2,a)$ with $a \in \{5,10,15\}$. The second half are  randomly sampled from $\text{Unif}(15,20)$.
In each replication of the simulation,
we generate $X \in \bbR^{n \times p}$ by drawing 
$n$ i.i.d.\ samples from the multivariate Gaussian distribution $N_p(0,\Sigma)$ with $(\Sigma)_{i,j}=0.6^{|i-j|}$,
and then sample $Y \sim N_n(X\beta,I)$. 
To construct knockoff statistic vector $W$ based on the generated data $(X, Y)$, 
we employ the model-X approach using ``equicorrelated" Gaussian knockoffs.
Subsequently, we apply elastic net \citep{zou2005regularization} with mixing parameter $0.9$ and regularization parameter determined by taking the 90th percentile of the regularization parameter vector returned by the Python function ``enet\_path" in the ``sklearn" package.
We use the tuning parameter vector $\bv$ with $v_m < 300$.
The simulation results are shown in Figures~\ref{Fig:highDim}.

\subsection{More simulation results for KJI and KCT in Section~\ref{sec:simu:Algo1AndCT}}\label{supp:simu-KJI and KCT}

\begin{figure}
	\centering
	\includegraphics[width=8.5cm, height=8.5cm]{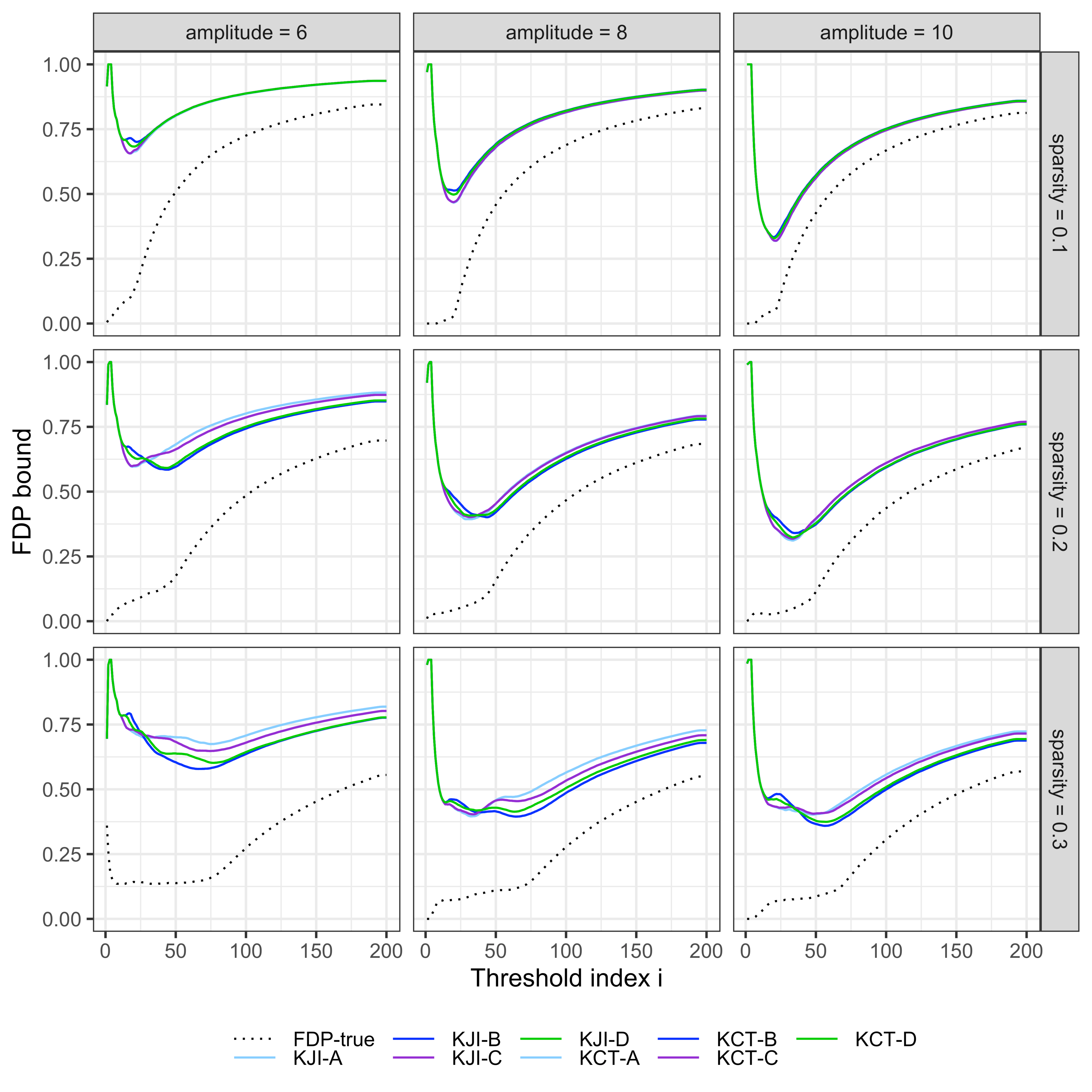}
	\caption{Simultaneous FDP bounds of KJI and KCT with four types of $(\bv, \bk)$ in the linear regression setting with $p=200$ and $n=500$. The dashed black line indicates the true FDP. All FDP bounds are the average values over $200$ replications.}
	\label{Fig:linear-p200-KCT}
\end{figure}

\begin{figure}
	\centering
	\includegraphics[width=8.5cm, height=8.5cm]{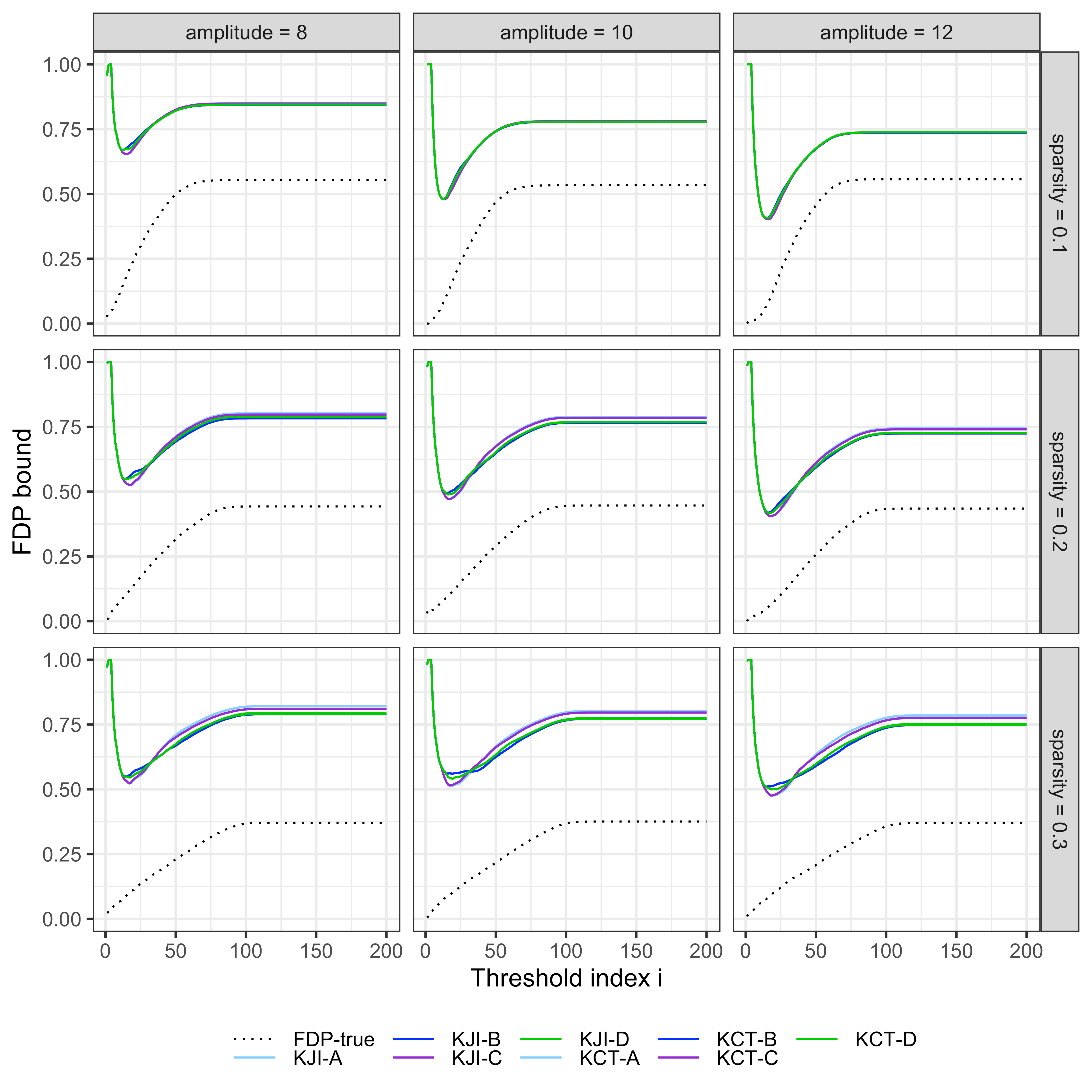}
	\caption{Simultaneous FDP bounds of KJI and KCT with four types of $(\bv, \bk)$ in the logistic regression model with 
		$p=200$ and $n=500$. The dashed black line indicates the true FDP. All FDP bounds are the average values over $200$ replications.}
	\label{Fig:logistic-p200-KCT}
\end{figure}

Figures~\ref{Fig:linear-p200-KCT} and \ref{Fig:logistic-p200-KCT} show the simulation results of KJI and KCT with four types of $(\bv, \bk)$ in the same linear and logistic regression settings as in Section~\ref{sec:simu:Algo1AndKR} with $p = 200$ and $n = 500$. 
By looking at the plots, the FDP bounds of KJI and KCT with the same $(\bv, \bk)$ are basically coincides (there are numerical differences but can hardly be seen from the plots). 

\subsection{Simulations to compare the closed testing based approaches with different local tests} \label{sec:simu:CTandRankCT}

In this appendix, we examine the performance of the closed testing method with the rank-generalization local test proposed in Section~\ref{sec:UniformlyImprovementClosedTesting}, and compare it to KCT (see Section~\ref{sec:simu:Algo1AndCT}).
Specifically, we consider the following rank-generalization approach:
\begin{itemize}
	\item KCT-rank: Our proposed method to obtain simultaneous FDP bounds based on closed testing using local test statistic is \eqref{formula:weighted-sum-test} with $w^I_{i,j}= r^I_j \Ind_{r^I_j > |I|-b^I_i}$, where the four types of $b^I_i$ are the same as KCT, and the critical value is approximated based on \eqref{formula:criticalvalueLocaltest}. 
	We denote them as KCT-rank-A, KCT-rank-B, KCT-rank-C and KCT-rank-D.
	We use the shortcut Algorithm~\ref{Algo-ct-shortcut} for the implementation of closed testing.
\end{itemize}

\begin{figure}
	\centering
	\includegraphics[width=8cm, height=8cm]{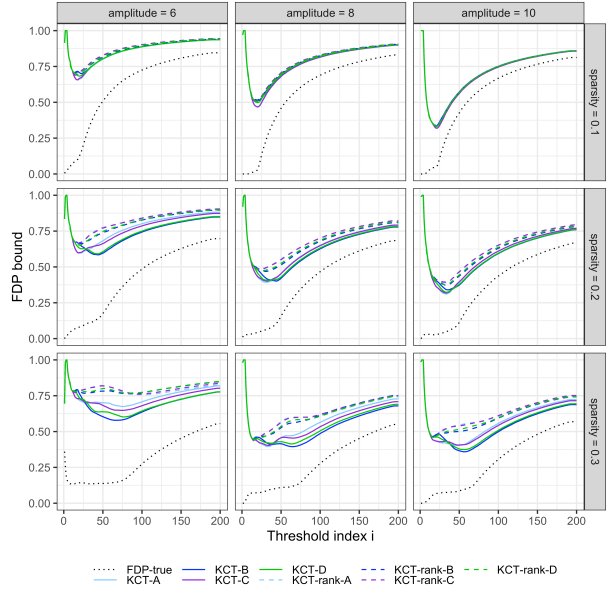}
	\caption{Simultaneous FDP bounds of KCT and KCT-rank in the linear regression setting with $p=200$ and $n=500$. The dashed black line indicates the true FDP. All FDP bounds are the average values over $200$ replications.}
	\label{Fig:linear-rankCT-p200}
\end{figure}

\begin{figure}
	\centering
	\includegraphics[width=8.5cm, height=8.5cm]{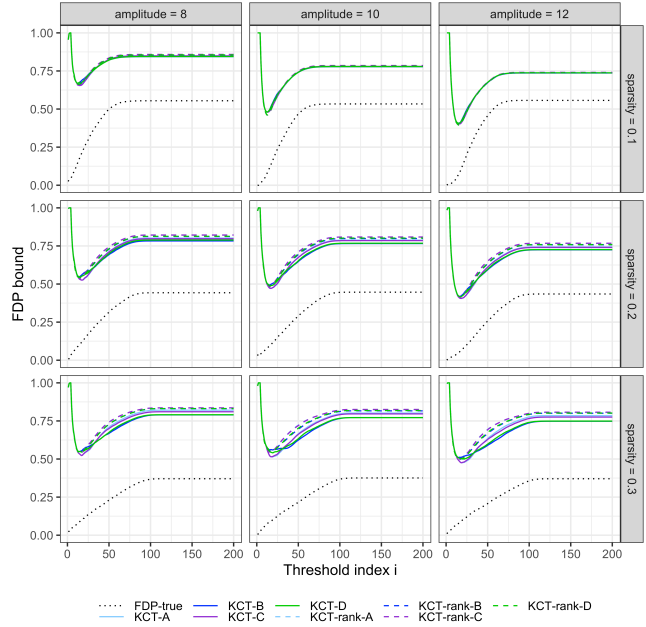}
	\caption{Simultaneous FDP bounds of KCT and KCT-rank in the logistic regression setting with $p=200$ and $n=500$. The dashed black line indicates the true FDP. All FDP bounds are the average values over $200$ replications.}
	\label{Fig:logistic-rankCT-p200}
\end{figure}

We first implement KCT and KCT-rank in the same linear and logistic regression settings as in Section~\ref{sec:simu:Algo1AndKR} with $p=200$ and $n=500$.
The simulation results are shown in Figure~\ref{Fig:linear-rankCT-p200} and \ref{Fig:logistic-rankCT-p200}. 
One can see that KCT is generally better than KCT-rank.
Possible explanation for this observation is that KCT also makes use of the information of rank because $|W_i|$'s are ordered.
So the rank-generalization version might not give much better FDP bounds when the information of rank is good. 
On the other hand, when the information of rank is bad, it will give worse FDP bounds than KCT because it is influenced more by using rank as weight in the local test statistic.

\begin{figure}
	\centering
	\includegraphics[width=8.5cm, height=8cm]{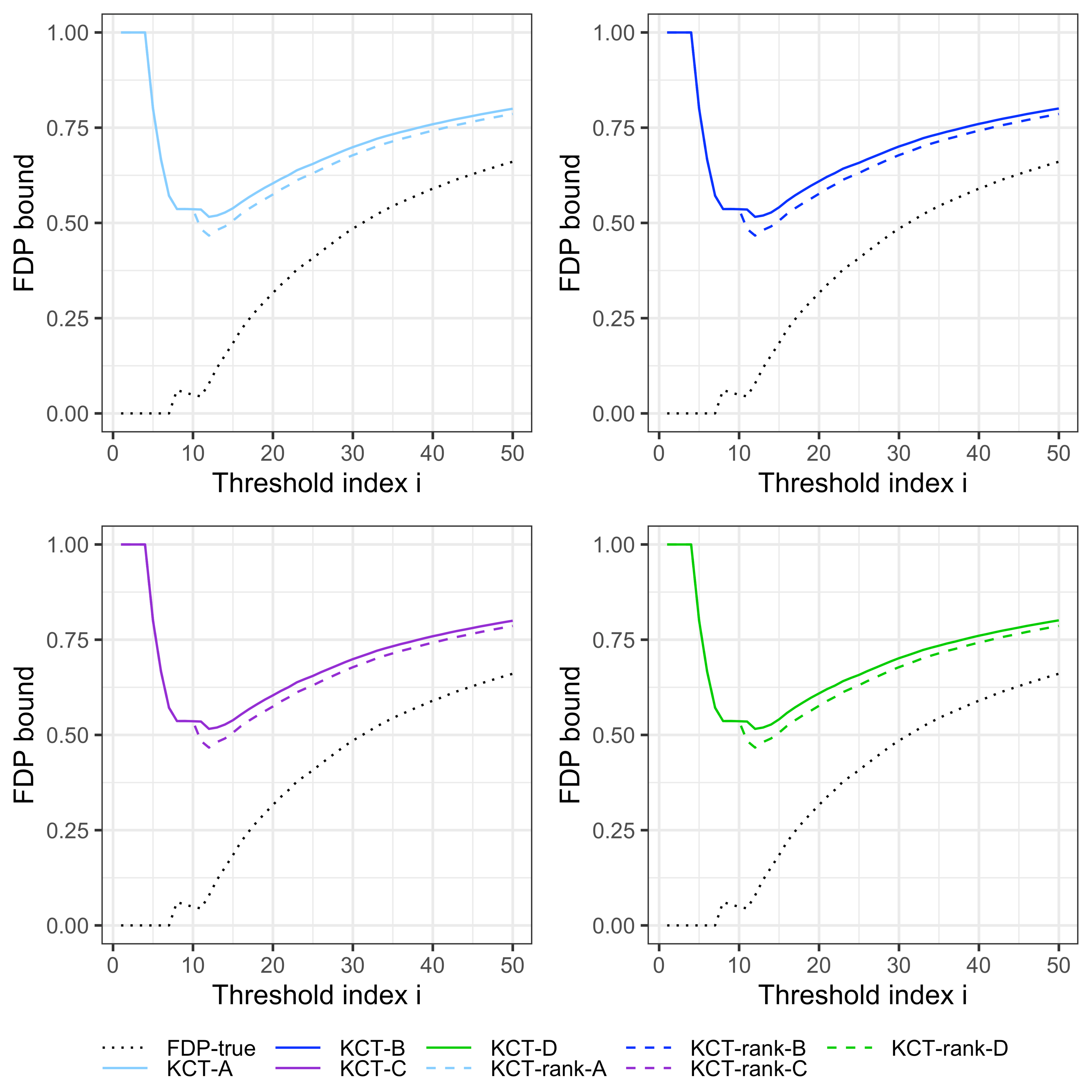}
	\caption{Simultaneous FDP bounds of KCT and KCT-rank in the simulation setting where the knockoff statistic vector $\bW \in \bbR^{50}$ is directly generated.
	The dashed black line indicates the true FDP. All FDP bounds are the average values over $200$ replications.}
	\label{Fig:rankCT}
\end{figure}

To show that KCT-rank can be better than KCT in certain cases, 
we consider a simulation setting directly generating knockoff statistic vector $\bW$.
Specifically, we consider a setting with $p=50$ and null variable set $\mN = \{8, 12,13,\dots,50 \}$.
For the knockoff statistics, we set $|W_i|=50-i+1$, 
$\text{sign}(W_i)=1$ for $i \not\in\mN$
and $\text{sign}(W_i) \stackrel{i.i.d.\ }{\sim} \{-1,1\}$ with same probability $1/2$ for $i\in\mN$.
It is clear that this $\bW$ satisfies the coin-flip property,
so it is a valid knockoff statistic vector.
Figure~\ref{Fig:rankCT} shows the simulation results in this setting,
and one can see that KCT-rank is better than KCT.

\end{document}